\documentclass[letterpaper,useAMS,usenatbib]{mn2e}
\usepackage{graphicx}
\usepackage{natbib}
\usepackage[pass]{geometry}

\title[Resolving structure around HD100546]{Resolving structure of the disk around HD100546 at 7 mm with ATCA}
\author[C.M. Wright, et al.]
{C. M. Wright$^{1}$\thanks{E-mail: c.wright@adfa.edu.au}, 
S. T. Maddison$^{2}$, D. J. Wilner$^{3}$, M.G. Burton$^{4}$, D. Lommen$^{5}$, 
\newauthor E. F. van Dishoeck$^{6}$, P. Pinilla$^{6}$, T. L. Bourke$^{7}$, F. Menard$^{8}$ and C. Walsh$^{6}$\\
$^{1}$School of Physical, Environmental and Mathematical Sciences, UNSW@ADFA, Canberra ACT 2600, Australia\\
$^{2}$Centre for Astrophysics \& Supercomputing, Swinburne University, PO Box 218, Hawthorn, VIC 3122, Australia\\
$^{3}$Harvard-Smithsonian Center for Astrophysics, 60 Garden Street, Cambridge, MA 02138, USA\\
$^{4}$School of Physics, University of New South Wales, Sydney, NSW 2000, Australia\\
$^{5}$Hwa Chong Institution, 661 Bukit Timah Road, Singapore 269734\\
$^{6}$Leiden Observatory, Leiden University, P. O. Box 9531, 2300 RA Leiden, The Netherlands\\
$^{7}$SKA Organisation, Jodrell Bank Observatory, Lower Withington, Macclesfield, Cheshire SK11 9DL, UK\\
$^{8}$Millenium Nucleus 'Protoplanetary Disks in ALMA Early Science', Universidad de Chile, Casilla 36-D, Santiago, Chile\\
}
\begin{document}

\date{version: 26 June 2014}

\pagerange{\pageref{firstpage}--\pageref{lastpage}} \pubyear{2009}

\maketitle

\label{firstpage}

\begin{abstract}

There is much evidence that planet formation is occurring in the disk around the 
Herbig Be star HD100546. To learn more about the processes occurring in this disk 
we conducted high resolution imaging at 43/45 GHz with the Australia Telescope 
Compact Array (ATCA). Multiple array configurations were used, providing a best 
spatial resolution of $\sim$ 0.15\arcsec, or 15~AU at HD100546's distance of 
$\sim$~100~pc. Significant structure is revealed, but its precise form is dependent 
on the $u-v$ plane sampling used for the image reconstruction. At a resolution of 
$\leq$ 30~AU we detected an inner gap in the disk with a radius of $\sim$ 25~AU and 
a position angle approximately along the known disk major axis. With different 
weighting, and an achieved resolution of $\sim$ 15~AU, emission appears at the centre 
and the disk takes on the shape of an incomplete ring, much like a horseshoe, again 
with a gap radius of $\sim$ 25~AU. The position angle of the disk major axis and 
its inclination from face-on are determined to be $140^{\circ}\pm5^{\circ}$ and 
$40^{\circ}\pm5^{\circ}$ respectively. The $\sim$ 25~AU gap radius is confirmed 
by a null in the real part of the binned visibilities at 320$\pm$10~k$\lambda$,
whilst the non-axisymmetric nature is also confirmed through significant structure 
in the imaginary component. The emission mechanism at the central peak is most 
likely to be free-free emission from a stellar or disk wind. Overall our data 
support the picture of at least one, but probably several, giant planets orbiting 
HD100546 within 25~AU.  

\end{abstract}


\section{Introduction}

As well as being the reservoir of material accreting onto young stars, disks also 
provide the raw material for the building of a planetary system. A diverse range 
of physical and chemical processes must however occur before the dilute mixture of 
gas and dust is transformed into discrete rocky and gas planets. Yet despite the 
discovery of thousands of extrasolar planets over the last few decades, with an 
exotic suite of orbital and physical properties, plus detailed studies of our own 
solar system, our understanding of the planet formation process remains incomplete.  

In order to learn more about the early stages of planet formation we have 
undertaken an extensive observing program from the millimetre to centimetre range of 
several Herbig star+disk systems, including HD100546, using the Australia Telescope 
Compact Array (ATCA)\footnote{The Australia Telescope Compact Array is part of the 
Australia Telescope which is funded by the Commonwealth of Australia for operation as 
a National Facility managed by CSIRO.}. Observations at millimetre wavelengths are 
especially important as they probe the colder regions of the disk -- the outer parts 
and midplane -- where the bulk of the mass resides (e.g. Beckwith \& Sargent 1991). 

HD100546 is a Herbig Be star which displays a range of phenomena suggesting it may be 
the host of a planetary system in the early stage of its formation and evolution. Its 
age is not well determined, but is likely to be in the range of 3.5--10 Myr. The lower 
end comes from Manoj et al. (2006) and the upper from van den Ancker et al. (1997).  
The old and new Hipparcos distances are 103$^{+7}_{-6}$~pc \citep{vandenAncker-etal1997} 
and 97$\pm$4~pc (van Leeuwen 2007), but we simply use 100~pc throughout this paper.
It is situated on the edge of, and associated with, the dark cloud 
DC296.2--7.9 \citep{Hu-etal1989, Vieira-etal1999}. HD100546 and its environs have been 
extensively studied over the last 20 years as its potential similarity to the early solar 
nebula has been gradually revealed. Its close proximity allows many phenomena to be 
resolved at the few to few tens of AU scale. 

An optical and near-IR scattering disk has been detected from about 15~AU out to 500~AU by 
many authors, including Pantin et al. (2000), Augereau et al. (2001) and Quanz et al. (2011). 
Spiral and dark lane structure was also observed within this scattering disk -- from about 
150 to 300~AU -- by Grady et al. (2001), Ardila et al. (2007), Boccaletti et al. (2013) and 
Avenhaus et al. (2014a). Thermal dust emission out to a few tens of AU was detected by 
mid-infrared nulling interferometry and direct imaging observations of the disk by Liu et al. 
(2003), Leinert et al. (2004) and Mulders et al. (2011). 

Besides the several hundred AU radius disk, an extended envelope has also been detected out 
to about 1000~AU via dust scattering in the optical regime by Grady et al. (2001) and 
Ardila et al. (2007). The latter find the colours of the envelope to be similar to those of Kuiper 
Belt Objects (KBOs) in our own solar system. They suggest that the HD100546 disk may be transitional 
between a pre-main sequence disk and one dominated by planetesimal collisions, making it an 
important target for our overall understanding of disk evolution.

HD100546 has a remarkable 2--45 $\mu$m spectrum measured by the Infrared Space Observatory 
(ISO, Malfait et al. 1998). Its spectrum resembles that of comet Hale-Bopp, with a wealth 
of features indicative of a large fraction of crystalline silicates, predominantly forsterite. 
Mulders et al. (2011) estimate a crystalline mass fraction of 40--60\%, located predominantly 
between 13 and 20~AU. This suggests that a process has occurred in the HD100546 disk to 
crystallise the dust, possibly common to that which occurred in our own solar system during 
its formation. 

Perhaps not surprisingly, independent observations have indicated the presence of at least 
one massive body. Grady et al. (2005) detect a central cavity extending about 13~AU from the star, 
much larger than can be explained by dust sublimation and with a centroid offset from the star 
by 5~AU along the disk major axis. They interpret this cavity as having been dynamically sculpted 
by one or more bodies, favouring a giant planet. Liu et al. (2003, 2007) also favour a giant 
(proto)planet based on their inference of a large inner gap at less than 10~AU from mid-infrared 
imaging and nulling interferometry. A 13~AU radius cavity has also been inferred from obervations 
of CO and OH infrared ro-vibrational transitions by several authors (van der Plas et al. 2009; 
Brittain et al. 2014, 2013, 2009; Liskowsky et al. 2012) and is also consistent with radiative 
transfer modelling of the near- to far-IR SED (Bouwman et al. 2003; Benisty et al. 2010; Mulders 
et al. 2011; Tatulli et al. 2011; Mulders et al. 2013a,b).

From spatially and velocity resolved [O I] 6300\AA ~data Acke \& van den Ancker (2006) infer the mass 
of the body clearing the cavity to be around 20 Jupiter masses (hereafter M$_J$), orbiting at about 
6.5~AU. This puts it in the brown dwarf rather than planet realm. Hydrodynamic simulations by Tatulli 
et al. (2011) instead find that a 1-8~M$_{\rm J}$ body placed at 8~AU can produce the 13~AU radius 
cavity within a timescale shorter than the age of the system. Most recently, using a combination of 
coronagraphy and polarimetric imaging Quanz et al. (2013a) claim to have directly detected a giant 
protoplanet at a radius of about 70~AU. Notably this could not be the body responsible for the inner 
clearing, and thus a picture of a planetary system begins to emerge.

There has been comparatively little work on the millimetre properties of the HD100546 disk. 
It was found to be a strong source of 1.3~mm dust emission by Henning et al. (1994, 1998) 
using the SEST single dish telescope. They inferred the presence of a disk -- embedded 
within a core-envelope structure -- since their spherical geometry radiative transfer model 
failed to reproduce the 1.3~mm flux by an order of magnitude. Direct evidence for a compact 
disk was provided by ATCA 3.4~mm observations of Wilner et al. (2003). More recently Walsh 
et al. (2014) and Pineda et al. (2014) presented ALMA observations of the disk at 870 $\mu$m
(plus 990 $\mu$m in Walsh et al.) using the same data set, which provided evidence for a central
disk cavity. 

The 7~mm data we present is part of a much larger observing campaign of HD100546 which we have 
conducted with ATCA over the last decade, ranging from a frequency of 4.8 to 95.5~GHz (wavelength of 
62.5 to 3.1~mm). The main body of this paper considers the high resolution 7~mm imaging, presenting 
the first millimetre picture of the hole within the HD100546 disk. Where appropriate, to support 
conclusions we will quote results from the larger data set, but which itself is confined to several 
appendices.

\section[]{Observations and data reduction}

We used the Australia Telescope Compact Array (ATCA) to observe HD100546 at multiple frequencies 
within the 3~mm, 7~mm and 16~mm bands, as well as at 4.8 and 8.64~GHz. Several array configurations 
were used from 31 May 2002 through to 27 June 2012. Table~\ref{TAB:hd100-obs7mm} contains a log 
of the 7~mm observations, which provided the best imaging data. Tables~\ref{TAB:hd100-obs3mm}, 
\ref{TAB:hd100-obs16mm} and \ref{TAB:hd100-obs36cm} list the 3~mm, 16~mm and 3+6~cm observational 
details. Before 2009 the continuum mode of the ATCA correlator had a bandwidth of 128 MHz with 32 
channels to give 4~MHz per channel in two sidebands. In early 2009 a new correlator was commissioned 
at ATCA, the Compact Array Broadband Backend, or CABB (Wilson et al. 2011). For each sideband this 
provided a bandwidth of 2~GHz with 2024 channels of 1~MHz each. For the 7~mm CABB data the two 
sidebands were centred at 43~GHz and 45~GHz.

Up until May 2010 the complex gains were derived from observations of the quasar PKS B1057-797, 
separated by about 10 degrees on the sky from HD100546's position. In 2010 and 2012 the quasar 
j1147-653 was used, which is much closer on the sky to HD100546 and so provided a better 
representation of the phase conditions along the latter's line-of-sight. The gain calibrator 
was typically observed every 5 to 15~mins for between 1 and 3~mins duration, dependent on 
atmospheric conditions. Pointing checks were also made on the quasar every 60--90~mins. 

The bandpass response was determined from 15 minute observations of one or more of the 
quasars PKS B0537-441, PKS B1253-055 and PKS B1921-293. As seen in Table~\ref{TAB:hd100-obs7mm}
in most cases flux calibration was performed using Uranus, though Mars and ATCA's primary 
cm-band flux calibrator, the quasar 1934-638, were also used. Fifteen minute observations were 
typically made. The secondary calibrators 1057-797 or j1147-653 were used on occasion after 
boot-strapping to one of the aforementioned primary calibrators within a few days or weeks. 
Given their intrinsic (and typically non-periodic) variability this approach obviously has
some risk. See for instance Torniainen et al. (2005). All subsequent data reduction, including 
calibration and imaging, was performed with the Miriad software (Sault et al. 1995). 

Appendix A contains a more detailed description of flux calibration issues, but we point out here 
the first two entries in Table~\ref{TAB:hd100-obs7mm} are the same data set but with a different
planet observation -- taken about 10 hours before and after the science track -- for the flux 
calibration. This shows a change in total flux of between 0.5 and 0.8~mJy. We've chosen the 
second observation as being closer to the true flux since the planet was at a similar elevation 
to the science track. 

\begin{table*}
\caption{7~mm ATCA observations of HD100546.}
\begin{tabular}{llllllllll}
\hline
 Date	& $\nu_1$ & $F_1$ & $\nu_2$ & $F_2$ & Phase cal.         & Flux cal.,         & $T_{\rm int}$ & Config & Min, Max \\ 
 UT	  &  (MHz)  & (mJy) & (MHz)   & (mJy) & elev. ($^{\circ}$) & elev. ($^{\circ}$) & (mins)        &        & baselines (m) \\ 
\hline
 4 Oct 2007 & 41000 & 7.32$\pm$0.17 P & 43000 & 7.93$\pm$0.18 P & 35-40 & Uranus, 65   & 140 & H75 & 31, 89 (4408) \\
            &       & 7.27$\pm$0.30 G &       & 7.97$\pm$0.36 G &       &              &     &     &        \\
 4 Oct 2007 & {\bf 41000} & {\bf 7.84$\pm$0.18 P} & {\bf 43000} & {\bf 8.58$\pm$0.20 P} & 35-40 & Uranus, 50   & 140 & H75 & 31, 89 (4408) \\
            &             & {\bf 7.77$\pm$0.36 G} &             & {\bf 8.80$\pm$0.40 G} &       &              &     &   & \\
 5 Oct 2007 & 41000 & 4.88$\pm$0.31 P & 43000 & 4.96$\pm$0.33 P & 40    & Uranus, 50   & 60  & H75 & 31, 89 (4408) \\
           &        & 5.57$\pm$0.73 G &       & 6.40$\pm$0.83 G &       &              &     &     &        \\
27 Jun 2008 & 42944 & 9.02$\pm$0.41 G & 44992 & 9.61$\pm$0.42 G & 37-40 & Mars, 28     & 260 & 1.5B  & 31, 4301 \\
30 May 2009 & 43000 & 6.10$\pm$0.08 P & 45000 & 6.69$\pm$0.10 P & 38-40 & Uranus, 42   &  60 & H214D &  82, 247 (4500) \\
            &       & 7.94$\pm$0.20 G &       & 8.25$\pm$0.25 G &       &              &     &       &          \\
31 May 2009 & {\bf 43000} & {\bf 7.99$\pm$0.12 P} & {\bf 45000} & {\bf 8.36$\pm$0.16 P} & 35-31 & Uranus, 27 & 60 & H214D & 82, 247 (4500) \\
            &             & {\bf 8.80$\pm$0.27 G} &             & {\bf 8.79$\pm$0.34 G} &      &        &    &      & \\
 1 Aug 2009 & 43000 & 5.47$\pm$0.05 G & 45000 & 5.86$\pm$0.07 G & 40-30 & 1934-638, 48 & 280 & 1.5A &  153, 4469 \\
 3 Aug 2009 & 43000 & 6.81$\pm$0.26 G & 45000 & 7.50$\pm$0.46 G & 25-22 & Uranus, 55   & 80  & 1.5A &  153, 4469 \\
15 Aug 2009 & 43000 & 5.89$\pm$0.17 G & 45000 & 6.67$\pm$0.23 G & 30-22 & Uranus, 42   & 190 & 6D &  77, 5878 \\
26 Aug 2009 & 43000 & 6.56$\pm$0.15 G & 45000 & 7.04$\pm$0.19 G & 28-34 & 1057-797$^{a}$     & 120 & 6D &  77, 5878 \\
 3 May 2010 & 43000 & 3.65$\pm$0.13 G & 45000 & 3.63$\pm$0.16 G & 24-52 & 1934-638, 55 & 383 & 6A &  337, 5939 \\
23 May 2012 & 43000 & 8.42$\pm$0.28 G & 45000 & 9.73$\pm$0.40 G & 13-52 & j1147-653    & 545 & 6D &  77, 5878$^{b}$ \\
20 Jun 2012 & 43000 & 6.73$\pm$0.07 G & 45000 & 7.36$\pm$0.09 G & 12-52 & Uranus, 40   & 495 & 6D &  77, 5878 \\
\hline
\end{tabular}
 \begin{list}{}{}
 \item
$^{a}$ Flux used for 1057-797 was 1.945 Jy at both frequencies from observation on 15 August 2009; 
$^{b}$ The 23 May 2012 observations did not have antenna's 2 and 3 available. 
P and G refer to Point and Gaussian source models in Miriad's UVFIT.
Numbers in bold are our best estimates of the 41, 43 and 45 GHz total fluxes 
of HD100546. A flux of 8.5$\pm$0.5~mJy is recommended at 44~GHz, dominated 
by thermal dust emission but with a minor contribution from free-free emission.
See Appendix C for details.
 \end{list}
\label{TAB:hd100-obs7mm}
\end{table*}


In order to combine the CABB 7~mm data from all array configurations we first determined 
offsets from the phase centre using the UVFIT routine in Miriad and an elliptical gaussian 
model. Such offsets were the result of several factors, including purposefully setting the 
phase centre off the nominal stellar position by some fraction of a synthesised beamwidth 
(to avoid artefacts due to DC offset and/or sampler harmonics errors, as stipulated in the 
ATCA Users Guide\footnote{www.narrabri.atnf.csiro.au/observing/users\_guide/html/atug.html}), 
proper motion of HD100546, and telescope pointing and/or tracking errors. The respective data 
sets were then corrected to a common phase centre using Miriad's UVEDIT. That the input offset 
was successful in each case was demonstrated by the residual offsets all being less than 
$\sim$ 0.04\arcsec, and mostly $\sim$ 0.01\arcsec, when again fit with an elliptical gaussian 
in UVFIT. A similar methodology for combining separate tracks has recently been applied to 
other disks, such as that of TW Hya in Andrews et al. (2012) and LkH$\alpha$ 330 in Isella 
et al. (2013).

Treating each sideband separately the respective configurations were combined with Miriad's UVCAT, 
fourier inverted, deconvolved with CLEAN and restored with a synthesised beam to produce a final 
image. The two independent frequency bands produced extremely consistent images so were combined 
to enhance the signal-to-noise, and the 'effective' frequency is referred to here as 44~GHz. As a 
further consistency check the same process was also applied to all the individual data sets. The 
longest baseline was almost 6 km, and the achieved resolution $\sim$ 0.15\arcsec--0.50\arcsec. This 
corresponds to $\sim$ 15-50 AU at the distance of HD100546, and depended on the weighting applied 
during the inversion. This weighting was varied from natural through to uniform and super-uniform, 
which apply progressively more weight to longer baselines to give higher resolution but at the 
cost of S/N. Our data provides some of the highest resolution millimetre images ever obtained of 
a circumstellar disk.

\section{Results}

\subsection{Fluxes}

Derived fluxes for HD100546 in both sidebands are presented in Table~\ref{TAB:hd100-obs7mm}, 
obtained in the $u-v$ plane with Miriad's UVFIT and point or elliptical Gaussian source models. 
Note that in October the 2007 the frequency pair was 41/43 GHz, whilst subsequently it was 
43/45~GHz. There are some differences in flux between tracks based solely on the statistical 
uncertainties, but it is unlikely to be due to intrinsic source variability. Whilst we have 
observed short term (i.e. hourly, factor of 2--3) flux variability for HD100546 at 4.8 and 
8.64~GHz, we have not detected variability at 19~GHz or 90--95~GHz (see Appendices A and B). 

Thus, the different 7~mm fluxes seen in Table~\ref{TAB:hd100-obs7mm} probably reflect one or 
more of the absolute flux calibration accuracy, 'missing' extended flux in the long baseline 
configurations due to inadequate $u-v$ plane sampling, and/or a Gaussian not being a good 
representation of the source morphology. Specifically note that the fluxes measured on 
5 October 2007 and 3 May 2010 are about half that typically measured. But these observations 
had the worst overall phase stability and in the case of 3 May 2010 the longest minimum 
baseline length.

With these considerations in mind the fluxes measured in the H75 and H214 compact hybrid 
configurations on 4 October 2007 and 31 May 2009 respectively probably represent the best total 
flux estimates. These are in boldface in Table~\ref{TAB:hd100-obs7mm}. The sky conditions were 
good and importantly stable over these relatively short observations, and HD100546 remained 
unresolved. The latter is demonstrated by the fact that there is little difference, within
the respective uncertainties, between the fluxes measured using a point source or elliptical 
Gaussian. A 44~GHz flux of 8.5$\pm$0.5~mJy is recommended. 

Of the extended configurations, the 20 June 2012 track was conducted in the best conditions, 
which were also relatively stable over the course of the observation. Indeed, this track was 
conducted as part of the commissioning of ATCA's water vapour radiometers (WVR), but the 
WVR-based phase correction provided no substantive improvement over that of the quasar-based 
correction (Jones, Indermuehle \& Burton 2012) This data set produced extremely good high resolution 
images, which were clearly superior to those from other tracks.

\subsection{Images}

\begin{figure*}
\includegraphics[scale=0.80]{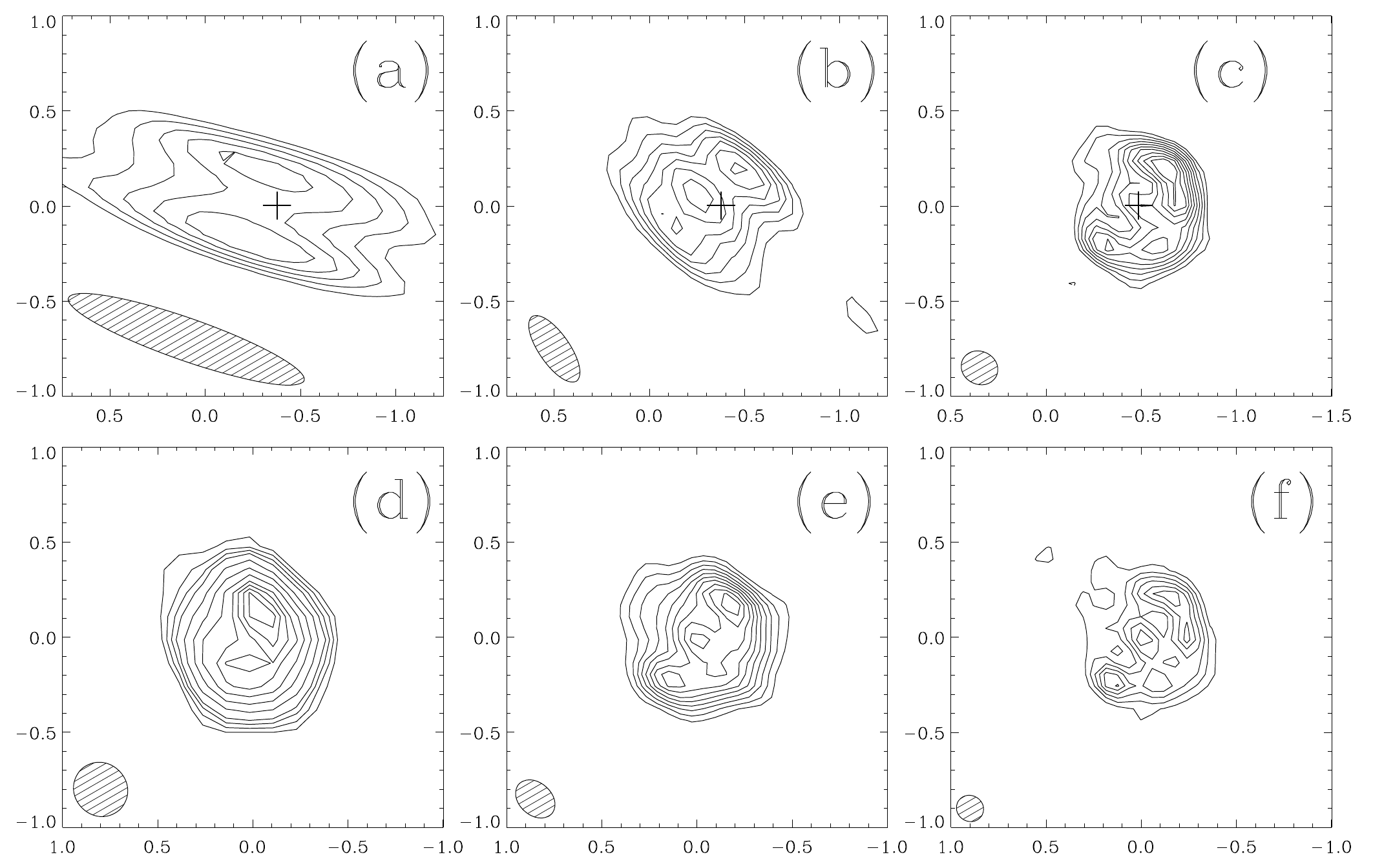}
\caption{Images of HD100546, with offsets in arcseconds. North is up and east to the left. For the top row 
images the offsets are from the J2000 position. The cross on these images shows the expected position of 
HD100546 at the time of the observation, using its established proper motion over 9.7 years for (a) and (b) 
and 12.5 years for (c). For the bottom row images the (0,0) position is made to coincide with the central 
emission peak.
a) Combined 43 and 45 GHz, naturally weighted in the 6D configuration on 26 August 2009, with a 
synthesised beam of 1.31\arcsec$\times$0.22\arcsec at PA=70.7$^{\circ}$. Contours are scaled by the RMS of 
0.054~mJy/beam, with levels at 7--15$\sigma$ in steps of 2. 
b) Combined 43 and 45 GHz, uniformly weighted in the 1.5A configuration on 1 August 2009, with a
synthesised beam of 0.44\arcsec$\times$0.17\arcsec at PA=34.6$^{\circ}$. Contours are scaled by the RMS of
0.039~mJy/beam, with levels at 7, 9, 11, 13, 15, 17.5, 20$\sigma$. 
c) Combined 43 and 45 GHz, uniformly weighted in the 6D configuration on 20 June 2012, with a
synthesised beam of 0.20\arcsec$\times$0.18\arcsec at PA=74.5$^{\circ}$. Contours are scaled by the RMS of 
0.030~mJy/beam, with levels at 6--24$\sigma$ in steps of 2. 
d) 43 GHz super-uniformly weighted from all CABB configurations combined, but without antenna 6 baselines. 
The synthesised beam size is 0.29\arcsec$\times$0.28\arcsec at PA=44.0$^{\circ}$. Contours are scaled by the 
RMS of 0.029~mJy/beam, with levels at 5, 7, 9, 11, 15, 20, 25, 30, 33, 36, 39$\sigma$.
e) 43 GHz uniformly weighted from all CABB configurations combined, with a synthesised beam of 
0.23\arcsec$\times$0.17\arcsec at PA=48.2$^{\circ}$. Contours are scaled by the RMS of 0.024~mJy/beam, 
with levels at 5--23$\sigma$ in steps of 2.
f) 43 GHz super-uniformly weighted from all CABB configurations combined, with a synthesised beam of 
0.15\arcsec$\times$0.14\arcsec at PA=59.2$^{\circ}$. Contours are scaled by the RMS of 0.032~mJy/beam, 
with levels at 3--13$\sigma$ in steps of 2.}
\label{FIG-7mm-normcal}
\end{figure*}

Figures~\ref{FIG-7mm-normcal}-a,b,c respectively show combined 43/45~GHz images acquired from single 
tracks on 26 August 2009 in the 6D configuration, 1 August 2009 in 1.5A, and 20 June 2012 in 6D. 
Figures~\ref{FIG-7mm-normcal}-d,e,f instead show 43~GHz images derived from combining the data 
from all array configurations with CABB. In Figures~\ref{FIG-7mm-normcal}-a,b,c the offsets are 
in arcseconds from the nominal J2000 position of RA=11:33:25.44058 and $\delta$=-70:11:41.2363. 
The centroid in these images is clearly shifted west of this position (with little or no N-S 
offset), which is accounted for by the known proper motion of HD100546 of $\sim$ -39~mas/yr 
in RA and 0.29 mas/yr in declination (position marked by a cross, van Leeuwen 2007). Being combined 
images, Figures\ref{FIG-7mm-normcal}-d,e,f have instead been shifted to a common position of 
the central emission peak.

Only for the 20 June 2012 data set in Figure~\ref{FIG-7mm-normcal}-c, with the best phase correction 
and $u-v$ coverage, does the cross coincide with the emission centroid. For the 1 and 26 August 2009 
data sets in Figures~\ref{FIG-7mm-normcal}-b and \ref{FIG-7mm-normcal}-a respectively the offset of 
HD100546 is smaller than would be predicted from its proper motion over 9.7 years. But in both cases 
the beam is highly elongated, and the offset between the centroid and the cross is along the direction 
of the elongation. We thus conclude in these cases that the centroid is consistent with the expected 
proper motion.

Both single-track and combined images are shown because they demonstrate the important point that 
the same structure is observed, thus providing high confidence in the fidelity of the combination, 
especially the crucial aspects of relative astrometry and flux calibration. These considerations 
become relevant for the self-calibration process presented in the next section.

Perhaps the most obvious features of the HD100546 disk demonstrated by Figure~\ref{FIG-7mm-normcal} 
is that it is very compact but with significant internal structure. Figure~\ref{FIG-7mm-normcal} 
shows unequivocally that, depending on the beam size and shape (i.e. the resolution along particular 
directions) the structure observed in the HD100546 disk at 7~mm is noticeably different. For instance, 
immediately apparent from all the images is a hole in the disk, but the presence or otherwise of a 
central peak is dependent on resolution (or equivalently the $u-v$ plane coverage). 

At a resolution of $\sim$ 0.20\arcsec--0.30\arcsec along the known disk major axis, as in 
Fig.~\ref{FIG-7mm-normcal}-a and \ref{FIG-7mm-normcal}-d, or 20--30~AU at the distance of HD100546, 
two peaks are observed, separated by $\sim$ 0.50\arcsec. The midpoint of the axis joining the 
two peaks is coincident with the stellar position, and the NW emission peak is brighter than the 
SE peak. But at $\sim$ 0.15\arcsec resolution along the disk major axis, or 15~AU, as in 
Fig.~\ref{FIG-7mm-normcal}-b,c and \ref{FIG-7mm-normcal}-e,f, there appears a central peak, 
$\sim$ 0.25\arcsec from each outer peak and concident with the stellar position. The axis 
joining all three peaks is aligned along a position angle (PA) of $140^{\circ}\pm5^{\circ}$. 
This is consistent with values of $130^{\circ}-160^{\circ}$ previously reported in the literature 
from the optical through to near-IR (e.g. Pantin et al. 2000, Augereau et al. 2001, Grady et al. 2001, 
Ardila et al. 2007, Tatulli et al. 2011, Quanz et al. 2011, Avenhaus et al. 2014a), and in the 
submillimetre regime with ALMA (Pineda et al. 2014, Walsh et al. 2014). The relative brightness 
of the central peak compared to the NW and SE peaks is not well constrained, but again the NW 
side of the disk is brighter than the SE. 

Also at $\sim$ 15~AU resolution, but with a more circular beam, as in Figs.\ref{FIG-7mm-normcal}-c,e,f, 
another type of asymmetry appears in the disk. Especially in Fig.~\ref{FIG-7mm-normcal}-c and 
\ref{FIG-7mm-normcal}-f the disk displays an almost horseshoe-shaped structure, with more emission in 
the SW than in the NE. The contrast between points on opposite sides of, and equidistant from, the 
stellar position is around a factor of 3, whilst even the horseshoe itself appears clumpy. 

Finally, we note that the total flux measured in the 6~km array configuration on 20 June 2012 is 
1--2~mJy less than that measured in compact hybrid configurations on 4 October 2007 and 31 May 2009 
(see Table~\ref{TAB:hd100-obs7mm}). This possibly indicates that there is a significant extended dust 
emission component which is filtered out by the interferometer for baselines longer than 80--100 m. 
This is supported by the 27 June 2008 track, where the flux on the shortest baseline of 31~m was 
significantly larger than would be extrapolated from a Gaussian fit to the flux on all other baselines 
(as determined from the annularly binned visibilities obtained with Miriad's UVAMP task). 

\subsection{Amplitude versus $u-v$ distance}

\begin{figure}
\includegraphics[scale=0.510]{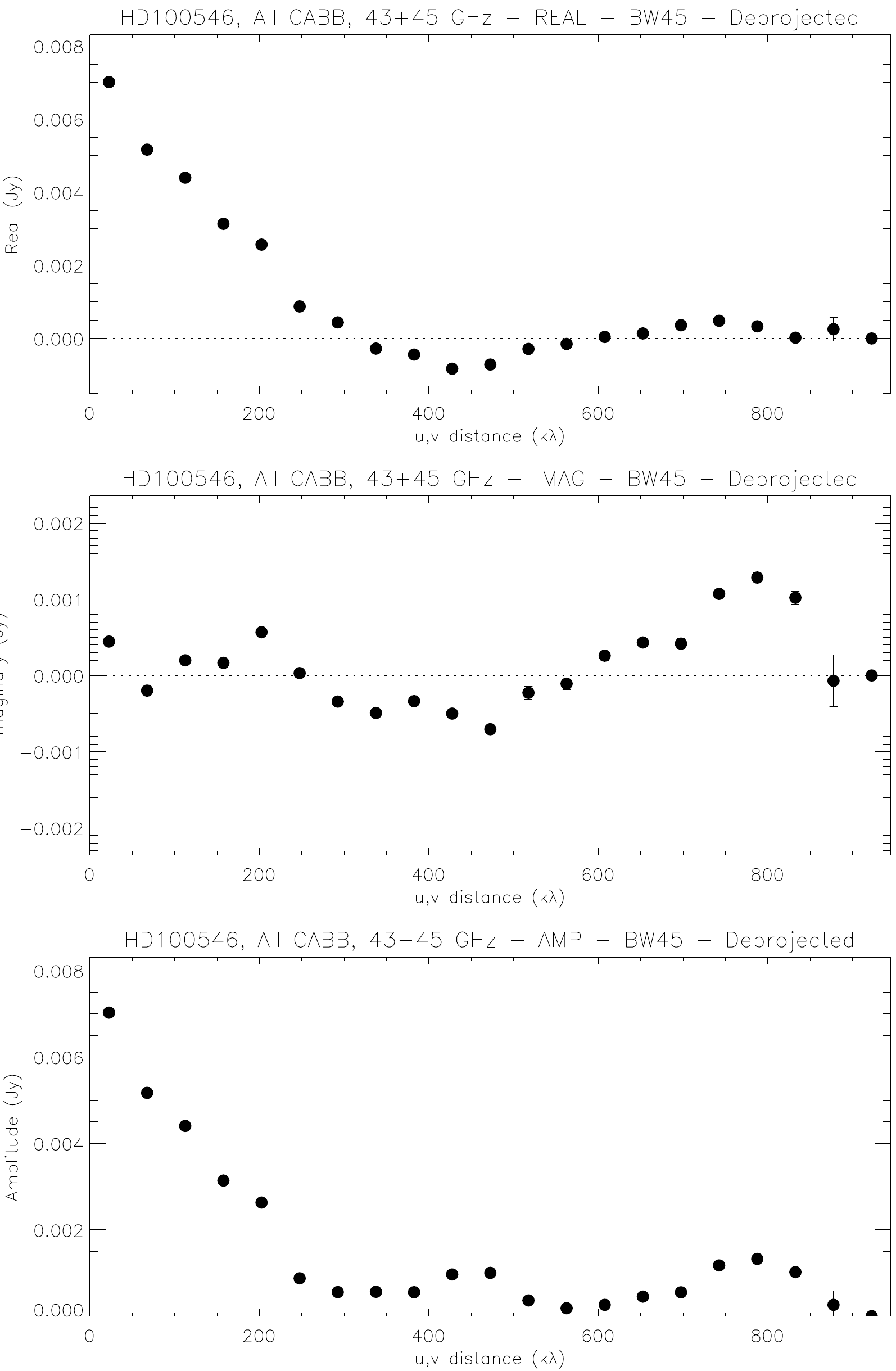}
\caption{Real, imaginary and total visibility amplitudes versus de-projected $u-v$ distance computed 
using all the CABB data sets and combining both frequency sidebands. A bin width of 45~k$\lambda$ 
has been used.}
\label{FIG-AllCABB-Vis}
\end{figure}

The images themselves in Figure~\ref{FIG-7mm-normcal} show that the gap radius in the HD100546 disk is 
$\sim$ 25~AU, as judged from the separation between the emission peaks along the disk major axis. Another 
estimate can be obtained from the visibilities, as depicted in Figure~\ref{FIG-AllCABB-Vis}. This shows the 
real and imaginary parts as a function of deprojected $u-v$ distance from the disk centre, following the 
treatment of Hughes et al. (2007), and equivalently Berger \& Segransan (2007). For the deprojection we have 
used our own directly derived values for the major axis position angle of 140$^{\circ}$ and the disk inclination 
from face-on of 40$^{\circ}$ (see next section). 

The plot shows a null in the real part of the visibility at 320$\pm$10~k$\lambda$, determined using a linear fit
to the binned visibilities in the region of $\sim$ 250--400~k$\lambda$. A similar feature has been observed in 
ten or so disks in recent years, and is indicative of a central hole or gap in the disk where there is a sharp 
change in the dust density (e.g. Andrews et al. 2011a, Brown et al. 2009, Andrews et al. 2009; Hughes et al. 2007). 
Whilst all the CABB data has been used to construct Figure~\ref{FIG-AllCABB-Vis} to enhance the S/N, the same null 
position is obtained -- within the uncertainty range -- separately for each frequency and each configuration with 
baselines sufficient to adequately sample the relevant $u-v$ space. Further, a similar value is found for 93.504~GHz 
data obtained in June 2008 (See Appendix A, Figures~\ref{FIG-3716mm-normcal}-b and \ref{FIG-3mmJun08-Vis}). On the
other hand, recently published ALMA data at 0.9~mm by Walsh et al. (2014) instead show a null in the deprojected 
visibilities at 290$\pm$5~k$\lambda$.

Using equation A11 in Hughes et al. (2007) the null at 320~k$\lambda$ corresponds to a gap radius of 24.7~AU, in 
good agreement with the value inferred from the reconstructed images. Equation A11 assumes that the emission occurs 
from a relatively thin ring of constant brightness, where the width of the ring is much less than its radius from 
the star, and the visibility function is a zeroth-order Bessel function. This may not necessarily be the case for 
HD100546. Equation A9 of Hughes et al. takes account of this, but requires knowledge of the sum of the power law 
indices $p+q$ for the disk surface density and temperature radial gradients. Typical values of the surface density 
and temperature exponents are $\sim$ 1 and 0.5 respectively (e.g. d'Alessio et al. 1999). This would give a hole 
radius of $\sim$ 13.1~AU, close to the value inferred from the optical, near-IR and mid-IR observations. Otherwise, a 
value of $p+q$ between 1 and 3 is reasonable (and required for Equation A9 to be valid) and gives radii of 11.5, 14.8 
and 18.1~AU for $p+q$ = 1, 2 and 3 respectively.

In addition to the null at $\sim$ 320~k$\lambda$, a second and possibly third null are seen at between 550 and 
600~k$\lambda$ and $\leq$ 850~k$\lambda$ respectively. The existence of a second null is similar to the cases of 
SAO~206462, SR~21 and LkCa~15 in Andrews et al. (2011a,b). Assuming access to a telescope with even longer baselines 
(or equivalently the same telescope but with higher frequency receivers) we would expect this pattern of nulls to 
continue, as is the case for the circumbinary disk around the T Tauri system GG~Tau in Guilloteau et al. (1999). 
Such a 'ringing' pattern of the visibility amplitude in Figure~\ref{FIG-AllCABB-Vis} is characteristic of a 
structure with relatively sharp boundaries between regions with and without emission.

\subsection{Self-calibration}

\begin{figure*}
\includegraphics[scale=0.80]{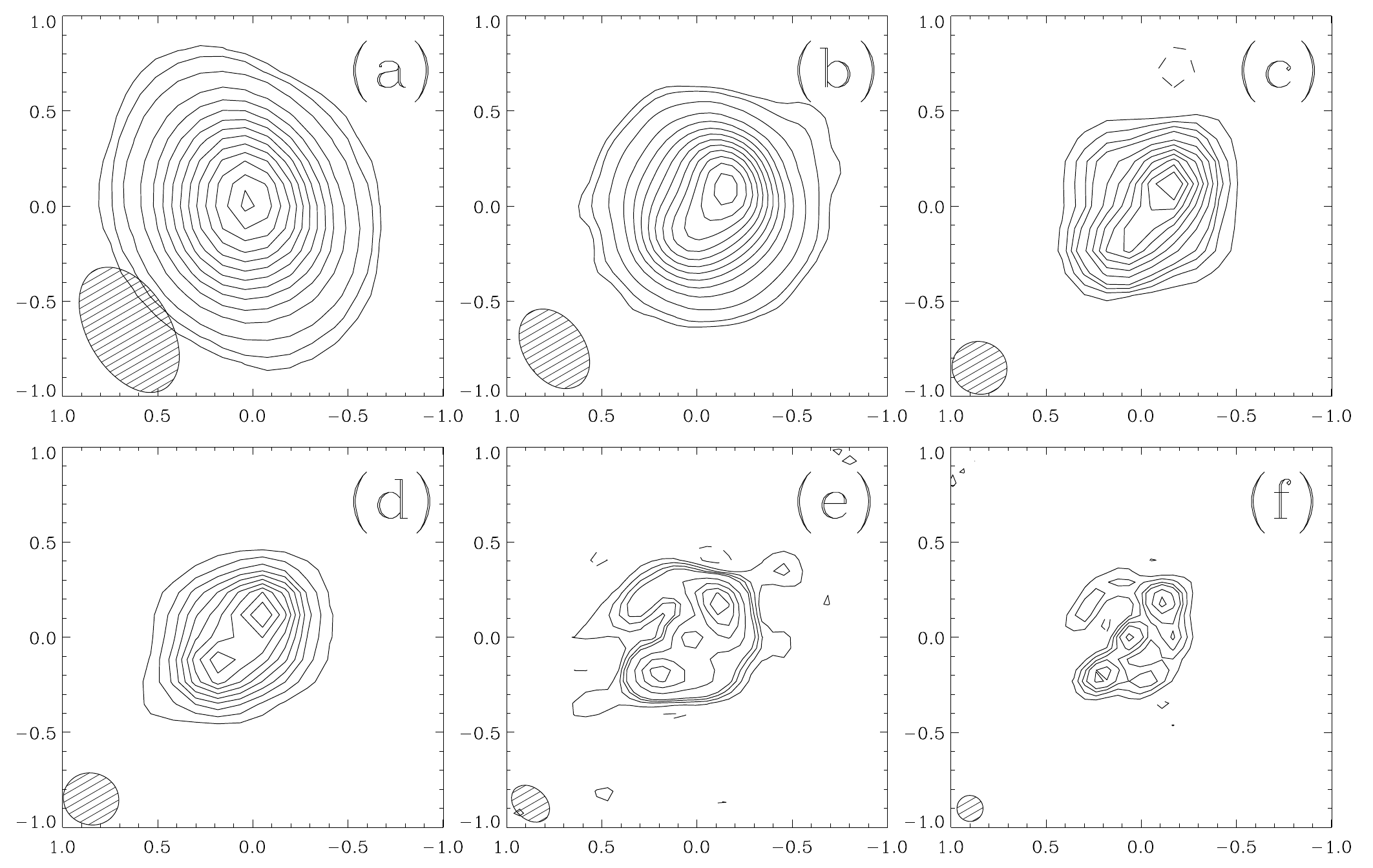}
\caption{Self-calibrated images -- phase only, two iterations -- of HD100546 at 44 GHz using all the configurations 
for which CABB data was obtained. North is up and east to the left. The resolution increases from a--f, being: 
a) 0.72\arcsec$\times$0.43\arcsec at PA=30.9$^{\circ}$ (natural weighting excluding antenna 6 baselines), with 
contours at 5, 10, 20, 40, 60, 80, 100, 120, 140, 160, 180, 200, 220, 235 times the RMS of 0.0198~mJy/beam.
b) 0.46\arcsec$\times$0.32\arcsec at PA=35.1.5$^{\circ}$ (natural weighting including all baselines) with contours 
at 5, 10, 25, 50, 75, 100, 125, 150, 175, 200, 225, 250, 275 times the RMS of 0.0119~mJy/beam. 
c) 0.29\arcsec$\times$0.28\arcsec at PA=68.1$^{\circ}$ (super-uniform weighting excluding antenna 6 baselines) with 
contours at 5, 10, 15, 20, 30, 40, 50, 60, 70, 80, 90 times the RMS of 0.0273~mJy/beam.
d) 0.29\arcsec$\times$0.27\arcsec at PA=73.5$^{\circ}$ (super-uniform weighting excluding antenna 6 baselines and 
20 June 2012 data), with contours at 5--50 times the RMS of 0.0402~mJy/beam, in steps of 5$\sigma$.
e) 0.23\arcsec$\times$0.16\arcsec at PA=47.9$^{\circ}$ (uniform weighting including all baselines) with contours at
5, 10, 15, 20, 40, 60, 80, 100 times the RMS of 0.0180~mJy/beam.
f) 0.138\arcsec$\times$0.136\arcsec at PA=62.4$^{\circ}$ (super-uniform weighting including all baselines) with 
contours at 5, 10, 15, 20, 30, 40 times the RMS of 0.0359~mJy/beam.}
\label{FIG-7mm-selfcal}
\end{figure*}

Although not shown in Figure~\ref{FIG-7mm-normcal}, signatures of phase errors, such as finger-like 
radial extensions, were seen in several tracks and in the combined data at levels of several times 
the RMS noise. Their presence, along with the fact that we could be confident of having a good input 
model -- given the extensive $u-v$ plane coverage of our data -- motivated us to attempt self-calibration. 
The results are shown in Figure~\ref{FIG-7mm-selfcal}.

Our procedure was to self-calibrate each frequency separately to serve as a consistency check. We 
also compared self-calibration results using the 'raw' 2048$\times$1~MHz channel data averaged into 
32$\times$64~MHz channels (using Miriad's UVAVER routine). Solution intervals of 15 and 30 seconds 
were trialled. All combinations of input parameters produced results which were extremely consistent 
for the two sidebands. Further, two initial input models were tried, a clean component and a point 
source, and only two phase-only iterations were performed. Results were robust in the sense that the 
gross features in the final images were similar for the two initial models, though for the point 
source model the central peak dominates over the two others on either side.
 
The self-calibration produced images entirely consistent with those in Figure~\ref{FIG-7mm-normcal}, though 
with a much improved dynamic range, and showed similar behaviour as a function of $u-v$ plane coverage. For 
instance, the presence or otherwise of a central peak within the inner disk gap is once again shown to be 
dependent on spatial resolution via the sampling and/or weighting of the visibility spacings. Also, the NW-SE 
brightness asymmetry is maintained, such that as the spatial resolution increases the NW side of the disk 
becomes prominent before the SE peak appears, and remains brighter at all scales. 

Perhaps the best outcome from the self-calibration is that it 'sharpened' the image of the disk, in the
process revealing part of the NE side of the disk in Figures 3e and 3f, which had only been hinted at 
in Figure~\ref{FIG-7mm-normcal}-c. However the disk is still incomplete, being horseshoe-shaped with a 
large opening in its eastern side. Assuming the disk to be circular if viewed face-on, and fitting an 
ellipse to those images where the synthesised beam is close to circular, e.g. Figures~\ref{FIG-7mm-selfcal}-d,f, 
the disk inclination is estimated to be $40^{\circ} \pm 5^{\circ}$. This is consistent with values of 
$40^{\circ} - 50^{\circ}$ previously reported in the literature from the optical through to near-IR 
(e.g. Pantin et al. 2000, Augereau et al. 2001, Grady et al. 2001, Ardila et al. 2007, Tatulli et al. 2011, 
Quanz et al. 2011, Avenhaus et al. 2014a), and in the submillimetre regime with ALMA (Pineda et al. 2014, 
Walsh et al. 2014).

\section{Discussion}

Our observations present direct evidence at millimetre wavelengths for a $\sim$ 25~AU radius cavity in the 
disk of HD100546, from the reconstructed images as well as the deprojected and binned visibilities. Although 
the ALMA sub-mm observations of Walsh et al. (2014) and Pineda et al. (2014) also showed a null in the 
deprojected visibilities versus baseline, only Walsh et al. interpreted this in terms of a cavity within the 
disk. Neither author presented images showing the cavity, as their chosen antenna baseline weightings only 
gave spatial resolutions of $\leq$ 1\arcsec along the disk minor axis and 0.4\arcsec--0.5\arcsec along the 
major axis. In this paper we show that a resolution of better than about 0.3\arcsec is needed to see the cavity 
in reconstructed images. However our images show not only the cavity, but also the existence of several 
asymmetries in the disk. We discuss each of these in turn below, as well as some of the inferred physical 
properties of the disk, such as its mass, overall size and dust population. Finally, we speculate on the type 
of planetary system that may be hosted by HD100546.

\subsection{Cavity radius and variation with wavelength}

Millimetre continuum observations trace the location principally of large dust grains, hundreds of microns 
to millimetre and even centimetres in size via their thermal emission within the disk mid-plane. On the 
other hand, the UV--optical--near-IR--mid-IR observations trace smaller dust grains, sub-micron to several 
microns in size, which scatter radiation off a disk 'atmosphere', and/or thermally emit from near a flared 
disk surface layer where they intercept direct stellar radiation. So it is perhaps not surprising that the 
two wavelength regimes typically produce somewhat different pictures of the disk, e.g. overall disk size 
(including the outer radius) and spiral patterns which have so far been difficult to (unambiguously) detect 
in the millimetre dust continuum. 

But an interesting facet of our HD100546 data is that the radius of the inner disk gap -- or perhaps more 
accurately the distance between the star and the peak of the 7~mm thermal dust emission -- is around 25~AU. 
This is significantly larger than the 14$\pm$2~AU radius inferred at much shorter uv--IR wavelengths, best 
seen in the surface brightness cuts of Avenhaus et al. (2014a) and Quanz et al. (2011). These display a clear 
maximum at $\sim$ 14~AU, and certainly well short of 25~AU. Our figure is extremely robust, being consistent 
amongst several different data sets. 

This cavity size differential cannot be easily explained by the mid-plane versus surface/atmosphere locations of 
the respective dust populations, but it is a finding that is becoming increasingly common for other transition 
disks (e.g. Dong et al. 2012). Examples include SR~21 (Follette et al. 2013; Andrews et al. 2011a), MWC~758 (Grady 
et al. 2013; Andrews et al. 2011a) and SAO~206462 (HD135344B, Garufi et al. 2013; Andrews et al. 2011a). 
Further, in several cases including HD100546 (Brittain et al. 2014, 2013, 2009; Liskowsky et al. 2012), the atomic 
and/or molecular gas approaches closer to the central star. Other examples include J160421.7-213028 (Zhang et al. 2014) 
and those studied by Pontoppidan et al. (2011, 2008). Thus, the overall picture is one where the gas and small dust 
grains are well mixed and have a similar spatial distribution, but which is different for the large dust grains. 

As well as the different cavity radius, the small dust grains and/or molecular gas in several transition disks 
(for which the respective data exists) also form a complete ring around the star, unlike the millimetre emission 
which only forms an arc or horseshoe (e.g. Oph IRS48 in van Der Marel et al. 2013; HD142527 in Casassus et al. 
2013 and Fukagawa et al. 2013). Both phenomena are well explained in a scenario where the large dust grains are 
caught (or trapped) at a pressure bump (Regaly et al. 2011; Pinilla, Benisty \& Birnstiel 2012; Birnstiel, 
Dullemond \& Pinilla 2013), across which the gas and small grains can diffuse. 

Our inferred central cavity radius of the HD100546 disk of $\sim$ 25~AU finds a parallel in the modelling 
of HD100546's disk by Mulders et al. (2013b). In order to reproduce mid-IR interferometric data these authors 
find that the boundary of the inner disk wall (near 13~AU) cannot be sharp, but must instead be significantly 
rounded. The best-fit radius they find -- at which a steep drop-off of the disk surface density begins -- is 
29~AU, with a range of 26 to 35~AU. Similarly, modelling the disk as a ring with a Gaussian radial brightness 
distribution, Walsh et al. (2014) found that their ALMA 0.9~mm observations could be well fit with a width of 
21~AU centered at 26~AU. 

In addition to the cavity size differential inferred from the large and small grain tracers, our 7~mm data also
suggests a difference in gap radius inferred fom long and short millimetre wavelengths. Specifically, the 7~mm 
visibility null determined here of 320$\pm$10~k$\lambda$ is significantly different to that measured at 0.9~mm 
by Walsh et al. (2014) of 290$\pm$5~k$\lambda$. With a very fine coverage of the $u-v$ spacings the ALMA 
value is extremely well determined. Whilst our data does not have an equivalent `fineness' in the $u-v$ coverage 
to determine the null position with the same level of precision, our signal-to-noise is still sufficent to have 
high confidence in our value. Thus we are very confident that this is a real difference. To further check 
this we have also used the same deprojection formula and values for the inclination and position angle used by 
Walsh et al. This did reduce our null to 315$\pm$10~k$\lambda$, but obviously still maintained the existence of 
the difference and approximately its magnitude as well. To our knowledge this difference is one of the first 
such findings for a transitional disk, and certainly the first across such a wide wavelength range of 1-to-7~mm. 
Recently van der Marel et al. (2015) found a similar effect for the first visibility nulls at 345~GHz and 690~GHz 
for several transition disks, including HD135344B, SR~21 and RX~J1615-3255, and also interpreted this as an increased
cavity size at the longer wavelength.

The fact that the null at 7~mm is higher than at shorter millimetre wavelengths is in agreement with particle 
trapping theory induced by a massive planet, which predict different radial dust distributions for different grain 
sizes (Pinilla et al. 2012; de Juan Ovelar et al. 2013; Pinilla et al. 2014). According to these models, larger 
grains are expected to be more concentrated in the pressure maximum than smaller grains, implying a slight difference 
in the disk radial extension observed at 7~mm than at sub-mm wavelengths. Thus, the larger particles to which our 
observing wavelength of 7~mm is biassed are located slightly radially inward of the smaller particles traced at 
0.9~mm by the ALMA data. More detailed modelling of both the ALMA and ATCA visibilities, in the context of sequential
planet formation in the disk, is presented in Pinilla et al. (submitted).

\subsection{Disk size (outer radius)}

From our best 7~mm data on 20 June 2012 the FWHM of the mm-emitting region is $\sim$ 50--60~AU, from an 
elliptical gaussian fit using Miriad's UVFIT. This is similar to the FWHM of $\sim$ 55~AU found at 16~mm 
(also using UVFIT; Appendix A, Figure~\ref{HD100-16mm-uvamp}), and is little or no different to the outer 
radius at 0.87~mm of 40--60~AU quoted by Pineda et al. (2014), or $\leq$ 100~AU (and nearer $\sim$ 50~AU) 
at 0.87--0.99~mm in Walsh et al. (2014). The visibilities at 3~mm presented in Figure~\ref{FIG-3mmJun08-Vis} 
also imply a very similar source size (see also Figure~\ref{FIG-3716mm-normcal}-b). We have not used our 
other 3~mm data (Figure~\ref{FIG-3716mm-normcal}-a) to estimate a disk size due to the relevant data set
having a poor phase correction. Perez et al. (2010) have shown how such an inadequate correction can lead 
to an unreliable estimate of the disk size. 

Since each wavelength is most sensitive to emission from dust grains with size similar to that wavelength, 
then within the spatial resolutions achieved by the various above-mentioned data sets there is 
no evidence for a large-scale radial gradient in the `characteristic' particle size (at least for 
millimetre-sized grains) within the HD100546 disk. As noted by Pineda et al. (2014) the outer radius of 
a dust disk should be the same when observed at different (sub-)millimeter wavelengths for the case of a 
planet--disk interaction, while for pure radial drift the outer disk radius inferred from different 
wavelengths would be different (see also Birnstiel \& Andrews 2014). 

The 1--10~mm observations of the AS~209 and UZ Tau~E disks presented in Perez et al. (2012) and Andrews (2013) 
respectively demonstrate excellent examples where radial drift is likely to be occurring, and dominating the
segregation of different particle sizes. Their visibility curves progressively broaden with increasing 
wavelength, indicating more and more compact emission for ever larger grains. For example, for AS~209 the FWHM 
of the deprojected and binned visibilities approximately doubles between 1--3~mm and 8--10~mm, whilst for 
UZ Tau~E it increases by about a factor of five. Assuming Gaussian source structures these represent concomitant 
changes in the size of the disk. This stands in contrast to the corresponding plots for HD100546, presented here 
at 3, 7 and 16~mm and at 1~mm in Walsh et al. (2014). All these have similar -- though as alluded to above not 
necessarily precisely equal -- widths, and thus provide yet more evidence of one or more planets already existing 
within the cavity of the HD100546 disk. 

Of course there is a large difference in the disk outer radius between the 1--16~mmm observations and the 
uv--IR observations which are sensitive to much smaller sized grains. Similarly, the disk size measured in 
molecular gas tracers, such as CO (Walsh et al. 2014; Pineda et al. 2014; Panic et al. 2010) and HCO+ (Appendix
B), is significantly larger than inferred from the mm-emitting dust. However, the data are insufficient to 
discriminate between scenarios where this contrast is due to radial migration of larger grains, which decouple 
more effectively from the gas, or whether the grains instead grew in situ within the inner disk. In the latter 
case the dust density may be naturally higher, potentially providing a more conducive environment for grain 
growth. For instance, the dust density would increase radially inward, so that the collision timescale between 
grains would be shorter, and/or -- as strongly suggested above -- there could be local density enhancements 
arising from a pressure maximum, created within the inner disk via interaction with an orbiting body.

Interestingly, Quanz et al. (2011) present evidence from near-IR scattering data on 10--140~AU scales that there 
may be a change in the relative fractions of large versus small grains at a disk radius of 40--50~AU. Though only 
probing the surface layers of the disk, and still only sensitive to particles in the micron size range, that the 
larger ones dominate at radii encompassed by the total extent of the mm-emitting disk found here is suggestive of 
a relation between the two data sets.

\subsection{NW--SE brightness asymmetry}

The sense of the brightness asymmetry between the two sides of the disk at 7~mm is consistent with what 
is observed at 3~mm (Appendix A, Figure~\ref{FIG-3716mm-normcal}-b), but opposite to that found by Pineda 
et al. (2014) at 0.87~mm with ALMA. However, their result is based purely on the residuals of a model fit, 
in which the model assumes a fixed inner disk radius of 14~AU, consistent with the value found from 
optical/near-IR scattering data, but less than the star-to-peak emission radius of $\sim$ 25~AU found in 
the millimetre regime here and in Walsh et al. (2014). So our observed asymmetry is a more direct finding, 
and thus we believe more reliable. 

Notably, the major axis brightness asymmetry we find is opposite to that seen in near-IR scattered 
light by Avenhaus et al. (2014a), who instead find the SE peak to be brightest, which they tentatively
interpret as being due to it being closer to an orbiting companion. Also, from Hubble Space Telescope 
far-ultraviolet observations Grady et al. (2005) found that their inferred $13.0\pm2.5$~AU radius 
cavity was not centred on the star, but instead $\sim$ 5$\pm$3~AU to the south-east along the major 
axis, where there was a local maximum in the reflection nebulosity, and implying an eccentricity of
0.38$\pm$0.24. If the finding by Grady et al. is true then that would also mean the NW side of the 
disk is physically closer to the star. Thus, the dust on that side would be warmer leading to enhanced 
thermal emission, precisely what we observe. 

But for this scenario to work -- known as pericenter glow and also seen in the HR4796A debris disk 
(Telesco et al. 2000; Wyatt et al. 1999) -- requires an elliptically orbiting body either inside or 
outside the disk. The elliptical orbit enforces an eccentricity to the orbits of dust particles,
with the consequence that the inner rim of the disk also becomes elliptical, a prediction that finds 
support, in the case of HD100546, in the modelling of OH ro-vibrational spectra by Liskowsky et al. 
(2012) and the direct observation of Grady et al. 

Also, for HD100546 the Grady et al. observation would predict the orbiting body's apocenter, i.e. its 
furthest distance from the star, to be on the SE side of the disk. For an elliptical orbit the apocenter 
is also the direction in which the body spends most of its time, since its velocity is lower than when 
it approaches close to the star. Thus, from a statistical perspective, for any one system we would expect 
to find an orbiting body closer to its apocenter than its pericenter. In the timeframe of our observations 
this does appear to be the case for HD100546, at least for the postulated inner body (Brittain et al. 
2014, 2013). Further, it is probably this body that would most influence the inner regions of the disk due 
to both its relative proximity (10--15~AU compared to 50--70~AU for the outer body) and its so far higher 
estimated mass range (Mulders et al. 2013b, Quanz et al. 2013a).

Whilst we propose here that an elliptically orbiting planet imposes the eccentricity onto the disk, this 
is not actually necessary. The hydrodynamic simulations of Kley \& Dirksen (2006) and Ataiee et al. (2013) 
show that a $\geq$ 3~M$_{\rm J}$ planet in a circular orbit around a solar mass star can also induce a 
significant eccentricity on the disk, e.g. up to $e$=0.25, via an instability. But given that 3 of the 4 
gas giants in our solar system have an eccentricity of $\sim$ 0.05, and that the distribution of extrasolar 
planet eccentricities peaks between $e$=0.1 and 0.4 (Wright et al. 2011), then it is reasonable to presume 
that these 'end state' planet orbits are a reflection of the situation in their precursor disk stage. Though 
obviously there could be orbital evolution of any one particular planet, it seems credible to anticipate that 
any planet(s) around HD100546 could also be in an eccentric orbit.

\begin{figure}
\includegraphics[scale=0.425]{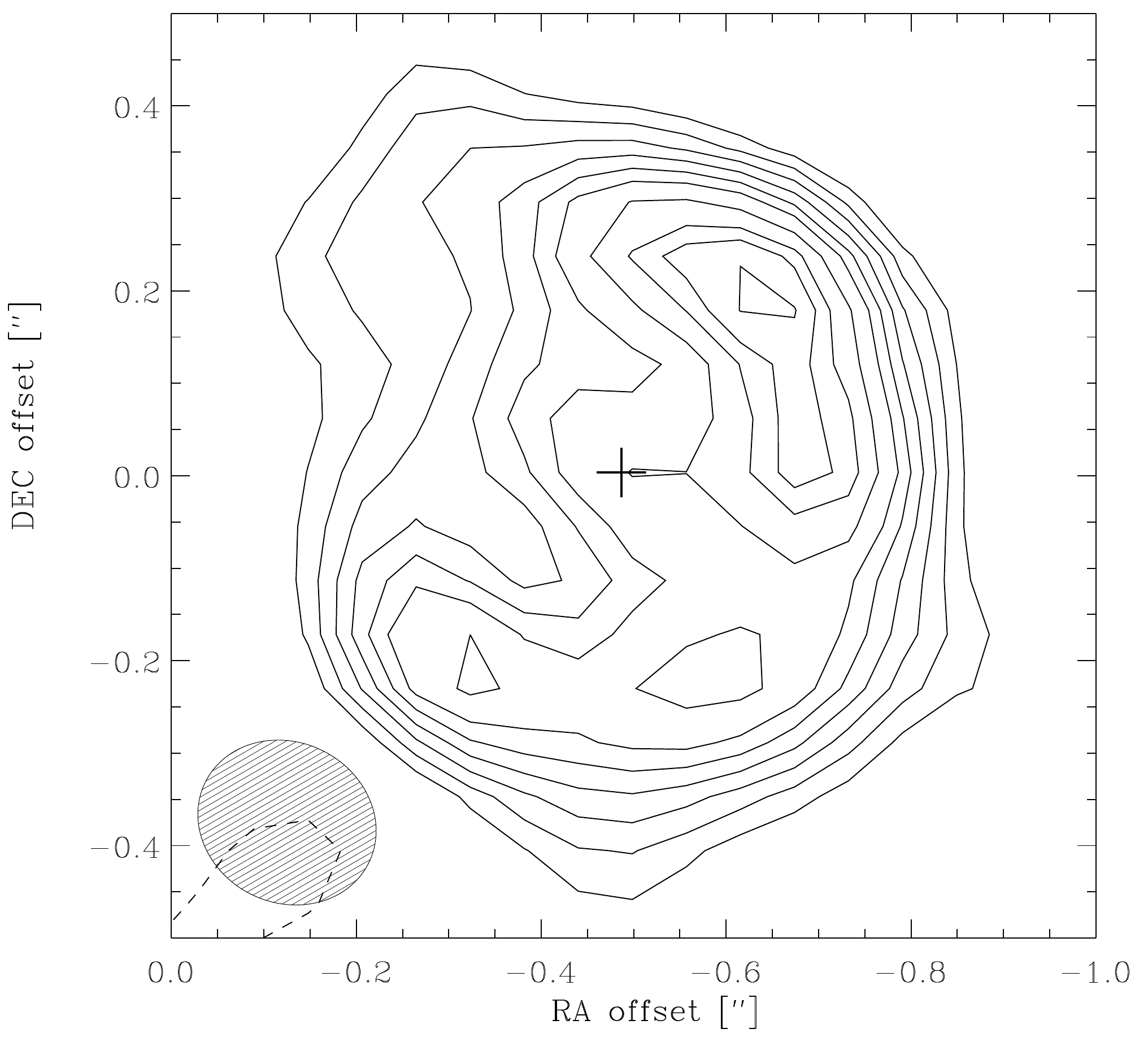}
\caption{Reconstructed image at 44~GHz from 20 June 2012, using uniform weighting and contour levels at 
5, 7, 9, 11, 13, 15, 18, 21, 23, 25.7 times the RMS of 0.03034~mJy/beam. Synthesised beam is 
0.20\arcsec$\times$0.18\arcsec at PA=74.5$^{\circ}$.}
\label{FIG-Pericenter}
\end{figure}

Our best data set, that of June 2012, provides direct support for the 'pericenter glow' scenario, and thus 
very strongly support the existence of one or more inner orbiting bodies. Figure~\ref{FIG-Pericenter} shows 
an expanded view of the 44~GHz image in Figure~\ref{FIG-7mm-normcal}, where the cross marks the position of 
the star after accounting for its proper motion of $-38.93\pm0.36$~mas/yr and $0.29\pm0.38$~mas/yr in RA and 
declination (van Leeuwen 2007) from its J2000 position over a 12.5 year period. Whilst the NW and SE peaks 
are aligned along a similar axis through the stellar position, the NW peak is slightly closer to the star. 
The individual 43 and 45~GHz images show a consistent trait. The implied eccentricity is only about 0.06, 
less than that implied by the Grady et al. (2005) result of 0.38$\pm$0.24 or the lower limit of Liskowsky 
et al. (2012) of $\geq$ 0.18 (0.07-0.30 at 99.7 per cent confidence). However, these latter two estimates are 
for smaller disk radii, $\sim$ 13~AU, whilst ours is for further out in the disk at $\sim$ 25~AU. Assuming 
a 1/r$^{2}$ falloff in the eccentricity (Liskowsky et al. 2012 and references therein) the expected value 
at 25~AU is $\geq$ 0.05, or 0.11$\pm$0.05, figures much more in line with our determination.

Finally, we point out that the 3.81~$\mu$m NACO/VLT Sparse Aperture Masking (SAM) images of HD100546 presented 
in Marino et al. (2014a,b) also suggest that the $\sim$ 13~AU radius cavity (at this wavelength) around HD100546 
is eccentric. As at 7~mm here, the SAM data detect both the star and the disk in the same image, which shows 
a slight but distinct offset of the star north west from the centre of an elliptical ring, and along its major 
axis. This is similar to the sense observed by Grady et al. (2005), but the magnitude of $e$ at $\sim$ 0.15 is 
less than half their best estimate, though within the uncertainty range and close to the figure of 0.18 from
Liskowsky et al. (2012). In the interest of completeness we note that, from near-IR polarimetric differential 
imaging, Avenhaus et al. (2014a) exclude an eccentricity of $\geq$ 0.133 at 99.8 per cent confidence for their
14$\pm$2~AU radius inner disk rim. 

\subsection{NE--SW azimuthal asymmetry}

In addition to the major axis asymmetry discussed above, there is also a minor axis asymmetry apparent
in our 7~mm images. Such an azimuthal asymmetry, in the shape of an arc or horseshoe, has been found in
the millimetre regime around several Herbig stars with transition disks. The factor of about 3 contrast 
for HD100546 is not nearly as dramatic as values of $\geq$ 130 for Oph IRS48 (van der Marel et al. 2013) 
or 30 for HD142527 (Casassus et al. 2013; Fukagawa et al. 2013), but is similar to those for other transition 
disks like LkH$\alpha$ 330 (Isella et al. 2013) and SR~21 and SAO~206462 (HD135344B, Perez et al. 2014). 
These asymmetries, typically interpreted recently in the context of a pressure bump/dust trap model 
mentioned above, may well be the result of on-going planet formation in the disks (e.g. Ataiee et al. 2013; 
Birnstiel et al. 2013; Fu et al. 2014).

Interestingly, the arc of millimetre emission in HD100546 is on the same side as a 'dark lane' seen in 
scattered light in the H and K bands by Avenhaus et al. (2014a). As Figures~3 and 4 in Avenhaus et al. 
show, the dark lane begins at about 20~AU from the star, similar to the millimetre emission, but its 
radial extent is apparently larger. Taken together, the bright millimetre thermal emission but faint 
near-IR scattered light is consistent with the SW being the near side of the disk, as also inferred by 
other authors (e.g. Quanz et al. 2011 and references therein). In other words, the dense, mid-plane of 
the disk closer to Earth shadows and obscures part of the far side.

As further support for this non-axisymmetry seen in the image plane, if the HD100546 disk was instead 
axisymmetric then with a suitable choice of the disk inclination and major axis position angle the 
deprojected imaginary component of the visibilities should be close to zero. This is clearly not the 
case in Figure~\ref{FIG-AllCABB-Vis}, and no other choice of inclination and position angle (consistent 
with literature values) makes it so. Thus, as the reconstructed images in Figure~\ref{FIG-7mm-normcal} 
suggest, the structure in the imaginary component is proof that the HD100546 disk is not axisymmetric 
at 7~mm. This is similar to the case of the LkH$\alpha$ 330 disk at 1~mm in Isella et al. (2013). 
Leveraging off their work, from Figure~\ref{FIG-AllCABB-Vis} the imaginary component appears to oscillate 
around zero with an amplitude of around 1~mJy at $\geq$ 200~k$\lambda$. Assuming a total integrated flux 
of around 7~mJy for the compact disk (20 June 2012 data in Table~\ref{TAB:hd100-obs7mm}) this implies 
that only about 15\% is contributed by the asymmetric part of the disk. Such an implied 'average' contrast 
is notably higher than the point-to-point brightness contrast of $\sim$ 3 for the azimuthal asymmetry 
inferred from the reconstructed images.

The imaginary component in Figure~\ref{FIG-AllCABB-Vis} approximates to zero at baselines less than 
about 200~k$\lambda$. This is even more obvious when narrower bin widths are used. It may suggest that 
the disk is axisymmetric on large spatial scales. This appears to be supported by the ALMA 300--350~GHz 
data of Walsh et al. (2014), where they find the imaginary component to be zero up to a baseline of 
300--350~k$\lambda$, at which point it rapidly increases above zero. Their $u-v$ coverage cuts off at 
$\sim$ 400~k$\lambda$ so a more complete comparison with our data cannot be made. We do however note 
that, whilst the respective data sets are qualitatively similar, where the imaginary component in our 
data first goes negative, in Walsh et al. it first goes positive. This does not appear to be due to 
the different deprojection formulae, of which we have used those in Hughes et al. (2007) and Walsh 
et al. used those in Berger \& Segransan (2007).

Interestingly, from mid-IR interferometry Panic et al. (2014) find a significant deviation of their
phases from zero, from which they infer an asymmetry in the disk wall at $\geq$ 10~AU from the star. 
They did not state the azimuthal position of the asymmetry, but since the bulk of their baselines 
were closer to the position angle of the disk minor axis suggests that the asymmetry is approximately
along that axis as well. A departure from centrosymmetry is also inferred by Lazareff et al. (2013) from 
near-IR interferometric data obtained with the PIONIER instrument. Finally, the reconstructed images
from the 3.8 $\mu$m SAM observations of Marino et al. (2014a,b), also appear to show a horseshoe-like
region, in which there is significantly more emission to the SW than NE at radii of 10--15~AU. 

Thus, three separate near- to mid-IR data sets suggest that the HD100546 disk is asymmetric, and probably in 
the same sense as inferred from our millimetre observations. If true this would imply similarly asymmetric 
distributions of the small and large dust grain populations, albeit at radii of $\sim$ 12 and 25~AU respectively. 
It remains to be seen whether such a scenario is consistent with the pressure bump/dust trap model, which typically 
implies that the smaller grains can diffuse through the trap, along with the gas. This does appear to be the 
case radially for HD100546, but apparently not azimuthally. Alternatively, it is perhaps as simple as the fact 
that the dust trap is a reservoir of material, in which the fragmentation products of collisions between large 
particles naturally lead to similar azimuthal spatial distributions for small and big dust grains.  

\subsection{Spectral energy distribution and dust grain size}

\begin{figure}
\includegraphics[scale=0.355]{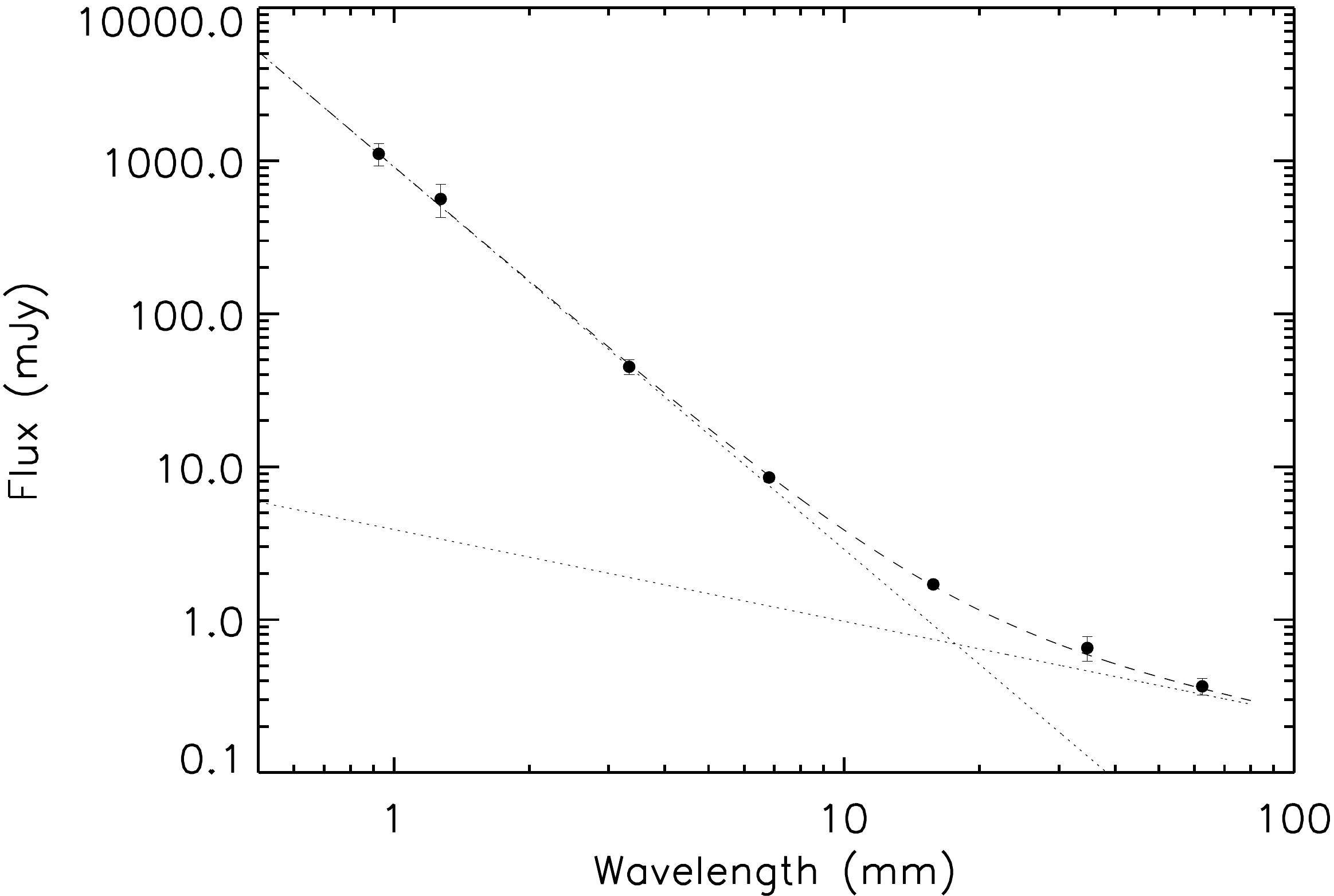}
\caption{Spectral energy distribution of HD100546 from $\sim$ 1--60~mm. Total 'average' fluxes at 
16~mm, 3.5~cm and 6.2~cm are 1.7$\pm$0.1, 0.65$\pm$0.12 and 0.37$\pm$0.05~mJy, obtained from the
tables in Appendix A. The dashed line is a composite `fit' to the SED assuming thermal dust emission 
with slope 2.45$\pm$0.05 dominates at 1 to 16~mm and free-free with spectral index 0.60$\pm$0.05 
dominates at 3.5 to 6.2 cm (both represented by dotted lines). The two respective contributions are 
`anchored' at 3~mm with a flux of 45$\pm$5~mJy and at 6.2~cm with a flux of 0.325$\pm$0.025~mJy. See 
Appendix C for further details.}
\label{FIG-SEDcompo}
\end{figure}

Using total fluxes of 1110$\pm$180~mJy (Walsh et al. 2014), 560$\pm$140~mJy (Henning et al. 1994, 1998), 
45$\pm$5~mJy (Table~\ref{TAB:hd100-obs3mm}) and 8.5$\pm$0.5~mJy (Table~\ref{TAB:hd100-obs7mm}), at 
respectively 'effective' frequencies of 324~GHz, 236~GHz, 90~GHz and 44~GHz, the best fit spectral index 
$\alpha$ to the SED is $\sim$ 2.45$\pm$0.08 (formal fit error). This is shown in Figure~\ref{FIG-SEDcompo}, 
where the entire spectral energy distribution from $\sim$ 1 to 60~mm is presented. The 3~mm, 16~mm, 3.5~cm 
and 6.2~cm data are presented in Appendix A and discussed in Appendix C, where respective contributions 
from thermal dust and free-free emission are discussed. 

With a spectral index from 0.9 to 7~mm of $\sim$ 2.45, and assuming optically thin emission, the 
dust emissivity index $\beta$, where $\kappa \propto \nu^{\beta}$, is $\sim$ 0.45. A rough correction 
to this value can be made where there is some contribution from optically thick emission, which tends 
to increase $\beta$. From independent samples of T Tauri stars both Rodmann et al. (2006) and Lommen 
et al. (2007) inferred values of $\delta$, the ratio of optically thick to thin emission, of around 
0.2. Using the relation $\beta_c = (\alpha-2)(1+\delta)$ we therefore find the corrected value to be 
$\beta_c \simeq 0.54$. Since our 16~mm flux, which is spatially resolved, also falls nicely on a fitted 
SED slope of 2.45 -- after accounting for a free-free emission component (see Appendix C) -- the case 
for invoking an optically thick component is weakened. We also note the work here of Ricci et al. (2012). 
Thus, in the following we will assume a value of $\beta = 0.5\pm0.1$.

Such a relatively low value for $\beta$ suggests that the dust grains in the HD100546 disk have undergone 
significant growth from their original ISM sub-micron sizes, where $\beta$ is typically around 1.7 (Draine 2006). 
As noted by van der Marel et al. (2013), dust grains larger than $\sim$ three times the observing wavelength do 
not contribute to the opacity. Thus our 7~mm data suggests growth up to at least 2~cm. But given we have also
detected dominant thermal dust emission at 16~mm, and probably a significant dust emission component up to 3.5 
and even 6.2 cm (Fig~\ref{FIG-SEDcompo}, Appendix C), then we can say with high confidence that growth has 
occurred up to at least 5~cm and probably up to around 20~cm. 

We can potentially improve upon these rough estimates by appealing to cosmic dust models. Using the astrosilicate 
curve in Figure~6 of Draine (2006) our value for $\beta$ can only be compatible with a grain size distribution 
$dn/da \propto a^{-p}$ for $p$ around 3.2 and maximum size $a$ of 100~cm. A similar inference can be made from 
the porous icy grain model in Testi et al. (2014), composed of astronomical silicates, carbonaceous material and 
water ice, with relative abundances as in Pollack et al. (1994) and a porosity of 50. And similarly for the 
compact, segregated spheres model of Natta et al. (2004, 2007). In these cases the maximum grain size is again 
$\geq$ 100~cm for $p \leq 3.3$, or about 10~cm for $p$=2.5. 

Interestingly, no grain model considered above can produce a $\beta$ less than about 0.5, regardless of maximum 
grain size, unless the index $p$ in the grain size distribution is $\leq$ 3. This suggests that the dominant 
process governing the grain size distribution in the millimetre-emitting HD100546 disk is not one of fragmentation 
and shattering, as might be expected in a collisionally evolved disk (or the ISM where $p$=3.5). Instead, an index 
$p \leq 3$ is expected for the case in which coagulation and growth determines the size distribution (e.g. Pinilla 
et al. 2014; Birnstiel et al. 2010; Natta \& Testi 2004 and references therein). On the other hand, as noted by 
Dullemond \& Dominik (2005) this growth process itself does maintain a population of small grains -- via shattering 
collisions -- which would otherwise be lost to the system on a timescale much shorter than its few-to-several Myr 
age. The signature of these grains is at much shorter wavelengths, namely optical/near-IR scattering and mid-IR 
thermal emission. Indeed, from radiative transfer modelling where $p$ was a free parmaeter, Harker et al. (2005) 
found a best fit value of $p$=3.5 for the SED up to 100~$\mu$m.  

The fact that grains as large as several tens of centimetres exist at radii out to a few tens of AU is a specific 
prediction of the dust trap models previously mentioned. Given the age of the HD100546 system is a few to several 
Myrs, under the ordinary radial drift scenario such grains would have long ago spiralled into the star -- the 
so-called `metre-size barrier'. A long-lived mechanism must therefore be acting to keep such particles out there, 
and, given the size distribution index is $\leq$ 3, allow their growth to be ongoing. 
 
Finally, from models that combined hydrodynamical simulations with dust evolution of a planet-sculpted cavity, 
Pinilla et al. (2014) found a linear relation between the disk integrated 1--3~mm spectral index and the cavity 
radius for transitional disks with a dust trap. The correlation is such that the millimete SED slope $\alpha_{1-3~mm}$ 
increases with cavity radius, and provides reasonable agreement with the observed properties of 20 transition 
disks. It results from the fact that for disks where the pressure bump is located further from the star (i.e. wider 
cavities), the maximum and critical grain sizes -- where the latter is the particle size that is perfectly trapped 
by the pressure bump -- are expected to be smaller. The correlation is $\alpha_{1-3~mm}$ = 0.012$\times$R$_{cavity}$+2.15, 
which, using the mm-inferred cavity radius of 25~AU for HD100546, predicts a spectral index of 2.45, in precise 
agreement with the value reported here. 

\subsection{Disk mass and inferred gas-to-dust mass ratio}

The dust mass in the disk can be obtained from the deceptively simple equation 
$M_{dust}=(F_\nu~d^2)/(B_{\nu}(T_d)~\kappa_\nu)$, where $F_\nu$ is the observed flux at frequency 
$\nu$, $d$ is the distance, $B_{\nu}(T_d)$ is the Planck function at the dust temperature $T_d$ and 
$\kappa_\nu$ is the dust opacity. Whilst the distance and flux are well known, the derived mass can 
be poorly estimated due to inadequate knowledge of the dust temperature and most especially the dust 
opacity $\kappa$. In addition to composition, the latter has a strong dependence on grain size, 
decreasing as the size increases (e.g. Miyake \& Nakagawa 1993). So an overestimate of $\kappa$, 
e.g. by using a value for small grains for a disk with ostensibly large grains, would seriously 
underestimate the dust mass of the disk. 

Therefore, we use a range of $\kappa$ values at 7~mm for amorphous silicate spheres from Draine (2006) to give 
a sense of the uncertainties involved. These are 0.2, 0.09 and 0.03~cm$^{2}$~g$^{-1}$, appropriate for a grain 
size distribution with an index of $p$=3.5, minimum grain size $a_{min}$ of 3.5~\AA, and maximum size a$_{max}$ 
of 1, 10 and 100~cm respectively. These give dust masses of $\sim$ 0.33, 0.73 and 2.2~M$_{\rm J}$ respectively, 
assuming a temperature of 40~K. A temperature variation of $\pm$20~K changes these estimates by factors of 
$\sim$ 0.7 for 60~K and $\sim$ 2.0 for 20~K. Similar masses are obtained using the opacities for the grain 
compositions, structures and sizes given in Natta \& Testi (2004), and in this case account can be taken of a 
lower size distribution index $p$.  

If a relatively flat grain size distribution exists in the disk around HD100546, as discussed above, then it 
means that relatively more mass is contained within larger particles. The opacity $\kappa$ could therefore be even 
lower than assumed above, and the derived masses subsequently increase. Notably, these masses are at least an order 
of magnitude larger than the estimate of Walsh et al. (2014) from ALMA 1~mm data, who used both a higher $\kappa$ 
and higher $\beta$. 
 
Panic et al. (2010) observed various rotational transitions of CO in the HD100546 disk and estimated a mass of 
$\sim 10^{-3} - 10^{-2}$~M$_{\odot}$, or 1--10~M$_{\rm J}$, of molecular gas within a radius of approximately 
400~AU from the star. Without even accounting for the factor $\sim$ 8 difference in the disk outer radii of the 
molecular and millimetre continuum emission, there is a clear indication in the ATCA data that the gas-to-dust 
mass ratio in the HD100546 disk has decreased from its canonical value of $\sim$ 100 in the ISM and molecular 
clouds. Unless a significant amount of material is hidden from view because it is in bodies much larger than a 
metre or so in size, there is apparently little material available for further gas-giant planet building within 
the HD100546 disk. We refer to Bruderer et al. (2012) for a detailed discussion of the gas-to-dust mass ratio 
toward HD100546.

\subsection{Nature of the central emission peak}

Our observations have revealed a local maximum in emission at the position of the central star, with 
a similar flux level as that from the two peaks on either side along the disk major axis 
(Figures~\ref{FIG-7mm-normcal} and \ref{FIG-7mm-selfcal}). Isella et al. (2014) also found a central 
emission peak at 7~mm for the transition disk source LkH$\alpha$ 15. An obvious question to be posed 
concerns the physical mechanism behind the emission, options being the stellar photosphere, hot dust 
from an inner disk, free-free from a wind or some thermal shock, accretion or other non-thermal process 
where material is accreting onto the star. All these have an observational basis for being tested, 
since an inner dust disk was found by Panic et al. (2014), and both outflow and accretion signatures 
have been detected via various phenomena (e.g. Viera et al. 1999; Deleuil et al. 2004; Grady et al. 2005; 
Guimaraes et al. 2006). Indeed, our own radio observations, combined with IR recombination line data 
and discussed in Appendix C, suggest an outflow rate of $\geq 10^{-8}$ M$_{\odot}$/year. 

All of these processes probably contribute at some level. The inner dust disk from Panic et al. (2014) 
only extends to a maximum radius of $\sim$ 0.7~AU and probably even smaller, of the order of 
$\sim$ 0.3~AU (Mulders et al. 2013b). The total mass of dust in this disk was not estimated by these 
authors, but Benisty et al. (2010) infer it to be only $\sim 3\times10^{-7}$~M$_{\rm J}$ composed of 
only 0.1--5 $\mu$m sized grains. Also, its temperature would obviously be in the range of hundreds 
up to even a thousand Kelvin, much hotter than the colder and more massive disk responsible for the 
bulk of the millimetre emission. It thus seems reasonable to conclude that the lower dust mass, 
smaller grain size and higher dust temperature would all conspire to produce much less 7~mm emission 
than is observed for the central peak, especially given the latter's comparable flux level to those 
of the major axis peaks of the outer disk. 

We instead assess that the central emission seen at 7~mm arises from free-free emission from an 
outflowing wind. The radio emission we found at 4.8 and 8.64~GHz has a spectral index very close to 
the canonical value of $\sim$ 0.6 for a spherical wind (Appendix C; Panagia \& Felli 1975). 
Further, the fluxes themselves suggest that up to about 10\% of the total emission at 7~mm could 
originate from this wind. This implies that the central peak would have a flux of around 1~mJy, 
which seems consistent with our observations. 

Finally, we can almost certainly rule out thermal shock and/or accretion emission since, as noted by 
Skinner, Brown \& Stewart (1993) in their radio continuum survey of Herbig Ae/Be stars, the expected 
spectral index is $-0.1$. This is much different to the index found at 4.8--8.64~GHz. Also, based on 
the formulae given in Skinner et al. and using our own plus literature values for accretion and/or 
wind mass loss rates for HD100546, the expected 4.8--8.64~GHz fluxes are far above what was measured. 
See Appendix C for further details.

\subsection{Modelling the large grain population}

Several seemingly quite successful attempts have been made to model the HD100546 SED with radiative 
transfer calculations up to 1--3~mm, assumimg a disk extending from about 10 to 400--500~AU. For instance, 
using similar data sets Benisty et al. (2010) and Tatulli et al. (2011) find a disk dust mass of around 
0.5~M$_{\rm J}$ for a grain size distribution between 1~$\mu$m and 1~cm ($p$=3.5). Bouwman et al. (2003) 
and Doering (2008) instead find an order of magnitude lower mass, around 0.065~M$_{\rm J}$, with the 
former using grain sizes of 10--200 $\mu$m ($p$=2) and the latter 0.01~$\mu$m--1~cm ($p$=3.5). 

As noted by Espaillat et al. (2010) the largest uncertainty in any model is in the adopted mass opacity 
of the dust. In the case of HD100546 those models with the least mass in larger grains, and thus opacities 
which are too high, severely underestimate the disk mass calculated from the observed fluxes, yet can still 
produce a reasonable match to the SED up to 1~mm. But even so, in all cases where the model extends to 3~mm 
the flux is underpredicted. It is anticipated that if these models were extended out to 7 and 16~mm they 
would increasingly fail to predict the observed flux, necessitating such data as presented here. 

Since none of the models optimised the fit to the long wavelength data, i.e. at $\geq$ 3~mm, none 
attempted to assess the large grain population. Although we believe our data is sufficiently unique 
to stand on its merits, we have attempted to model the mm--cm fluxes using the irradiated accretion 
disk models of D'Alessio et al. (2005). These models do not consider the disk to be simply reprocessing 
stellar radiation, but also to be self-luminous through an accretion flow. A comprehensive model grid 
is available for download from the web\footnote{https://www.cfa.harvard.edu/youngstars/dalessio/}, 
covering a large range of parameter space which overlaps with 
those inferred from our data, such as the maximum grain size, power law size distribution, and the 
accretion rate. Further, proper stellar atmosphere (Kurucz) models are used as input, rather than the 
assumption of a blackbody.

The D'Alessio models have been used to successfully model the SED of Herbig Ae/Be stars. For example, 
Merin et al. (2004) produced a good fit to the SEDs of HD34282 and HD141569, both of which overlap in 
properties to HD100546. We follow the approach they used for HD34282, and instead of trying to match the 
entire SED only consider the disk midplane component, responsible for much or all of the millimetre to 
centimetre emission. We consider the data from 0.9~mm all the way up to 6~cm, corrected for free-free 
emission (see Appendix C for details of the correction). 

Some of the free parameters can be fixed or constrained to a narrow range given what is known about 
HD100546, e.g. the distance is fixed at 100~pc. A range of stellar effective temperatures and ages 
bracket the $\sim$ 2.4--2.8~M$_{\odot}$ mass and 30--100~L$_{\odot}$ luminosity of HD100546 reported 
in the literature (e.g. van den Ancker et al. 1997; Blondel \& Djie 2006), or even the figures of 
4$\pm$1~M$_{\odot}$ and log(L/L$_{\odot}$)=2.65$\pm$0.25 of Levenhagen \& Leister (2004). So we have 
examined models with T$_{\rm eff}$ from 5000--10000~K and ages 1--10~Myr. The HD100546 disk inclination 
is 40--50$^{\circ}$, between the two values of 30$^{\circ}$ and 60$^{\circ}$ offered in the grid, but 
which anyway had little effect on the fluxes and no detectable influence on the mm--cm SED slope. The 
outer radius is fixed to be 300~AU for the 6000-10000~K cases, closest to the observed radius of the 
millimetre-emitting disk of HD100546 of $\leq$ 100~AU, whilst the 5000~K model does have a 100~AU 
radius option. 

We stress here that we are not claiming that T$_{\rm eff}$ for HD100546 is as low as 5000--8000 K. As
a late B star it is certainly higher, up to 10500 K (van den Ancker et al. 1997) or even 11000--12000 K 
(Acke \& Waelkens 2004; Levenhagen \& Leister 2004), but those model grids are not available in the d'Alessio 
web archive. Rather, our aim here is simply to show that the long wavelength SED fluxes and slope are 
essentially independent of the stellar properties, through effective temperature and mass, but instead 
principally determined by respectively the luminosity (stellar plus accretion) and grain opacity (and thus 
disk mass).

\begin{table*}
\caption{Summary of D'Alessio models that can fit the HD100546 mm to cm emission}
\begin{tabular}{llllllllll}
\hline
T$_{eff}$ &  Age  & Mass          & Luminosity    & $\dot{M}$               & R$_{in}$ & R$_{out}$ & p & $a_{max}$ & M$_{\rm dust}$ \\ 
   (K)    & (Myr) & (M$_{\odot}$) & (L$_{\odot}$) & (M$_{\odot}$~yr$^{-1}$) & (AU)     & (AU)    &  &  (cm) & (M$_{\rm J}$)  \\ \hline
\multicolumn{10}{c}{Disk mid-plane}   \\
 10$^{4}$         &   1  &  4.0  & 251.9  &  10$^{-7}$  & 1.34 &  300  & 2.5  &  10  &  4.3 \\   
 10$^{4}$         &  10  &  2.3  &  29.5  &  10$^{-7}$  & 0.48 &  300  & 2.5  &  10  &  5.7 \\   
 9$\times10^{3}$  &   1  &  4.0  & 210.0  &  10$^{-7}$  & 1.24 &  300  & 2.5  &  10  &  4.6 \\   
 9$\times10^{3}$  &   3  &  2.7  &  71.0  &  10$^{-7}$  & 0.72 &  300  & 2.5  &  10  &  5.1 \\   
 9$\times10^{3}$  &  10  &  2.0  &  17.1  &  10$^{-7}$  & 0.38 &  300  & 2.5  &  10  &  6.0 \\   
 8$\times10^{3}$  &  1   &  4.0  &  165.1 &  10$^{-7}$  & 1.09 &  300  & 2.5  &  10  &  5.1 \\   
 7$\times10^{3}$  &  1   &  4.0  &  130.4 &  10$^{-7}$  & 0.97 &  300  & 2.5  &  10  &  5.4 \\   
 6$\times10^{3}$  &  1   &  3.5  &  59.1  &  10$^{-7}$  & 0.65 &  300  & 2.5  &  10  &  6.1 \\   
 5$\times10^{3}$  &  1   &  3.0  &  14.9  &  10$^{-7}$  & 0.34 &  300  & 2.5  &  10  &  7.1 \\   
 5$\times10^{3}$  &  1   &  3.0  &  14.9  &  10$^{-7}$  & 0.34 &  100  & 2.5  &  10  &  2.7 \\   
\hline
\end{tabular}
\label{TAB-Alessmods}
\end{table*}

\begin{figure}
\includegraphics[scale=0.380]{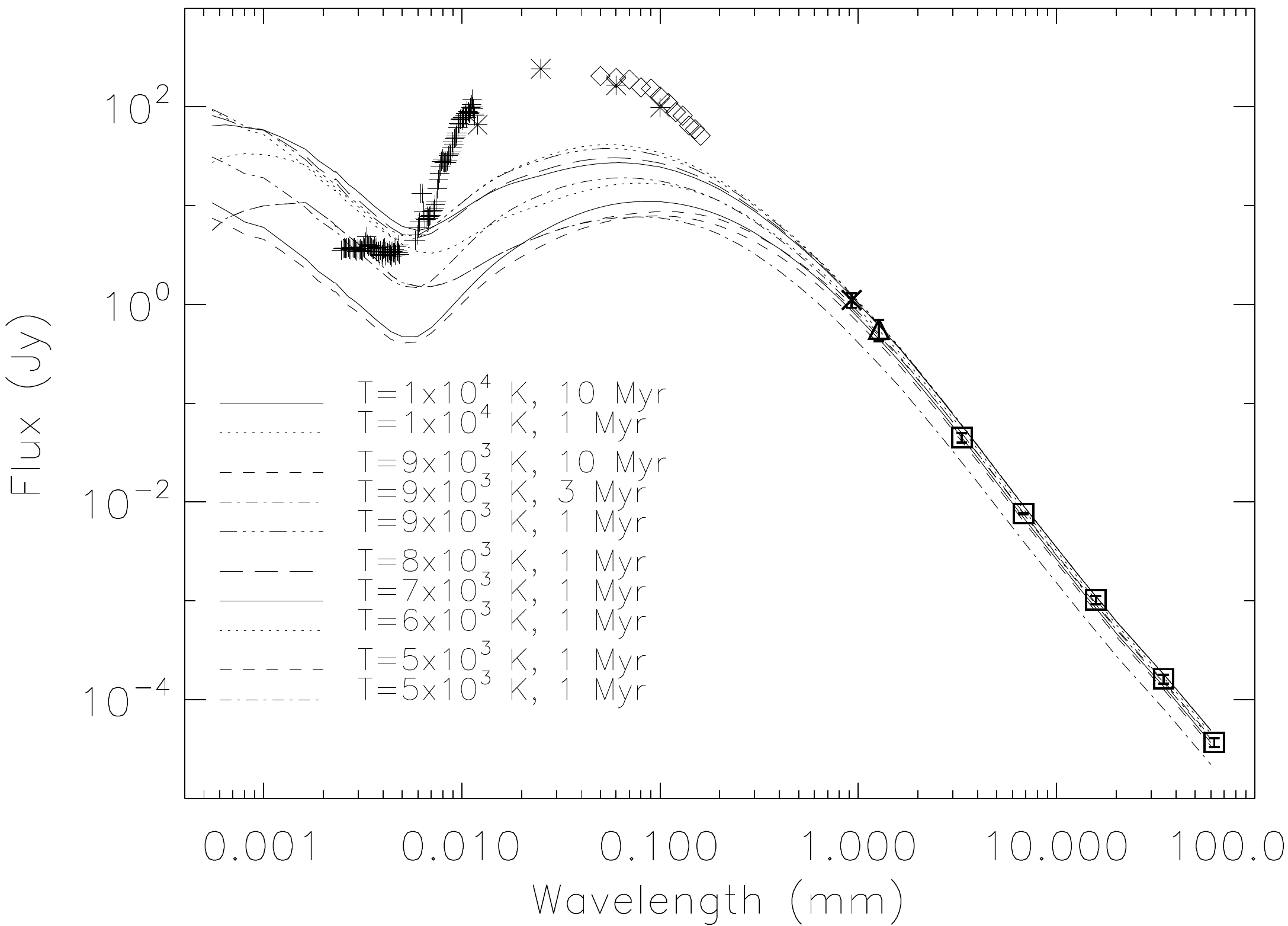}
\caption{Spectral energy distribution of HD100546 from the near-infrared to centimetre regime, with 
fits to the disk midplane emission from the D'Alessio irradiated accretion disk model as listed in 
Table~\ref{TAB-Alessmods}. The 2.5--11.5~$\mu$m (plus signs) and 50--200~$\mu$m (diamonds) data are 
respectively from ISOPHOT and ISO-LWS (binned in 10~$\mu$m intervals) taken from the ISO Data 
Archive.
Other points (asterisks) are the IRAS 12, 25, 60 and 100~$\mu$m fluxes. The ALMA 900~$\mu$m flux (cross)
is from Pineda et al. (2014) and Walsh et al. (2014), whilst the SEST flux (triangle) is the mean of 
the two values reported by Henning et al. (1994) and Henning et al. (1998). ATCA data reported here 
are shown as squares, corrected for free-free emission.}
\label{HD100-Alessio}
\end{figure}

Indeed, experience showed that the main input parameters in the models which affected the SED were the 
mass accretion rate $\dot{M}$, the maximum grain size $a_{max}$ and the index $p$ of the power law grain size 
distribution. Available in the grid were four values of $\dot{M}$ from $\dot{M}=10^{-9}$ to 
10$^{-6}$~M$_{\odot}$~yr$^{-1}$, six values of $a_{max}$ from 1~$\mu$m to 10~cm, and two values of $p$, 2.5 and 
3.5 representing dominant growth and fragmentation processes respectively. We considered variations of these 
parameters for multiple combinations of the stellar properties, shown in Table~\ref{TAB-Alessmods}, which 
also summarises the parameter combinations that could reasonably fit the mm-to-cm spectrum 
(see Figure~\ref{HD100-Alessio}).

We found that no combination of stellar properties with an $\dot{M}$ of 10$^{-6}$ or 
10$^{-9}$~M$_{\odot}$~yr$^{-1}$ could match the observed fluxes. The former produced far too much 
emission whilst the latter provided far too little. An $\dot{M}=10^{-7}$~M$_{\odot}$~yr$^{-1}$, along 
with $a_{\max}=10$~cm and index $p=2.5$ provided by far the best match to the data. If $a_{max}$ was 
reduced to 1~cm then an accretion rate of $\dot{M}=10^{-8}$~M$_{\odot}$~yr$^{-1}$ could almost fit the 
fluxes up to 3~cm, but bigger grains are needed for the 6~cm data point. For $p=3.5$ with $a_{max}$ of 
either 1~cm or 10~cm the spectral slope was clearly too steep. The models definitely needed more of the 
available mass in the bigger grains (or equivalently less in the small grains and thus less emissivity 
at shorter wavelengths). 

Figure~\ref{HD100-Alessio} shows the models from Table~\ref{TAB-Alessmods}. We have chosen not to include
shorter wavelengths (uv, optical, near-IR) as that part of the SED has been adequately treated elsewhere 
(e.g. Mulders et al. 2011; Malfait, Bogaert \& Waelkens 1998), and our intention is to instead treat the mm--cm 
component which has not been addressed in detail before. As previously noted, and now demonstrated in 
Figure~\ref{HD100-Alessio}, the fits in the mm--cm regime are insensitive to the stellar properties 
(T$_{\rm eff}$, mass), and instead are particularly sensitive to the dust mass in the disk via the relative 
proportions of small and large particles. The average mass of around 5~M$_{\rm J}$ for the 300~AU radius disks 
compares favourably with our observationally derived value of $\sim$ 1~M$_{\rm J}$ discussed previously. The 
only model in Figure~\ref{HD100-Alessio} which does not fit the fluxes -- but still reproduces the slope -- 
is for a disk radius of 100~AU, and thus a factor of two lower dust mass. 

As well as those papers cited above, other authors have optimised their radiative transfer models to fit shorter 
wavelength data, i.e. up to only a few hundred microns, and especially to constrain the dust mineralogy (e.g. 
Mulders et al. 2011; Harker et al. 2005; Elia et al. 2004; Dominik et al. 2003; Malfait, Bogaert \& Waelkens 1998). 
Using the D'Alessio irradiated accretion disk prescription we were unable to find a model to fit the predominantly 
surface layer origin near-, mid- and far-infrared emission of HD100546 that was physically consistent with the 
millimetre--centimetre model. This is apart from a general result that it almost certainly requires a separate 
population of much smaller grains, with a maximum size between 1 and 10~$\mu$m, in agreement with the aforementioned 
earlier studies. 

But given the differences between some of the (fixed) model parameters and real properties of the disk, 
the inability to find a physically consistent model is not surprising. For example, the model and observed 
1-16~mm disk outer radii differ by a factor of around 3. Most crucially the model assumes the disk extends 
all the way in to the dust sublimation radius, typically $\leq$ 1~AU, whereas there is a $\geq$ 10~AU wide 
gap between the compact inner disk of HD100546 (which generates the near-IR emission) and the extended outer 
disk (the inner wall of which produces most of the mid-IR emission). Though not completely cleared of gas 
and dust, the much reduced density in the gap will obviously result in significantly less emission than the 
models would otherwise predict. 

These apparently have a secondary effect on the mm--cm SED. Indeed, it is impressive that the d'Alessio models 
do so well in fitting the HD100546 mm--cm data. Presumably the emission from the colder outer disk and its 
midplane has little 'memory' of the region (e.g. its density and temperature structure) in which original 
stellar and accretion ultraviolet radiation is processed into infrared radiation which eventually heats the dense 
disk interior. In conclusion, we find that the D'Alessio models can reproduce the mm--cm SED of the HD100546 disk 
with a set of parameters, specifically $\dot{M}$, $p$ and $a_{max}$, which had previously been inferred 
directly from the data. 

\subsection{Conjecture on the type of planetary system hosted by HD100546}

\subsubsection{General review of exoplanet statistics}

As some indication of what may be expected of the planetary make-up of transition disk systems we can look 
to the statistics of confirmed exoplanets and/or brown dwarf companions around solar analogues. Though a 
generalisation, the logic behind this is simply that the end result of planet formation should at least partly
be a reflection of what occurred during the formation phase, though of course keeping in mind the possibility 
of perturbing processes such as scattering and migration during the intervening period. 

There are now a sufficient number of confirmed exoplanets, with well defined properties, that we can begin to 
speculate on what the planetary system of their evolutionary precursors could (or should) look like. For this 
purpose we use as one resource the Exoplanet Orbit Database (EOD) of Wright et al. (2011), which includes bodies 
up to 24~M$_{\rm J}$. Further, numerous massive exoplanet and/or brown dwarf companion direct imaging surveys 
toward sun-like stars have been of sufficient sensitivity that statistical estimates can be made for the 
frequency of such bodies at various orbital radii.

At the time of writing there are just over 1500 confirmed exoplanets in the EOD, discovered primarily using 
transit and radial velocity techniques, but also including a few tens from direct imaging. These have their 
own -- in many cases different -- sensitivities to planets of different masses at various orbital radii (or 
equivalently orbital periods), and thus there is a risk of succumbing to selection bias when extracting 
statistical data. Whilst taking on board the specific caution of Wright et al. in this regard, here we only 
infer gross properties and check their robustness against the different discovery techniques, the specific 
biases of which can partially compensate each other.   

\begin{figure}
\includegraphics[scale=0.10]{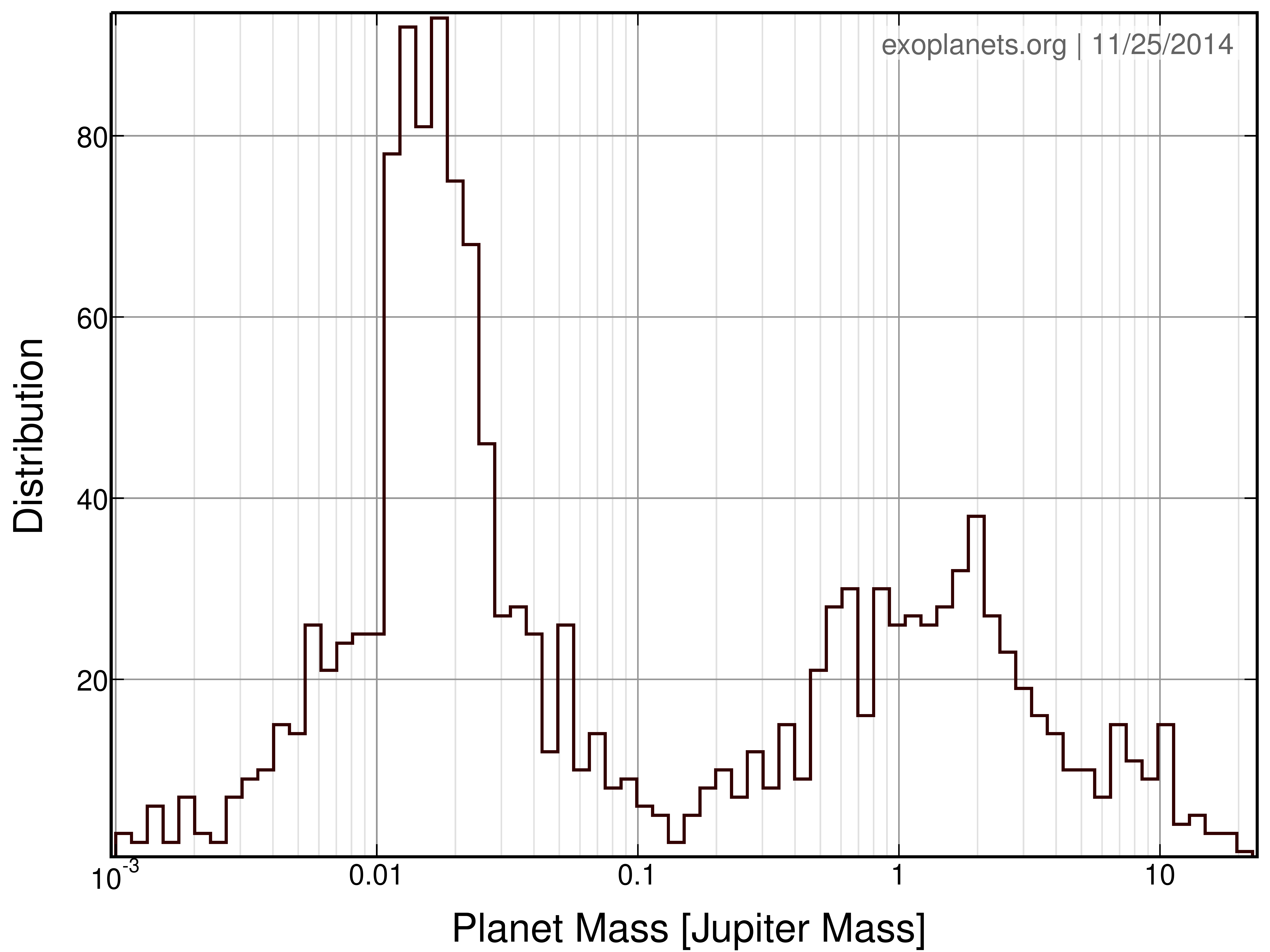}
\caption{Histogram of confirmed exoplanet masses. The first peak at 0.01-0.03~M$_{\rm J}$ is mainly contributed 
by transit detections, and the second peak at 0.5-3~M$_{\rm J}$ mainly by radial velocity (RV) detection. Each 
detection method does reveal both peaks, though at slightly different positions. For the RV method the peaks are 
at 0.04--0.05~M$_{\rm J}$ and 2~M$_{\rm J}$, whilst for the transit method they are at 0.01--0.02~M$_{\rm J}$ 
and 1~M$_{\rm J}$. This plot was generated using the database of Wright et al. (2011).}
\label{plmass-histo}
\end{figure}

With the caveat that only a handful of confirmed exoplanets orbit host stars with mass $\geq$ 2~M$_{\odot}$, 
the log$_{10}$(Mass) distribution of exoplanets has two broad but distinct peaks, one at 0.01--0.03~M$_{\rm J}$ and 
the other at 0.5--3~M$_{\rm J}$. See Figure~\ref{plmass-histo}, which only includes RV and transit discoveries. 
With respectively about 1000 and 500 planets, the dominant population is reversed between transit and radial 
velocity discovery techniques, probably due to their different mass and orbital radii sensitivities. But 
nevertheless both show the two peaks, and even more clearly a distinct minimum at 0.1-0.2~M$_{\rm J}$. There 
is possibly a third minor peak at 7--10~M$_{\rm J}$, which, despite its questionable significance, is also the 
mass range in which most exoplanets found via direct imaging lie. 

Most interestingly for the purpose of this paper is that there are only a handful of exoplanets with mass 
$\geq$ 10~M$_{\rm J}$. Despite the good sensitivity of the radial velocity technique to bodies in this mass 
range, it is possible that it has not fully sampled the parameter space where these bodies might exist beyond 
a radius of about 5~AU, the region of interest for HD100546. However, many imaging surveys of solar-like stars, 
with ages spanning tens to hundreds of Myr, have been sensitive enough to detect such bodies at separations beyond 
5~AU, and their rarity has essentially been confirmed (e.g. Chauvin et al. 2014; Ma \& Ge 2014; Biller et al. 
2013; Janson et al. 2013; Nielsen et al. 2013; Wahhaj et al. 2013; Evans et al. 2012; Nielsen \& Close 2010; 
Metchev \& Hillenbrand 2009; Apai et al. 2008; Kraus et al. 2008; Nielsen et al. 2008; Kouwenhoven, Brown \& Kaper 
2007; Lafreniere et al. 2007; Lowrance et al. 2005). Indeed, most of these studies have found results consistent 
with the statement of Brandt et al. (2014), namely that with high confidence only a few percent of stars host 
planets with mass 5--70~M$_{\rm J}$ at radii 10--100~AU. The paucity of such bodies has been termed the ''brown 
dwarf desert'', the driest part of which was determined to be around 30$\pm$20~M$_{\rm J}$ by Grether \& Lineweaver 
(2006). 

\subsubsection{HD100546 planets -- limits and putative detections}

For HD100546 itself, from HST STIS data Grady et al. (2005) concluded they would have easily detected a young 
chromospherically active low mass stellar companion, of spectral type earlier than about M5, within 2.5--10~AU. 
It is difficult to put hard constraints on the mass of a companion they could or could not have detected. On the 
one hand they used AT Mic as a template, an 8--20~Myr binary whose components are flare stars with spectral types 
of M4-4.5 Ve and masses of 0.16-0.31~M$_{\odot}$. On the other hand their upper bound to HD100546's chromospheric 
and transition region flux meant they could not have detected an 8–-10~Myr 0.06~M$_{\odot}$ brown dwarf. Thus, 
the suggestion is that they might have detected a companion with a mass between about 60 and 150~M$_{\rm J}$, 
especially if its age was less than 5--10~Myr.
 
Outside of the cavity, Quanz et al. (2013a) and Currie et al. (2014) directly imaged a putuative giant protoplanet 
(HD100546 b) at 3.8~$\mu$m (L' filter) at an observed separation of 47$\pm$4~AU and position angle of $\sim$ 10$^{\circ}$. 
This was deprojected to 53~AU by Quanz et al. (2015) using the ALMA-inferred disk inclination and position angle 
from Pineda et al. (2014), or 68~AU by Quanz et al. (2013a) using the NIR scattering disk figures from Quanz et al. 
(2011). The emission consists of a point (or slightly extended) source embedded in a finger-like region extending along 
a similar PA. Assuming the point source to arise from photospheric emssion of a massive planet or substellar object, 
the authors infer a mass of around 15--20~M$_{\rm J}$ for an age of 5--10~Myr. Quanz et al. (2015) recovered the 
putative planet at 3.8~$\mu$m as well as at 4.8~$\mu$m (M' filter), but despite having the requisite sensitivity they 
failed to detect it at 2.1~$\mu$m (K$_{\rm S}$ filter), and similarly for Boccaletti et al. (2013) at 2.15~$\mu$m. 

Interestingly, the 3.8~$\mu$m photometry of Quanz et al. (2013a) and Quanz et al. (2015), despite using the same instrument 
and filter, resulted in significantly discrepant magnitudes, namely 13.2$\pm$0.4 and 13.92$\pm$0.10 respectively. Our own 
mass estimates from the photometry, using the AMES COND models (see below) suggest that for L'=13.2 the mass is 
16--21~M$_{\rm J}$ for an age of 5--10~Myr, consistent with 17--21~M$_{\rm J}$ implied by the M'=13.33 measurement but 
inconsistent with the upper limit of 9--12~M$_{\rm J}$ for K$_{\rm S}$ $>$ 15.43. On the other hand L'=13.92 implies a 
mass of 11--14~M$_{\rm J}$, consistent with the K$_{\rm S}$ upper limit mass but inconsistent with the M'-implied mass. 
Such inconsistencies, plus other properties such as its morphology, prompted all authors to favour a model wherein the
source is a planet still in the process of formation, and where the point or slightly extended component consists of the
planet plus its circumplanetary disk. The luminosity provided by the circumplanetary disk means that the planet mass could 
be much less than 10~M$_{\rm J}$ and perhaps as low as 1~M$_{\rm J}$ (Currie et al. 2014).

In addition to the finger-like extension surrounding HD100546 b, all authors also found another emission feature at 
a similar separation from the star and with a similarly extended morphology -- though without an obvious point source 
component -- $\sim$ 90$^{\circ}$ away on the southeast side of the disk. It is almost certainly due to thermal emission, 
rather than scattering, as none of Boccaletti et al. (2013), Avenhaus et al. (2014a) or Quanz et al. (2015) detect it 
at K$_{\rm S}$, and it appears to brighten from L' to M'. Currie et al. (2014) suggest it may be an additional spiral 
arm of the HD100546 disk, though it would differ from the other known spiral arms which are only seen in scattered 
radiation (and thus are disk 'surface' features, e.g. Grady et al. 2001 and Ardila et al 2007). 

At $\sim$ 50~AU from the central star the emission cannot be reprocessed stellar radiation and so its heating mechanism is 
currently unknown. By extrapolation we also suggest that the origin of the extended emission surrounding HD100546 b is also 
unknown, though Quanz et al. (2015) suggest it arises from localised compressional heating of the HD100546 circumstellar 
disk. Perhaps even more controversially, but given the unknown physics behind the extended emission, we suggest that the 
nature of the feature identified as the protoplanet HD100546 b remains open to interpretation. Having said that, the fact 
that it is co-moving with HD100546 means it is certainly physically associated with it, e.g. a feature within the 
circumstellar disk. Notably, no sign of either feature is seen in our mm-wave data (see Section 4.11).   
 
Finally, the work of Ardila et al. (2007) using HST and Boccaletti et al. (2013) provides stringent constraints on the 
presence of other planetary-mass bodies from 90~AU to 1200~AU and 50~AU to 400~AU respectively. Essentially they were 
able to exclude the presence of 5--10~Myr old objects with mass $\geq$ 20--30~M$_{\rm J}$ at around 50~AU, with progressively 
lower limits with increasing separation, down to only $\sim$ 2~M$_{\rm J}$ at 150~AU to 400~AU and 4~M$_{\rm J}$ beyond 400~AU.

\subsubsection{Single companion to HD100546 interior to 13~AU?}

As previously noted, the mere existence of a gap -- or cavity in the case of HD100546 -- may suggest the 
presence of a massive orbiting body, whilst the gap width (cavity radius) may be used to infer the body's mass. 
For instance, Mulders et al. (2013b) performed hydrodynamic modelling, and found that a 20--80~M$_{J}$ body, with 
a preferred value of 60~M$_{J}$ and orbiting at 8--10~AU, well reproduced the surface density profile of the 
rounded wall. This puts the gap-sculpting body firmly in the brown dwarf category.

A similar mass can be found without recourse to hydrodynamic simulations. Dodson-Robinson \& Salyk 
(2011) found that, within 0.5~Myr, the maximum gap opened by a planet would not exceed about five 
Hill radii, where the Hill radius is expressed as $R_{H}=r_{p}\times(M_{P}/3M_{*})^{1/3}$, with 
$r_{P}$ and $M_{P}$ being the planet's orbital radius and mass respectively, and $M_{*}$ the stellar 
mass. For a 50~M$_{\rm J}$ body orbiting at 13~AU from a 2.4~M$_{\odot}$ star $R_{H}$ is $\sim$ 2.4~AU, 
so that a gap is opened from 1 to 25~AU, consistent with the 1--7~mm data for HD100546. 

There is however a potential problem with such a massive single body being present in the HD100546 disk 
cavity. It is very close to the lower limit found by Grady et al. (2005) discussed above, and may even 
exceed it assuming a younger age than 8--10~Myr. For instance, according to the Baraffe et al. (2003) 
evolutionary models, at 1 Myr a $\sim$ 25~M$_{\rm J}$ body has about the same luminosty of 
$\sim$ 10$^{-4}$~L$_{\odot}$ as a 50~M$_{\rm J}$ body at 5 Myr or 70~M$_{\rm J}$ at 10 Myr. There is 
in fact good evidence that HD100546 could be much younger than 10~Myr, essentially based on the presence 
of an outflow and a relatively high mass accretion rate (see below and Appendix C).

For HD100546 the 'single companion' situation can be dramatically improved by using the work of Pinilla, 
Benisty \& Birnstiel (2012). They found that by including grain evolution in their hydrodynamical simulations 
the location of the pressure bump, i.e. where the large dust grains pile up, can instead be at about 
10 Hill radii for a body with mass $\geq$ 5~M$_{\rm J}$. In this case a 10~M$_{\rm J}$ planet orbiting at 
13~AU has an $R_{H}$ of 1.4~AU, and thus can open the gap out to 25~AU seen in HD100546.

Yet this also poses a problem given that such bodies are exceedingly rare, as discussed above. Also, it is 
still only a single body. The example of our own solar system tells us there are four gas giants orbiting 
within about 30~AU, and around two thirds of the $\sim$ 1500 currently known and confirmed exoplanets are 
part of a multi-planet system (Wright et al. 2011). So why would HD100546 only host a single planet within 
its disk cavity? Whilst there is still disagreement in the literature about how large a cavity a single body 
can open, there is overall agreement that multiple (Jupiter mass) bodies will each open their own gap, which 
could overlap to produce a much broader cavity (e.g. Zhu et al. 2011). 

In this context the case of the LkCa~15 planet is instructive. It is so far the only directly imaged exoplanet 
candidate within a transitional disk cavity, whose cavity radius is $\sim$ 50~AU (Kraus \& Ireland 2012; Andrews 
et al. 2011a,b). The suspected planet orbital radius and mass are $\sim$ 16~AU and 6~M$_{\rm J}$ respectively, and 
from its spatially resolved morphology is thought to be still in the process of formation. Second epoch observations 
presented in Ireland \& Kraus (2013) show rotation by an amount consistent with that expected, whilst new M-band 
data allowed a refined mass estimate to be made of 1--3~M$_{\rm J}$. Using a stellar mass of 1.2~M$_{\odot}$, and 
even taking the higher mass estimate of 6~M$_{\rm J}$ at an orbital radius of 16~AU, gives a Hill radius of 
$\sim$ 1.9~AU, insufficient to clear the entire cavity even using the 10$R_{H}$ constraint. The situation is made 
even worse if a lower mass for LkCa~15b is assumed. Furthermore, given its presumed youth, i.e. less than 1 Myr, 
there has probably been sufficient time to clear a gap but not necessarily to clear an entire cavity all the way 
to the central star (e.g. Dodson-Robinson \& Salyk 2011).

\subsubsection{Multiple companions to HD100546 within 13~AU?}


Along with the arguments presented above, the hypothesis that HD100546 may host a multi-planet system 
within its disk cavity is driven primarily by the requirement to pass a potentially significant amount of 
material from the outer to inner disk and then onto the star, whilst still maintaining the 'appearance' of 
an empty cavity from $\sim$ 1--13~AU. There are various estimates in the literature for the mass accretion 
rate of HD100546, but it is the lower figure of less than a few times 10$^{-9}$ M$_{\odot}$/yr, and possibly 
an order of magnitude lower, from Grady et al. (2005), that appears to have gained most acceptance. But how 
reliable is it?
 
Firstly, this estimate was only indirect, based principally on a comparison of the HD100546 HST STIS 
ultraviolet spectrum to that of two other Herbig Ae/Be stars with `known' accretion rates (HD163296 
and HD104237). On the other hand, modelling of IUE spectra produced estimates as high as 
$\sim 5 \times 10^{-7}$ M$_{\odot}$/yr (Talavera et al. 1994; Blondel \& Djie 2006). Most recently, an 
accretion rate of log~M$_{\rm acc}$=$-7.23\pm0.13$, or $\sim (6\pm2)\times10^{-8}$~M$_{\odot}$/yr, was 
determined by Pogodin et al. (2012). Finally, our own 3 and 6~cm radio observations, coupled with an 
analysis of ISO hydrogen recombination line data, suggests a mass outflow rate of 
a few $\times$10$^{-8}$ M$_{\odot}$~yr$^{-1}$, and thus an accretion rate around ten times higher 
(applying the canonical conversion factor, e.g. Calvet 2004). Appendix C contains further details 
on the HD100546 accretion rate.

We thus conclude that the mass accretion rate of HD100546 is at least $5\times10^{-8}$~M$_{\odot}$/yr,
which then puts it in better agreement with correlations of accretion rate with stellar mass for Herbig 
Ae/Be stars, e.g. in Donehew \& Brittain (2011) and Garcia Lopez et al. (2006). But this presents a 
significant problem when trying to unite it with other properties of HD100546. Most specifically, the 
total mass of gas and dust of the compact inner disk is at most about 10$^{-8}$~M$_{\odot}$, with estimates 
including a few $\times$10$^{-10}$ M$_{\odot}$ of small dust grains from $\sim$ 0.25 to 4~AU (Tatulli et al. 
2011; Benisty et al. 2010), 5.1$\times$10$^{-9}$ M$_{\odot}$ of small dust grains from 0.3--9.8~AU, (Bouwman 
et al. 2003), and $\leq$ 1.5$\times$10$^{-8}$ M$_{\odot}$ of gas within the entire cavity (Brittain 
et al. 2009).

The obvious implication is thus that the compact inner disk, and even the cavity up to 13~AU, would be 
completely drained of material in much less than a year, or at best on the timescale of about a year. 
So they must evidently be continually replenished from the outer disk. But how? Dodson-Robinson \& Salyk 
(2011) find that the interconnecting tidal streams from multiple planets provide a way for an abritrary 
amount of mass transfer to occur, through an apparently optically thin hole, than would otherwise be 
the case for a single planet. For instance, they present a model in which at least a factor of two 
times more small dust particles can exist within a cavity if it is 'hidden' in geometrically thin but 
optically thick streams (i.e. with a low filling factor for its thermal emission). Inclusive of solid 
mass potentially being cloaked in much larger particles, such as pebbles, rocks, boulders and planetesimals, 
there could be much more material in the cavity than could be seen with current observational techniques.

This scenario has significant attraction for HD100546, but not merely to account for the relatively high
mass accretion rate and apparently (almost) empty cavity. It may also bear on the conjecture by Grady et al. 
(1997) and Vieira et al. (1999) that much of the accreting gas is associated with star-grazing bodies, 
possibly comets or asteroids (although this model has been questioned, but not excluded, by Beust et al. 
2001 and Deleuil et al. 2004). In this context, Dodson-Robinson \& Salyk (2011) note that their multi-planet 
model for transition disks predicts that the observed small dust (e.g. in the mid-IR) is non-primordial, 
and instead the product of collisional grinding of planetesimals. This is a very real possibility for 
HD100546 given the almost identical nature of its mid-IR spectrum to that of Comet Hale-Bopp in our solar 
system (Malfait et al. 1998). 

Also, that such a large amount of material must be crossing the cavity to refuel the inner disk and 
subsequently accrete onto the star, possibly episodically (or at least not in a continuous manner), 
may account for the numerous examples of variability seen for this system. Such variability is 
observed over different timescales of minutes, hours, days, months and years. Examples include accreting 
and outflowing gas profiles (Pogodin 1995; Grady et al. 1997; Vieira et al. 1999; Deleuil et al. 2004; 
Giumaraes et al. 2006)), optical polarization (Yudin \& Evans 1998; Clarke et al. 1999) and radio
emission (Appendix D). In particular, Vieira et al. (1999) detected an Algol-like $uvby$ photometric 
minima coincident with a H$\alpha$ line profile variation. Such Algol-like minima for Herbig stars are 
commonly interpreted as extinction variations, related to discrete events of localised material passing 
in front of the star.

\subsection{HD100546 in the context of other transition disk systems}

Consistent with the dearth of massive exoplanet or brown dwarf companions at large orbital radii toward main sequence 
stars, and despite the requisite sensitivity being achievable, it is proving difficult to find 10--30~M$_{\rm J}$ companions 
orbiting transition disk host stars, especially within but also exterior to the cavity. For instance, although statistically 
limited Andrews et al. (2011a) noted that for at least half the then known (12 or so) transition disks -- for which a search 
had been conducted -- there were no companions found with mass $\geq$ 20--30~M$_{\rm J}$ interior to the cavity, and thus 
that could be solely responsible for its sculpting. See also Kraus \& Ireland (2010) where the limits are expressed as 
5--8~M$_{\rm J}$ at 5--30~AU.

\subsubsection{Known planet or low mass brown dwarf companions}

As previously noted, only LkCa~15 has a directly imaged protoplanet candidate within the disk cavity. Otherwise, the 
$\sim$ 140~AU radius cavity of HD142527 hosts a stellar mass companion with a mass of 0.1-0.4~M$_{\odot}$ at a radius of 
about 13~AU (Biller et al. 2012; Close et al. 2014; Rodigas et al. 2014). But Casassus et al. (2012) found no planetary 
mass companions $\geq$ 12~M$_{\rm J}$ at radii of 14-35~AU, whilst Rameau et al. (2012) excluded brown dwarfs or the most 
massive giant planets at radii $\geq$ 40-50~AU, and planets $\geq$ 6-10~M$_{\rm J}$ or $\geq$ 3-5~M$_{\rm J}$ at radii 
beyond $\sim$ 100~AU or 200~AU respectively. Also, a companion brown dwarf candidate was inferred in the cavity of the 
T Cha disk by Huelamo et al. (2011), but was not confirmed despite multiple attempts (e.g. Olofsson et al. 2013, 
Sallum et al. 2015) and has recently been shown by Cheetham et al. (2015) to be a scattering feature of the disk 
itself.

A massive brown dwarf and a massive planet/low mass brown dwarf have also been claimed within the disk of HD169142 by 
Reggiani et al. (2014) and Biller et al. (2014). This disk includes a cavity extending to 20-25~AU, then a 15-20~AU thick 
ring of material followed by an annular gap from 40-70~AU (e.g. Osorio et al. 2014, Quanz et al. 2013b). However, both 
postulated bodies have potentially serious problems. The locations of the independent L' detections of the massive BD 
only agree at the limits of the respective uncertainties, 0.156$\pm$0.032 arcsec at PA=$7.4^{\circ}\pm11.3^{\circ}$ of 
Reggiani et al. and 0.11$\pm$0.03 arcsec at PA=$0^{\circ}\pm14^{\circ}$ of Biller et al. The Reggiani et al. location 
puts the body at 22.6$\pm$4.7~AU, either just inside or even within the 15-20~AU wide dusty ring. Furthermore, neither 
author detected anything at z', J, H and K$_{\rm S}$, despite the requisite sensitivity being easily achieved. Thus, 
much like the case of HD100546 b discussed previously the L' detection cannot be due to photospheric emission from a 
substellar or planetary mass companion, and instead is more likely to either be a planet in formation (Reggiani et al.) 
or a feature of the disk itself heated by an unknown mechanism (Biller et al.). Similarly, a second possible body 
detected by Biller et al. at H and K$_{\rm S}$ at a separation of $\sim$ 0.18 arcsec (26~AU) and PA $\sim$ 33$^{\circ}$ 
is also located within the ring of material and lacks a concomitant detection at z'.   

\subsubsection{Summarizing stellar and disk properties of transition systems}

Since no such compendium exists to our knowledge, in Table~\ref{TAB:ExoBD-Limits} we have summarised the properties 
of most known transition disk systems with a mm-resolved cavity, plus the observational limits on the mass and location 
of companions. In addition to objects already 'catalogued' as transition disks we have included several other 
sources based on possible nulls in their deprojected visibilities. These include DS Tau (Pietu et al. 2014) as well 
as BP Tau, CI Tau and FT Tau (Guilloteau et al. 2011). 

\begin{table*}
\tiny
\centering
\caption{Companion limits to transitional disk host stars}
\begin{tabular}{llllllllllllll}
\hline
 Star &  d   &  Age  &   Mass        & $\dot{M}$        & Null,R$_{11}$,R$_9$  & Gap R & Gap R & Gap R   & Gap R & Mass & Mass & Sep. & Refs. \\
      & (pc) & (Myr) & (M$_{\odot}$) & ($\times10^{-8}$ & (k$\lambda$, AU, AU) & [mm]  & [SED] & [N/MIR] & [gas] & req'd & limit & (AU) & \\    
      &      &       &               & M$_{\odot}$/yr)  &              & (AU)  &  (AU) &  (AU)   &  (AU) & (M$_{\rm J}$) & (M$_{\rm J}$) & & \\ \hline      
 HD100546   & 97  & 3.5-10 & 2.4 & 6-50   & 320 [25,18] & [25]   &  10  & 12-14   & 13      & 60,7.5 &  60-150     & 2.5-10 & 1 \\
            &     &      &     &        & 290 [27,20] & {\bf 26} &      &         &         &        &  100-160    & 30     & 2 \\
            &     &      &     &        &             &          &      &         &         &        &  9-12       & 48     & 3 \\
            &     &      &     &        &             &          &      &         &         &        &  6-9        & 70     & 2 \\
            &     &      &     &        &             &          &      &         &         &        &  4-6        & 90     & 2\\
            &     &      &     &        &             &          &      &         &         &        &  2-4        & $\geq$110 & 2,4  \\
 LkCa~15    & 140 & 2-5 & 1.1 & 0.14-0.40   & 145 [76,56] & {\bf 50},46 & {\it 39}       &  50 & 13-23   & 28,3.5 & 16-20 & 14  & 5  \\    
            &     &     &     &             &             & [61]        &  46,58         &     &         &       & 13-17 & 16.1 & 5 \\      
            &     &     &     &             &             &             &                &     &         &       & 10-14 & 18.2 & 5 \\      
            &     &     &     &             &             &             &                &     &         &       & 6-12  & 28-37.8 & 5,6 \\ 
            &     &     &     &             &             &             &                &     &         &       & 4-9   & 42-56   & 5,6 \\
            &     &     &     &             &             &             &                &     &         &       & 4-6   & 56-77   & 5,6 \\  
            &     &     &     &             &             &             &                &     &         &       & 2.5-5.5 & $\geq$80 & 5,6 \\ 
            &     &     &     &             &             &             &                &     &         &       &  20-30  & 2.8-5.6 & 7 \\ 
            &     &     &     &             &             &             &                &     &         &       &  10-20  & 5.6-22.4 & 7 \\ 
            &     &     &     &             &             &             &                &     &         &       &  19-30  & 22.4-44.8 & 7 \\ 
 HD142527   & 145 & 1-10 & 2.2 & 6.9-20    & 78 [148,108] & [140]       & 130  & 143     &  90     & 50,6.3 & 27-66 & 29-44  & 8 \\
            &     &      &     &           &              &             &      &         &         &        & 20    & 77     & 9 \\ 
            &     &      &     &           &              &             &      &         &         &        & 6.5   & 101    & 9 \\  
            &     &      &     &           &              &             &      &         &         &        & 9     & 130    & 9 \\  
            &     &      &     &           &              &             &      &         &         &        & 5     & 200    & 9 \\ 
            &     &      &     &           &              &             &      &         &         &        & 2.5   & 218    & 9 \\  
HD169142    & 145 & 3-12 & 1.7 & 0.31-4.0  & NG          & [25-30]      &  23  & 20-25   &         &        & 28-52 &  26.1  & 10 \\
            &     &      &     &           &             & {40-70}      &      &         &         &        & 7-12  &  26.1  & 10 \\  
            &     &      &     &           &             &              &      &         &         &        & 14-20 &  49.3  & 11 \\  
            &     &      &     &           &             &              &      &         &         &        & 13-18 &  87    & 12 \\  
            &     &      &     &           &             &              &      &         &         &        & 6-11  &  145   & 12 \\  
 HD135344   & 142 & 8$\pm$6 & 1.7 & 0.54-2.0 & 170 [66,48] & {\bf 46},39 & 30,45 &  28   &  0.5   & 43,5.3 & 230      & 14   & 13 \\     
            &     &         &     &          &             &  [60-65]    &       &       &        &        & 85       & 36   & 13 \\      
            &     &         &     &          &             &             &       &       &        &        & 50       & 43   & 13 \\      
            &     &         &     &          &             &             &       &       &        &        & 21       & 57   & 14 \\      
            &     &         &     &          &             &             &       &       &        &        & 16       & 71   & 14 \\      
            &     &         &     &          &             &             &       &       &        &        & 12       & 85   & 14 \\      
            &     &         &     &          &             &             &       &       &        &        & 9        & 114  & 14 \\      
            &     &         &     &          &             &             &       &       &        &        & 7        & 142  & 14 \\      
 MWC~758    & 200 & 3-5  & 2.0 &  6-89  & 190 [83,61] & {\bf 73}  & $\leq$1 & $\leq$20 & $\leq$30 & 50,6.3 & $\leq$80 & 0-60  & 15 \\
            &     &      &     &        &             & [100]     &    &          &        &        & 80       &  30   & 15 \\
            &     &      &     &        &             &           &    &          &        &        & 45       &  40   & 15 \\
            &     &      &     &        &             &           &    &          &        &        & 15       &  50   & 15 \\
            &     &      &     &        &             &           &    &          &        &        & 8        &  100  & 15 \\
            &     &     &      &        &             &           &    &          &        &        & 3        &  200  & 15 \\
            &     &     &      &        &             &           &    &          &        &        & 2        &  $\geq$300 & 15 \\ 
 GM Aur     & 140 & 1-7 & 1.4 & 0.50-1.0  & 225 [49,36] & {\bf 28} [29] & 20-24 & 20.5 & 19$\pm$4 & 25,3.1 & 19-35     & 2.8-5.6 & 7 \\
             &    &     &     &           &             &               &       &      &          &        & 9-22      & 5.6-22.4 & 7 \\
             &    &     &     &           &             &               &       &      &          &        & 22-40     & 22.4-44.8 & 7 \\
 J160421.7  & 145 & 3.7-11 & 1.0 & $\leq$0.001 & 125 [90,66] & 72,{\bf 79} &    & 63      & 31      & 25,3.1 & 22-29 & 2.9-5.8 & 16 \\     
 -213028    &     &      &     &             &             &   [80]      &      &         &         &        & 15-19 & 5.8-23.2 & 16 \\   
            &     &      &     &             &             &             &      &         &         &        & 19-28 & 23.2-46.4 & 16 \\   
            &     &      &     &             &             &             &      &         &         &        & 5-73  & 300-60  & 17 \\      
 DM Tau     & 140 & 3-8 & 0.55 & 0.20-1.1  & 385 [29,21] & {\bf 19} [16] &  3   & $\leq$15.5 &      & 14,1.7 & 18-33     & 2.8-5.6 & 7 \\  
             &     &    &      &           &             &                 &      &          &      &        & 8-18      & 5.6-44.8 & 7 \\
 UX Tau A   & 140 & 1   & 1.5 & 0.10-4.7 & 330 [33,25] & {\bf 25} [22] & {\it 30}       & $\leq$23 &   & 38,4.7 &  31  & 2.8-5.6 & 7 \\
            &     &     &     &          &             &               & 56,71          &          &   &        &  17  & 5.6-22.4 & 7 \\
            &     &     &     &          &             &               &                &          &   &        &  35  & 22.4-44.8 & 7 \\
 RY Tau     & 140 & 0.5-8 & 2.0 &  2.2-25  & 625 [18,13] & [{\bf 14}] &  18   &       &           & 50,6.3 &  105   &  0.35-4.2  & 18 \\
            &     &       &     &          &             &            &       &       &           &        &  40-77 &  2.8-44.8  &  7 \\
 LkHa~330   & 250 & 3 & 2.4 & 0.16-1.6   & 215 [92,67] & {\bf 68},47,41 & 50   &         &  4      & 60,7.5 & 50      & 10    & 19 \\ 
            &     &   &     &            &             &  [77,100]      &      &         &         &        & 25      & Cavity & 20 \\  
 SR21 A     & 125 & 1-5 & 2.0 & $\leq$0.14-1.3   & 210 [47,34] & 37,{\bf 36},33 & 18 & $\leq$12 & 7      & 50,6.3 & 40-60  & $\geq$18 & 21 \\ 
            &     &     &     &            &             &  [35]          &    &          &        &        & 50     & $\geq$5  & 19 \\          
            &     &     &     &            &             &                &    &          &        &        & 21     & Cavity  & 20 \\        
 WSB~60     & 125 & 0.9-3 & 0.2  & 0.10-0.37 & 370 [27,20] & {\bf 15} [15]   &      &            &      & 6.3,0.8 & 95-160 & 2.5 & 22 \\ 
            &     &       &      &           & 350 [28,21] & 20              &      &            &      &         & 21-27 & 18.75 & 23 \\
            &     &       &      &           & 350 [28,21] & 20              &      &            &      &         & 15-21 & 62.5  & 23 \\ 
 Oph~IRS48  & 121 & 1   & 2.0  &  0.32    &   NG        & {\bf 63}  &      &   30-60    & 20-30 & 55,6.9  &  150  &  2.5    & 22 \\
            &     &     &      &          &             &           &      &            &       &         &  100  &  18.75  & 23 \\
            &     &     &      &          &             &           &      &            &       &         &  50   &  62.5   & 23 \\
 SR24 S     & 125 & 0.2-2.4 & 2.0 & 3.0-7.4 & 230 [43,31] & {\bf 29}      &      &          &        & 35,4.4 & 75   & 2.5 & 22 \\
            &     &         &     &         & 200 [49,36] & 32            &      &          &        &        & 100  & 12.5-100 & 24 \\
 DoAr~44    & 125 & 1-7   & 1.0 & 0.63-0.93 & 230 [43,31] & {\bf 30} [28] &  36  &          &        & 30,3.8 & 70-90 & 2.5 & 22 \\
            &     &       &     &           & 215 [46,34] & 33            &      &          &        &       & 80-120 & 12.5-62.5 & 23,24 \\
J1633.9      & 120 & 2   & 0.7 & 0.013 & 280 [34,25] & {\bf 23$\pm$2}      & 8$\pm$2 &     &   & 18,2.2 &     10     & 2.4-4.8 & 25 \\
-2442        &     &     &     &       &             & [23]                &         &     &   &        &      6     & $>$4.8  & 25 \\
 PDS70      & 140 & 5-10  & 0.8 &           & NG          & [80]    &       & 65    &           & 20,2.5 & 3-6      & 18.2  & 26 \\ 
            &     &       &     &           &             &         &       &       &           &        & 2-4      & 28    & 26,27 \\
            &     &       &     &           &             &         &       &       &           &        & 1.8-3.5  & 56    & 26,27 \\
            &     &       &     &           &             &         &       &       &           &        & 1.3-3.0  & 84    & 26,27 \\
            &     &       &     &           &             &         &       &       &           &        & 1.2-2.6  & $>$140 & 26,27 \\
 Sz91       & 200 & 5     & 0.5 &   0.008   & 135 [117,86] & {\bf 97}    &   & 65$\pm$4 & $\leq$28 & 13,1.6 & 80-100 &  30  & 30,31 \\
            &     &       &     &           &             & [86$\pm$25] &   &          &        &        &          &      & \\
 J1615-3255 & 185 &     & 1.2 & 0.04-0.32 & 280 [52,38] & {\bf 30} [27] & 2-4  &       &        & 28,3.5 &          &      & \\
 CQ Tau     & 100 & 5-10  & 1.5 & $\leq$0.50-11.2 & 280 [28,21] & 21      &       &       &           & 38,4.7 &          &      & \\
 DS Tau     & 140 & 4-7   & 0.7 & 0.16-4.1 & 300 [37,27] & 15$\pm$5 &       &       & 30$\pm$9  & 18,2.2 & 40-80    & 2.8-5.6 & 7 \\
            &     &       &     &          &             & 20$\pm$4 &       &       &           &        & 20-36    & 5.6-11.2 & 7 \\
            &     &       &     &          &             &          &       &       &           &        & 24-43    & 11.2-22.4 & 7 \\
            &     &       &     &          &             &          &       &       &           &        & 30-60    & 22.4-44.8 & 7 \\
 BP Tau     & 140 & 0.9-3.2 & 0.8 & 1.3-81  & 450 [25,18] & 18      &       &       &           & 20,2.5 & 40-45    & 2.8-5.6  & 7 \\
            &     &       &     &          &             &          &       &       &           &        & 21-31    & 5.6-22.4 & 7 \\
            &     &       &     &          &             &          &       &       &           &        & 44-51    & 22.4-44.8 & 7 \\
 CI Tau     & 140 & 2     & 0.8 & 1.5-6.5  & 330 [33,25] & 25       &       &       &           & 20,2.5 & 64-78    & 2.8-5.6  & 7 \\
            &     &       &     &          &             &          &       &       &           &        & 26-35    & 5.6-22.4 & 7 \\
            &     &       &     &          &             &          &       &       &           &        & 32-38    & 22.4-44.8 & 7 \\
 FT Tau     & 140 & 1.6  & 0.30 &  3.1     & 500 [22,16] & 16      &       &       &           & 7.5,0.9 & 29-34    & 2.8-5.6  & 7 \\
            &     &       &     &          &             &         &       &       &           &         & 12-20    & 5.6-44.8  & 7 \\
 T Cha      & 100 & 3-10  & 1.5 & 0.4      &  NG         & 20      & 7.5,15 & 12$\pm$2 &       & 38,4.7 &  28-55    & 1-10  & 28,29 \\
\hline
\end{tabular}
 \begin{list}{}{}
 \item
\noindent
References for the stellar and disk parameters are provided in Appendix E. Numbered references for the planet detection
limits are the following: 
(1) Grady et al. (2005) (2) Boccaletti et al. (2013) (3) Quanz et al. (2015) (4) Ardila et al. (2007) 
(5) Thalmann et al. (2010) (6) Bonavita et al. (2010) (7) Kraus et al. (2011)
(8) Casassus et al. (2012) (9) Rameau et al. (2012)
(10) Biller et al. (2014) (11) Reggiani et al. (2014) (12) Grady et al. (2007)
(13) Vicente et al. (2011) (14) Grady et al. (2009)
(15) Grady et al. (2013)
(16) Kraus et al. (2008) (17) Ireland et al. (2011)
(18) Pott et al. (2010)
(19) Brown et al. (2009) (20) Andrews et al. (2011a)
(21) Follette et al. (2013)
(22) Simon et al. (1995) (23) Ratzka et al. (2005)
(24) Ghez et al. (1993)
(25) Cieza et al. (2012)
(26) Hashimoto et al. (2012) (27) Riaud et al. (2006)
(28) Olofsson et al. (2013) (29) Huelamo et al. (2011)
(30) Romero et al. (2012) (31) Canovas et al. (2015)
\noindent
Note: NG means Not Given
 \end{list}
\label{TAB:ExoBD-Limits}
\end{table*}

The first five columns of Table~\ref{TAB:ExoBD-Limits} are the object name, assumed distance, and the ranges found 
in the recent literature for the stellar age, mass and accretion rate. In some cases the spread in these values is large,
even up to a factor of two in the mass and a decade or so in the age and accretion rate. The mass and age can be critically 
dependent on the particular stellar evolutionary model used, whilst the accretion rate may be dependent on the tracer 
used as well as in some cases being time variable. Sources for these stellar parameters are provided in Appendix E, and
similarly for the disk parameters in columns 6--10, described below.

The sixth column of Table~\ref{TAB:ExoBD-Limits} provides the deprojected $u-v$ distance at which the first null in 
the real part of the visibility occurs, as well as the inferred radii using equations A9 and A11 in Hughes et al. (2007). 
Whilst the null position is explicitly stated by some authors, in other cases it has been read directly off their 
visibility plots. Different authors may also have observed the same target and found a slightly different null position, 
the reason for which may simply be due to different choices for the disk position angle and inclination used in the 
deprojection. This is apart from the case of HD100546 itself, for which the null position almost certainly changes 
between the ALMA 340~GHz observation of Walsh et al. (2014) and our ATCA observations at 44~GHz.

Equation A9 in Hughes et al. is valid for a disk whose outer radius is significantly larger than the inner (i.e. cavity) 
radius, and where the addition of the power law indices for surface density $\Sigma \propto R^{-p}$ and temperature 
$T \propto R^{-q}$ is in the range $1 \leq p+q \leq 3$. A value of 3 has been assumed for all sources. This holds no 
particular physical significance, and indeed such a simple power law for the surface density is generally no longer 
considered appropriate for protoplanetary disks. See for example Andrews et al. (2009) for further details. Equation A11 
of Hughes et al. is instead strictly valid for a thin ring of constant brightness, where $\Delta R/R \ll 1$. 

Columns 7--9 give the various estimates for the cavity radius, derived from millimetre imaging, modelling of the 
infrared SED, and directly from near- and/or mid-infrared images, all from dust continuum. The mm-measured cavity 
radius is differentiated between that inferred from i) modelling within the respective literature of the interferometer 
visibilities in conjunction with the entire IR--mm SED, and ii) the peak-to-peak separation along the disk major axis, 
either stated by authors or derived from their images and given here in square brackets. Our preferred value is given in 
boldface. For some objects there are two or more estimates of the SED-inferred cavity size, which in cases like LkCa~15 
and UX Tau A is due to different assumptions by respective authors for the central star properties such as spectral 
type (Espaillat et al. 2010). The most recent -- and thus preferred -- SED values for these two stars are given in 
italics, from Espaillat et al. (2011), and thus resolves the otherwise discrepant (and probably unphysical) directly 
observed and model values. Column 10 instead gives the hole radius inferred using gas tracers, from either infrared 
ro-vibrational or millimetre rotational spectroscopy. 

Clearly neither equation A9 nor A11 of Hughes et al. (2007) is generally appropriate for deriving the 'true' cavity size. 
From a purely empirical perspective, when $p+q=3$ equation A9 does appear to provide an estimate in better agreement with 
the 'true' cavity size determined from the detailed modelling. However, in a few cases (e.g. HD100546 and HD142527) the 
assumption of a thin ring seems to hold true. This is perhaps not so surprising for HD142527 given its large cavity radius. 
Also, whilst it obviously indicates the presence of a cavity or gap, infrared SED modelling is not generally a reliable 
estimator of its size, at least compared to that measured at millimetre wavelengths. This may in some cases be due to 
inappropriate stellar parameters in the SED model, but in other cases is probably related to the fact that the IR SED is 
primarily sensitive to the small grain population whilst the millimetre regime is sensitive to larger grains. And as we have 
seen for  HD100546, but also for HD135344, SR~21A, GM Aur and perhaps also DM Tau, the near- and/or mid-IR images indeed 
show a cavity size more consistent with the SED value. 

\subsubsection{Observational limits of companion masses in transition disk systems}

Column 11 of Table~\ref{TAB:ExoBD-Limits} provides estimates of the mass of a single orbiting body capable of clearing 
the observed mm cavity. The two estimates are based on the assumption that the body could clear a gap out to 5 and 10 Hill 
radii, as postulated by Dodson-Robinson \& Salyk (2011) and Pinilla et al. (2012) respectively. The orbital radius of the 
body is placed in the middle of the cavity. For comparison, columns 12 and 13 of Table~\ref{TAB:ExoBD-Limits} provide 
observed limiting masses and separations of possible orbiting bodies within the disk, whilst column 14 provides references 
for the data to derive these companion limits. 

The limiting masses and separation were typically derived from either near-IR sparse aperture masking or coronagraphic 
imaging coupled with adaptive optics and angular differential imaging. Where such typically very recent data was not available 
we also considered lunar occulation or speckle imaging observations, which were much less sensitive to giant planets or low 
mass brown dwarves, but were at least useful in ruling out high mass brown dwarf or low mass stellar companions. Respective 
authors provided companion detection limits -- typically quoted to 3--5$\sigma$ -- as a function of radius in arcseconds from 
the central object, either in terms of a flux ratio or magnitude difference. These were converted to absolute magnitudes 
which were then compared to evolutionary models of the photospheric emission from massive exoplanets and/or brown dwarfs 
(see below). 

For consistency, where possible we have used the respective author's data to derive our own planet mass estimates, even 
though those authors may also have provided such estimates. This was not possible for LkH$\alpha$330 or SR21~A, for which Brown
et al. (2009) and Andrews et al. (2011a) quoted private communications for an upper limit of the secondary mass. We generally 
find good agreement between our own and literature values, with any discrepancies likely resulting from using different ages 
and/or central stellar mass (where the planet mass limit is expressed as a secondary-to-primary mass ratio). In only one case, 
RY Tau, do we find a significant difference. Kraus et al. (2011) quote a mass ratio of 0.009 within most of the cavity, which 
translates into a secondary mass of 14~M$_{\rm J}$ using their preferred primary mass of 1.46~M$_{\odot}$, or up to 
19~M$_{\rm J}$ assuming a stellar mass of 2~M$_{\odot}$ also found in the literature. However, using their observed H-band 
magnitude difference and an age even as young as 1~Myr, we instead find an upper limit on the secondary mass of 
$\sim$ 40~M$_{\rm J}$. 

We tested the model isoschrones found at http://phoenix.ens-lyon.fr/Grids/, including the DUSTY and COND variants described 
in Baraffe et al. (2003), Allard et al.(2001) and Chabrier et al. (2000), as well as the BT-SETTL grid described in Allard, 
Homeier \& Freytag (2012). We direct the reader to these papers for a full explanation of the photospheric physics and 
chemistry treated in the models. Briefly however the DUSTY and COND variants represent extreme cases wherein the formation 
of dust in the equation of state is included in both, but only in DUSTY is dust scattering and absorption in the radiative 
transfer equation taken into account, and where it is further assumed that the dust remains in the photosphere (i.e. there
is negligible settling). On the other hand the COND variant assumes all the dust has disappeared from the photosphere via 
gravitational settling, and thus neglects dust opacity in the radiative transfer. The BT-SETTL variant includes such
settling. In practice, and consistent with Boccaletti et al. (2013) in their study of HD100546, we found little difference 
between these model variants.  

These grids are of particular utility as they provide predicted absolute magnitudes in different filters for a wide variety 
of optical and infrared instruments at many different observatories around the world. Further, the grid spacing for age is
only 1~Myr in the age range of 1-10~Myr relevant for most of our sample, whilst the grid spacing for the secondary mass varies
between 0.001--0.01~M$_{\odot}$ in the mass range relevant to giant planets and brown dwarves. Of course, these grids are for
only one of the two main postulated modes of planet formation, namely gravitational instability within the disk rather than 
core-accretion, otherwise generally classified as hot- and cold-start models respectively. The formation mechanism of giant 
planets in the radius range of interest here is still a hotly debated topic, e.g. Spiegel \& Burrows (2012), Janson et al. 
(2011) and Dodson-Robinson et al. (2009). But what is certain is that core-accretion/cold-start giant planets are predicted 
to be much fainter -- by several magnitudes in the near-IR -- than their gravitational-instability/hot-start counterparts, and 
thus to not be generally detectable with current techniques (e.g. Vigan et al. 2012).  

Another assumption in our treatment is that there is no differential extinction between the primary and secondary bodies. This 
is probably unrealistic, but necessitated by the simple fact that assigning any such number would be a guess. If however the
putative planet suffers more (less) extinction than the star then the mass we derive would be lower (higher). Further, we 
have assumed that any planet would be the same age as the star itself. Whilst this seems like a reasonable assumption we note 
that the directly-imaged planet candidates so far discovered toward transition disk sources, be they internal (LkCa~15) or 
external (HD100546) to the cavity, or somewhere in between (HD169142), have all been assessed to be much younger than the 
star itself. 

With the aformentioned caveats in mind, and by way of example, in the case of MWC~758 any body within 60~AU of the star 
must have a mass less than the most massive brown dwarfs (Grady et al. 2013), otherwise it would have been detected. Also, 
at separations of 50, 100, 200 and 300~AU any body must be less massive than 15, 8, 3 and 2~M$_{\rm J}$ respectively (or 
alternatively a body more massive than 15~M$_{\rm J}$ would have been detected at any separation greater than 50~AU and 
similarly for the other mass/separation combinations). In several cases (e.g. HD100546, LkCa~15, HD142527) several data 
sets were available, which used either a different imaging technique and/or a different filter, but which gratifyingly 
gave consistent secondary mass limits.

Putting all the data together, Table~\ref{TAB:ExoBD-Limits} shows that in all cases where sufficient sensitivity has 
been attained (11 of 27) the mm-derived cavity could not have been carved by a single body if its clearance radius was 
only five Hill radii. Thus, under the 5R$_{\rm H}$ scenario these cases would likely require multiple giant planets to 
carve the disk gap. Further, in one of these cases, namely PDS~70, even a body which could clear out to 10R$_{\rm H}$ 
is excluded at the younger end of the age range for its parent cloud Centaurus.

However, perhaps the most striking feature of the compilation in Table~\ref{TAB:ExoBD-Limits} is that massive gas giant 
planets and/or low mass brown dwarves are rare at large orbital separation from their host star. For instance, for 22 
sources there was sufficient sensitivity to have detected a body of 50~M$_{\rm J}$ or less somewhere in the orbital range 
of greater than about 3 to 60~AU (dependent on the specific sensitivity of individual targets). But as previously noted, 
for only 3 of those 22 sources (i.e. LkCa~15, HD100546 and HD169142, or $\sim$ 14\%) has a planet 
detection been made. Without pursuing a rigorous statistical treatment this is remarkably similar to the direct imaging 
detection rates cited in 4.9.1 for typically much older systems. Interestingly it also agrees with the work of Cieza 
et al. (2012), who suggest that only $\leq$ 18\% in their sample of 74 Spizter-selected transition disks in nearby clouds 
are best explained by the dynamical interaction of recently formed giant planets.

Indeed, the likely direct descendents of transition disks are the gas-poor debris disks, whose dust population -- without 
ongoing planetesimal collisional replenishment -- would otherwise be depleted on relatively short timescales by inward spiral 
and radiative blow-out. Wahhaj et al. (2013) conducted a direct imaging survey for giant planet companions of 57 debris disks, 
finding none and concluding that at 95\% confidence $<$13\% have a $\geq$ 5~M${\rm J}$ planet beyond 80~AU, and $<$21\% have 
a $\geq$ 3~M${\rm J}$ planet outside of 40 AU, using hot-start evolutionary models.

Overall, we believe these results are best explained by a scenario wherein the centrally-cleared gaps and/or broad cavities 
in most -- not necessarily all -- transition disk and (by extension) debris disk systems are created by overlapping gaps 
cleared by multiple one-to-several-Jupiter-mass bodies rather than a single many-Jupiter-mass body. From a statistical 
standpoint then, it is unlikely that only a single body is responsible for carving out the $\sim$ 25~AU radius mm-wave 
cavity within the HD100546 disk. 

\subsection{Why no direct planet detection or second ring of emission at 7~mm?}

There is no evidence in Figure~\ref{FIG-7mm-normcal} or Figure~\ref{FIG-7mm-selfcal} for emission directly 
associated with an orbiting body, such as HD100546~b at $\sim$ 47~AU (Quanz et al. 2015, 2013a; 
Currie et al. 2014). HD100546~b has been conjectured by Quanz et al. (2015, 2013a) and Currie et al. (2014) 
to be a protoplanet, one still in the process of formation, and thus potentially still embedded in a dust and 
gas envelope. 

Further, apart from evidence for a fainter component of extended emission beyond the $\sim$ 50~AU radius 
compact circumstellar disk -- as indicated by the difference between compact and extended array fluxes in 
Table~\ref{TAB:hd100-obs7mm} -- we do not resolve this into a second ring of emission. The gap between
this second ring and the $\sim$ 50~AU radius disk could have been cleared by the 53--68~AU planet (disk
inclination dependent), as postulated by Walsh et al. (2014). Our 7~mm extended component could well be in 
the form of a second ring, but the data is insufficient to go beyond this simple statement. 

It is thus interesting to ask why neither the 53--68~AU planet (or rather its circumplanetary disk) nor a 
second ring were detected in our data set. 

\subsubsection{Upper limit on circumplanetary disk mass}

The best 7~mm 3$\sigma$ sensitivity we achieved is 0.075~mJy/beam in Figure~\ref{FIG-7mm-normcal}-e. For 
silicate dust properties similar to those used in Section 4.6 to calculate the disk mass -- i.e. a dust 
opacity at 7~mm from Draine (2006) with maximum sizes of 0.1, 1, 10 and 100~cm -- and a temperature of 40~K, 
our limiting sensitivity gives $\sim$ 2, 1, 2 and 6 Earth masses of dust (or 0.0065, 0.003, 0.0065 and 
0.02~M$_{\rm J}$; note that Figure 3 of Draine (2006) shows the dust opacity at a wavelength of 7 mm is the 
same when the maximum size is 1~mm or 10~cm). For a population of silicate grains with a smaller maximum 
size, i.e. 100~$\mu$m or less, the limiting value is $\sim$ 16 Earth masses (or 0.05~M$_{\rm J}$). 

A temperature of 40 K is chosen since Isella et al. (2014) shows that it is the 'asymptotic' temperature 
at radii $\geq$ 1~AU in a circumplanetary disk around a 10~M$_{\rm J}$ planet -- regardless of whether it 
formed by gravitational instability or core accretion -- and with a mass accretion rate of 
$\leq$ 10$^{-4}$ M$_{\rm J}$/yr. They found that for accretion rates $\geq$ 10$^{-6}$ M$_{\rm J}$/yr, 
required to build a planet within the disk dispersal timescale, the circumplanetary disk temperature
is set by viscous heating, and is essentially independent of the planet luminosity or irradiation from 
the central star. Thus, as we might have expected anyway given its large orbital radius, the temperature 
of the putative circumplanetary disk around the 50--70~AU protoplanet is independent of HD100546 itself. 

Using the VLA at 7~mm Isella et al. (2014) searched for circumplanetary disk emission from the candidate 
planet LkCa~15b within the cavity in the disk around LkCa~15, found by Kraus \& Ireland (2012). With a 
3$\sigma$ sensitivity about 7 times better than ours they were able to set an upper limit of 
$\sim$ 0.1~M$_{\rm J}$ for the circumplanetary disk mass, inclusive of dust and gas with a mass ratio of 
1:100, or simply 0.001~M$_{\rm J}$ of dust. Taking into account the different senstivities, this is 
gratifyingly close to our maximum mass of $\sim$ 0.0065~M$_{\rm J}$ for a disk with grains up to 1~mm in 
size, despite the use of dust opacities from different published works (inclusive of particular dust 
components and their relative abundances, as well as the minimum and maximum sizes).

Thus, our work is in very good agreement with that of Isella et al., namely that if there is a 
circumplanetary disk around forming gas giant planets in or outside the cavity of transition disks 
then their mass seems to be only a small fraction of the planet mass. In the case of the 50--70~AU planet 
of HD100546, assuming a mass of $\geq$ 5~M$_{\rm J}$ (Quanz et al. 2013a) then the fraction (dust only) 
is $\leq$ 1\%, and possibly as low as only several hundredths of a percent. If the gas-to-dust mass 
ratio of the circumplanetary disk is as evolved as we argue for the HD100546 circumstellar disk in 
section 4.6, i.e. nearer 1 than 100, then these values only increase by at most a factor of several.     

\subsubsection{A second ring of emission}

The outer ring of emission postulated by Walsh et al. (2014) has a maximum surface brightness of around 
3 mJy/beam at 302--346~GHz, judging from the residuals of an inner-ring-only fit in their Figure 4. Assuming 
a spectral slope similar to the value we find for the inner disk, i.e. $\alpha$=2.45, the inferred 44~GHz 
surface brightness is 0.023~mJy/beam, i.e. about the same as our best RMS in Figure~\ref{FIG-7mm-normcal}-e. 
Even a slope as flat as $\alpha$=2.0 only extrapolates to a 7~mm surface brightness of $\sim$ 0.055~mJy/beam. 
This is below our best 3$\sigma$ sensitivity of 0.075~mJy/beam and thus we could not have detected the outer 
ring in our high resolution images. 

However, if this second ring of 0.023~mJy/beam filled the beam of our H214 observation of 31 May 2009, 
which was about 50 times larger in area than the ALMA beam, then its total flux would be around 1.2~mJy. Within 
the uncertainties this is comparable to the flux difference between the H214 data set, in which HD100546 was a 
point source, and the 6~km configuration data of 20 June 2012 which resolved the compact disk but would 
probably have resolved out anything more extended. 


\section{Conclusions}

From a variety of observations over the last decade or so, spanning wavelengths from the ultraviolet 
to the infrared and spatial resolutions from sub-AU to tens of AU, the case for a planet orbiting 
at around 10~AU from the star HD100546 has been very persuasive. Although the planet itself has not
been directly detected, the new 7~mm observations reported here, at 15--50~AU resolution, make that 
case compelling, albeit from an abundance of circumstantial evidence. The effect of the planet on the 
dust emission from the circumstellar disk is consistent with previously published models of early
planetary system evolution. For instance, we find a hole in the disk with a radius of around 25~AU, 
but which is almost twice that inferred from uv--IR observations. Such a size discrepancy agrees
with expectation if a planet sculpts the cavity. Other findings from our observations -- including 
those presented in the appendices -- also support this scenario, including:\\
1) The dust grains in the disk have evolved significantly from what would be expected in the ISM. The
dust emission spectral index is 2.45$\pm$0.10, and under the assumption the emission is mostly optically 
thin -- supported by spatially resolved data down to 19~GHz -- the dust emissivity index is 0.5$\pm$0.1, 
compared to an ISM figure of $\sim$ 1.7. For any reasonable grain properties found in the literature, 
the dust has grown to sizes of at least 10~cm, and possibly up to 100~cm.\\
2) The dust mass of the disk within a radius $\leq$ 100~AU is of the order of a Jupiter mass. From other
published work the gas mass within a radius of $\sim$ 400~AU is of a similar magnitude, or at most ten 
times greater. Whatever the case, the gas-to-dust mass ratio has seemingly evolved from its ISM value of 
$\sim$ 100.\\
3) A brightness asymmetry exists between the two sides of the disk along its major axis. We interpret
this as a case of pericenter glow, whereby the NW side of the disk is closer to the star and thus is 
warmer with enhanced thermal dust emission. This in turn is a product of an elliptically orbiting 
massive body which spends most of its time near its apocenter on the SE side of the disk.\\
4) A brightness contrast is seen between the SW and NE sides of the disk, such that the disk overall 
takes the shape of a horseshoe, with the opening on the ENE side. Such an azimuthal asymmetry seen in 
the reconstructed image is further supported by non-zero imaginary components of the visibilities on 
small spatial scales. By analogy with similar transition disk systems the asymmetry is interpreted
as a sign of the large dust grains being trapped in a pressure maximum, in turn a by-product of 
previous planet formation.\\
5) The outer disk radius is wavelength independent from 1 to 16~mm, more consistent with a planet-disk 
interaction scenario than a radial drift process whereby larger grains -- probed by longer wavelength
emission -- migrate inward.

We have presented perhaps the most detailed ever single study of a Herbig Be star and its disk, showing the
value of conducting observations across a wide spectral range, from the infrared through to true radio regimes.
ATCA will thus remain an important facility to complement the flood of data that is already emerging from ALMA 
on transition (and other) disk systems. Indeed, ATCA will be crucial to both separate the dust and free-free 
(or other) emission components of these systems, as well as to study the parent star and/or the physical
processes occurring in the vicinity of the star-disk-outflow boundaries. 
 
In the case of HD100546 this approach has revealed that it does indeed have a significant outflow of a few
$\times10^{-8}$ M$_{\odot}$/yr -- evidenced by the 6~cm radio flux and IR hydrogen recombination lines 
presented here. A relatively high mass accretion rate, even up to a few $\times10^{-7}$ M$_{\odot}$/yr, 
must accompany this outflow, which means that the few tenths of AU radius inner disk, and even the entire 
cavity out to $\sim$ 10~AU, would be depleted on a timescale of less than a year. Significant amounts of 
material must then be passing across the cavity from the outer disk, which planet formation models suggest 
would be better sustained by a multi- rather than single-planet system within about 10~AU.

The passage of such a large amount of material in a relatively short time suggests that the inner regions of
the HD100546 disk are highly dynamic. This may be reflected in the many examples of variability phenomena 
seen for this system, including variable radio emission presented here over timescales as short as tens of 
minutes. Frequent collisions may also occur between planetesimal-sized bodies, a possible mechanism behind 
the production of comet-like dust first detected via ISO observations.  

Overall, the properties of HD100546 and its disk suggest that the system has two faces, one of relative youth 
and another of maturity. Its youth is reflected in parameters such as the inflow/outflow rates, presence of dense
molecular gas out to several hundred AU, and a grain population indicative of growth rather than fragmentation. 
Its maturity is reflected in the highly processed nature of separate grain populations, including crystalline, 
cometary-like silicates and Kuiper Belt Object-like colours, as well as the large inner cavity likely resulting 
from past planet formation. Whilst all of this perhaps leaves its age somewhat uncertain, by any (astronomical 
or other) definition of the word it is truly a 'transition' disk system. 

\bsp

\section*{Acknowledgments}

This research was supported by an Australian Research Council (ARC) Australian Research Fellowship 
awarded to CMW as part of Discovery Project DP0345227, and subsequently as an ARC Future Fellowship 
FT100100495. STM acknowledges support from the Swinburne Researcher Development Scheme and Swinburne 
Special Studies Program. We acknowledge the award of significant amounts of observing time for this 
project by the ATNF and NANTEN2 Time Assignment Committees. The assistance of Paul Jones and Balthasar
Indermuhle in conducting the May and June 2012 7~mm observations is greatly appreciated. We thank the 
ATNF staff at Narrabri for their hospitality, assistance and patience, especially Robin Wark, Bob Sault 
and Phil Edwards. We also thank Hiroaki Yamamoto for assistance during the period over which the NANTEN2 
observations were performed, and providing 'first look' spectra. We acknowledge the contribution of 
Tony Wong, particularly his very user-friendly Miriad script for reducing ATCA data. We thank Sean Andrews
for providing the IDL script to perform the visibility deprojections. Finally, we appreciate the patience 
of the referee in reviewing such a long paper. This research has made use of the Exoplanet Orbit Database 
and the Exoplanet Data Explorer at exoplanets.org. This research has made use of the SIMBAD database operated 
at CDS, Strasbourg, France.


\appendix

\section[]{Fluxes at 3, 16, 35 and 62 mm}

Tables~\ref{TAB:hd100-obs3mm}, \ref{TAB:hd100-obs16mm} and \ref{TAB:hd100-obs36cm} contain the observation 
logs at 3~mm, 16~mm and 3.5 \& 6.2~cm respectively. Apart from one 3~mm observation on 31 July 2009 all 
were taken with the old ATCA correlator. The tables are in the same format as Table 1. The observations 
from 16~mm to 6.2~cm are new and presented here for the first time, as are the bulk of the 3~mm observations. 
However, some of the 3~mm data from 2002 were presented in Wilner et al. (2003). This data was re-analysed, 
as well as data taken in an independent study and available in the ATCA archive (C1089 in 
Table \ref{TAB:hd100-obs3mm}, with flux quoted in Acke et al. 2004). The 2002 data has been included since 
it enabled a much better understanding of the absolute flux calibration issues at 3~mm with ATCA, although 
note that at that time only 3 antennas were equipped with 3~mm receivers. From this the 90~GHz flux quoted 
in Wilner et al. (2003) has been revised upward. Also note that the observing wavelength reported in Acke 
et al. (2004) is incorrect, being 3.5~mm instead of the quoted 2.9~mm. 

On several occasions HD100546 was observed in dual mode at 3~mm, in which the upper sideband was 
configured with 128~MHz and 33 channels, whilst the lower sideband contained 257 channels covering 
16~MHz. This was centred on the HCO$^+$ J=1--0 transition at 89.188518~GHz, with a channel spacing 
of 0.21~km~s$^{-1}$. As seen in Table~\ref{TAB:hd100-obs3mm} a continuum flux could also be extracted
from this sideband, though with reduced statistical precision. 

Calibrations and data reduction were performed using the same methodology as at 7~mm. At 3~mm Uranus 
and sometimes Mars was used to set the flux scale, but in some cases a planet was not available so 
it was necessary to use B1057-797 or even B0537-441. 

Absolute flux calibration at 3--7 mm can be problematic at ATCA, particularly in terms of the temporal stability 
of the atmospheric conditions and thus phases. Over a 12 hour -- and sometimes much shorter -- synthesis track 
the RMS path length error of the ATCA seeing monitor (Middleberg, Sault \& Kesteven 2006; Indermuehle \& Burton 2014)
can vary by an order of magnitude, e.g. between $\leq$ 100 $\mu$m (very good) and $\sim$ 1000 $\mu$m (very bad). 
This also means that the flux calibrator, e.g Uranus, may be observed under significantly different conditions than 
the science target, especially for times when it is only visible several hours before or after a science track. 
Thus, inclusive of issues such as phase decorrelation and non-uniform observing conditions, we estimate the absolute 
flux calibration to be no better than $\sim$ 20--30\% in most cases at 3 and 7~mm.

At 16~mm the primary flux calibrator PKS B1934-638 set the absolute flux scale on most occasions, 
though in a few instances either Mars or the gain calibrator B1057-797 itself was used instead. 
Accuracy is $\leq$ 10\%. At 3.5 and 6.2~cm the absolute flux scale was typically set using 1934-638, 
or the secondary calibrator PKS B0823-500. See the ATNF web site for further
details\footnote{www.narrabri.atnf.csiro.au/observing/calibrators}. Accuracy is as good as a few 
percent at these wavelengths.

\begin{table*}
\caption{3~mm ATCA observations of HD100546.}
\begin{tabular}{llllllllll}
\hline
 Date	& $\nu_1$ & $F_1$ & $\nu_2$ & $F_2$ & Elev.             & Flux cal.,         & $T_{\rm int}$  & Config & Notes \\ 
 UT	  & (MHz)   & (mJy) & (MHz)   & (mJy) & 1057 ($^{\circ}$) & elev. ($^{\circ}$) & (mins)         &        & \\ \hline
31 May 2002  & 89182 (l) & 44.0$\pm$13.0 P   & 89184 (c) & 51.2$\pm$3.6 P & 27-40 & Mars, 24   & 220 & EW352 &       \\
11 Jun 2002  & 85980   & 29.4$\pm$5.8$^a$ P  & 86243     & 33.4$\pm$4.2 P & 34-40 & Uranus, 68 & 130 & EW352 & C1089 \\
12 Jun 2002  & 85980   & 33.9$\pm$12.3$^a$ P & 86243     & 42.1$\pm$8.3 P & 35-40 & Mars, 35   & 50  & EW352 & C1089 \\
5 Aug 2002   & 85980     & 40.7$\pm$6.2 G    & 86243     & 40.1$\pm$6.8 G & 29-39 & Uranus, 73 & 120 & 750F  & C1089 \\
8 Aug 2002   & 89180 (l) & 63.6$\pm$16.2 P   & 89184 (c) & 50.9$\pm$3.6 P & 31-40 & Mars, 24   & 260 & 750F  &  \\
9 Jun 2007   & 89182 (l) & 48.0$\pm$17.2 P   & 91456 (c) & 36.8$\pm$4.1 P & 33    & 1057-797$^b$ & 30  & EW352 &  \\
2 Oct 2007   & {\bf 89182 (l)} & {\bf 41.5$\pm$4.0 P} & {\bf 91456 (c)} & {\bf 51.4$\pm$1.2 P} & 37-40 & Uranus, 62 & 150 & H75   &  \\
3 Oct 2007   & {\bf 89182 (l)} & {\bf 46.2$\pm$2.6} & {\bf 91456 (c)} & {\bf 45.0$\pm$0.7} & 36-40 & Uranus, 59 & 210 & H75   &  \\
23 Oct 2007  & 89182     & 34.7$\pm$1.6 P    & 91456     & 35.1$\pm$1.6 P & 33-40 & 1057-797$^c$ & 150 & H214  &  \\
7 May 2008   & 93504     & 41.5$\pm$1.0 G    & 95552     & 43.6$\pm$1.2 G & 25-40 & Uranus, 62 & 500 & 750A  &  \\
27 June 2008 & 93504     & 51.5$\pm$2.7 G    & 95552     & 40.3$\pm$2.4 G & 29-36 & 0537-441$^d$ & 125 & 1.5B  &  \\
30 July 2009 & 93000     & 52.2$\pm$2.5 G    & 95000    & 46.7$\pm$3.0 G & 22-30 & Uranus, 39 & 120 & 1.5A & CA03 offline \\
\hline
\end{tabular}
 \begin{list}{}{}
  \item[$^{a}$] Fluxes without baseline 3--4, which was corrupted; $^{b}$ Flux taken to be 2.5 Jy from 
  18 May 2007; $^{c}$ Flux taken to be 1.5 Jy from 2 \& 3 October 2007; $^{d}$ Flux taken to be 7.8 Jy 
  from observations in weeks prior to and after 27 June 2008; 
  Unless otherwise noted both channels were observed in continuum mode. l and c refer to line and continuum 
  channels, 16 and 128 MHz wide respectively. The line channel was set for HCO$^+$ J=1--0 at rest frequency 
  89.188518~GHz. P \& G refer to point and elliptical Gaussian model fits with UVFIT in Miriad. 
  Numbers in bold are our best estimates of total fluxes.
  A flux of 45$\pm$5~mJy is recommended at 90~GHz, completely dominated by thermal dust emission.
 \end{list}
\label{TAB:hd100-obs3mm}
\end{table*}

\begin{table*}
\caption{16~mm ATCA observations of HD100546.}
\begin{tabular}{llllllllll}
\hline
 Date & $\nu_1$ & $F_1$ & $\nu_2$ & $F_2$ & Elev.             & Flux cal.,         & $T_{\rm int}$ & Config & Notes \\ 
 UT		&  (MHz)  & (mJy) & (MHz)   & (mJy) & 1057 ($^{\circ}$) & elev. ($^{\circ}$) & (mins)        &        &       \\ \hline
 20 Aug 2005	& 18496 & 1.50$\pm$0.14 & 18624 & 1.55$\pm$0.15 & 33-40 & None$^{a}$ & 156 & H214 &  \\
  6 May 2006	& 18496 & 1.46$\pm$0.13 & 18624 & 1.84$\pm$0.13 & 36-39 & 1934-638, 43 & 72 & H214 &  \\
  8 May 2006	& 18496 & 2.00$\pm$0.42 & 18624 & 0.95$\pm$0.44 & 36    & Mars, 33     & 24 & H214 & CA03,04 offline \\
  9 May 2006	& 18496 & 2.07$\pm$0.45 & 18624 & 2.12$\pm$0.46 & 39    & 1057-797$^{b}$     & 20 & H214 & CA03,04 offline \\
  10 May 2006	& 18496	& 1.80$\pm$0.35 & 18624 & 1.63$\pm$0.36 & 39 & 1057-797$^{c}$ & 20 & H214 & CA04 offline \\
  10 Oct 2006	& {\bf 18496}	& {\bf 1.70$\pm$0.10} & {\bf 18624} & {\bf 1.59$\pm$0.10} & 33-40 & 1934-638, 46 & 170 & H214 &  \\
  1 May 2007  & {\bf 18496} & {\bf 1.61$\pm$0.06 G} & {\bf 19520} & {\bf 1.79$\pm$0.07 G} & 28-40 & 1934-638, 44 & 440 & 1.5C &  \\ 
\hline
\end{tabular}
 \begin{list}{}{}
  \item $^{a}$ No flux calibration observation was taken, but from many other obsevations of HD100546 and other
  targets at 16~mm the 'raw' flux from Miriad is typically within $\sim$ 10\% of the calibrated flux; $^{b,c}$ Flux
  taken to be 1.85 Jy from observations on 6 \& 8 May 2006. 
  G refers to an elliptical Gaussian model fit with UVFIT in Miriad.
  Numbers in bold are our best estimates of total fluxes. 
  A flux of 1.7$\pm$0.1~mJy is recommended at 19~GHz, with a dominant thermal dust emission component, 
  but significant free-free component. See text for details. 
 \end{list}
\label{TAB:hd100-obs16mm}
\end{table*}

\begin{table*}
\caption{3 and 6~cm ATCA observations of HD100546.}
\begin{tabular}{llllllllll}
\hline
 Date & $\nu_1$ & $F_1$ & $\nu_2$ & $F_2$ & Elev.             & Flux cal.,         & $T_{\rm int}$ & Config & Notes \\ 
 UT		&  (MHz)  & (mJy) & (MHz)   & (mJy) & 1057 ($^{\circ}$) & elev. ($^{\circ}$) & (mins)        &        &       \\ \hline
9 May 200	  & 4800 	& $\le$0.66 (3$\sigma$) & 8640 & 0.77$\pm$0.21   & 36-39 & 0823-500, 66 & 72  & H214 &  \\
9 Oct 2006  & 4800  & 0.34$\pm$0.06         & 8640 & 0.66$\pm$0.07   & 23-40 & 1934-638, 30 & 372 & H214 &  \\
11 Jul 2007 & 4800  & 0.42$\pm$0.10 G       & 8640 & 0.53$\pm$0.08 G & 20-40 & 1934-638, 48 & 478 & 6C   &  \\
31 Aug 2008 & 4800  & $\le$0.69 (3$\sigma$) & 8640 & 1.36$\pm$0.37   & 38-40 & None         & 75  & H75  & CA03 offline \\
31 Aug 2008 & 4800  & $\le$0.72 (3$\sigma$) & 8640 & $\le$1.32 (3$\sigma$) & 39.5 & 0823-500, 61 & 105 & H75  & CA03 offline \\
31 Aug 2008 & 4800  & $\le$1.38 (3$\sigma$) & 8640 & $\le$0.81 (3$\sigma$) & 38 & None     & 45  & H75  & CA02,03 offline \\
\hline
\end{tabular}
 \begin{list}{}{}
  \item G refers to an elliptical Gaussian model fit with UVFIT in Miriad. 
 \end{list}
\label{TAB:hd100-obs36cm}
\end{table*}

\subsection[]{Images}

\begin{figure*}
\includegraphics[scale=0.80]{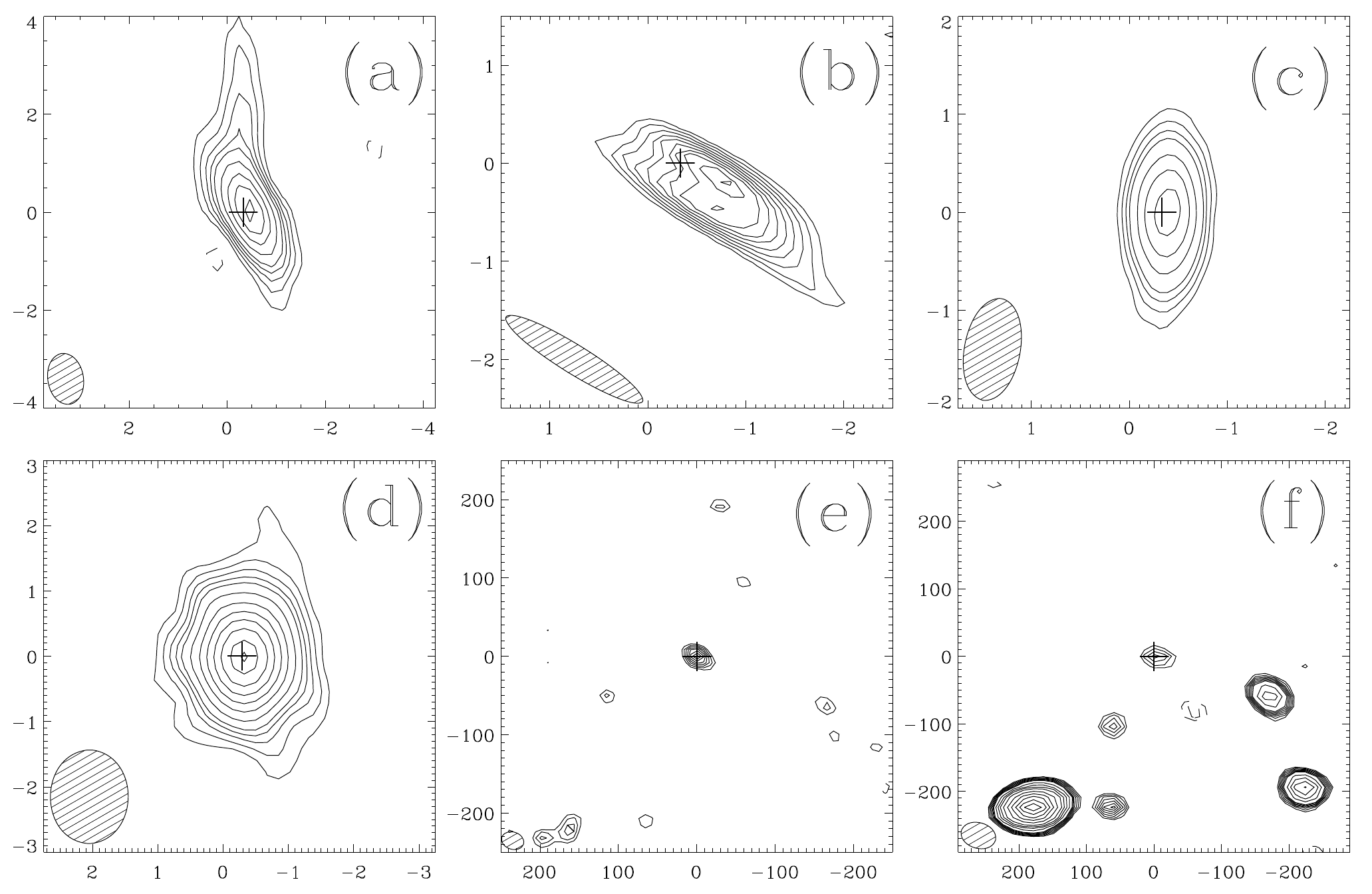}
\caption{Images of HD100546, with arcsecond offsets from the J2000 position. North is up and east 
to the left. The cross shows the expected position of HD100546 at the time of the observation, using 
its established proper motion over $\sim$ 8.35 years for (a), 8.5 years for (b) and (c), 7.33 years 
for (d) and 6.77 years for (e) and (f).
a) Combined 93.504 and 95.552~GHz, naturally weighted in the 750A configuration on 7 May 2008, 
with a synthesised beam of 1.05$\times$0.73\arcsec at PA=9.4$^{\circ}$. Contours are in terms 
of the RMS of 0.526 mJy/beam, with levels at 3, 5, 7, 9, 11, 15, 20, 25, 30$\sigma$. 
b) 93.504~GHz, naturally weighted in the 1.5B configuration on 27 June 2008, with a synthesised 
beam of 1.64$\times$0.29\arcsec at PA=58.2$^{\circ}$. Contours are in terms of the RMS of 
0.716 mJy/beam, with levels at 3, 4, 5, 6, 7, 8, 9, 10$\sigma$. 
c) Combined 42.944 and 44.992~GHz, naturally weighted in the 1.5B configuration on 27 June 2008, 
with a synthesised beam of 1.05$\times$0.57\arcsec at PA=$-10.7^{\circ}$. Contours are in terms of 
the RMS of 0.163 mJy/beam, with levels at 4, 6, 8, 10, 15, 20, 25$\sigma$. 
d) Combined 18.496 and 19.520~GHz naturally weighted in 1.5C on 1 May 2007. The synthesised beam 
size is 1.43$\times$1.20\arcsec at PA=$-1.0^{\circ}$. Contours are in terms of the RMS of 
0.032 mJy/beam, with levels at 3, 5, 7, 9, 11, 15, 20, 25, 30, 35, 40, 42.5$\sigma$.
e) 8.64~GHz naturally weighted in H214 (without antenna 6) on 9 October 2006, with a synthesised beam 
of 29.0$\times$22.1\arcsec at PA=72.0$^{\circ}$. Contours are in terms of the RMS of 0.053 mJy/beam, 
with levels at 3, 4, 5, 6, 7, 8, 9, 10$\sigma$.
f) 4.8~GHz naturally weighted in H214 (without antenna 6) on 9 October 2006, with a synthesised beam 
of 52.4$\times$38.1\arcsec at PA=74.1$^{\circ}$. Contours are in terms of the RMS of 0.061 mJy/beam, 
with levels at 3, 3.5, 4, 4.5, 5, 5.5, 6, 8, 10, 12, 15, 20, 25, 30, 35, 40$\sigma$.}
\label{FIG-3716mm-normcal}
\end{figure*}

Figure~\ref{FIG-3716mm-normcal} shows images of HD100546 from 3.2~mm to 6.2~cm. The cross in each panel 
corresponds to the expected position of HD100546 based on its proper motion from its J2000 position. 
In Fig.~\ref{FIG-3716mm-normcal}-b there is a relatively large offset between the emission centre and 
expected position. That it occurs along an axis similar to that of the synthesised beam elongation 
suggests it is primarily due to the short synthesis track. This is similar to the cases at 7~mm in 
Figure~\ref{FIG-7mm-normcal}-a,b. The finger-like extension at 3~mm in Fig.\ref{FIG-3716mm-normcal}-a 
is almost certainly due to inadequate phase correction. The 'seeing' for this data set also seems to 
have been quite poor, judging from the elongated nature of the main body of emission along the beam
major axis. 

In all cases apart from Fig.~\ref{FIG-3716mm-normcal}-b the cross and peak of emission closely coincide. 
At 3, 7 and 16~mm in Fig.~\ref{FIG-3716mm-normcal}-a,c,d respectively there is however a consistent small 
shift of the peak west of the expected position. In all these cases the emission is resolved, as judged 
from a decrease in flux with increasing $u-v$ distance. So the slight positional offset could feasibly 
be due to the disk being brighter to the west, as found in the high resolution 7~mm images in 
Figures~\ref{FIG-7mm-normcal} and \ref{FIG-7mm-selfcal}, as well as at 3~mm in Fig.~\ref{FIG-3716mm-normcal}-b. 

The 3.5~cm and especially 6.2~cm images in Fig.~\ref{FIG-3716mm-normcal}-e,f respectively show a cluster of 
radio sources in the field around HD100546. That they appear to have negative spectral indices, i.e. increasing 
flux toward lower frequencies (longer wavelengths) suggests they are almost certainly extragalactic (although
the image size at 3.5~cm had to be expanded beyond the HPBW to see most of these sources). The source at offset
position (180,-225) has a double lobed structure, confirmed by the 6C 6.2~cm observation of 11 July 2007 not
shown here, and is probably a radio galaxy jet. 

\begin{figure}
\includegraphics[scale=0.515]{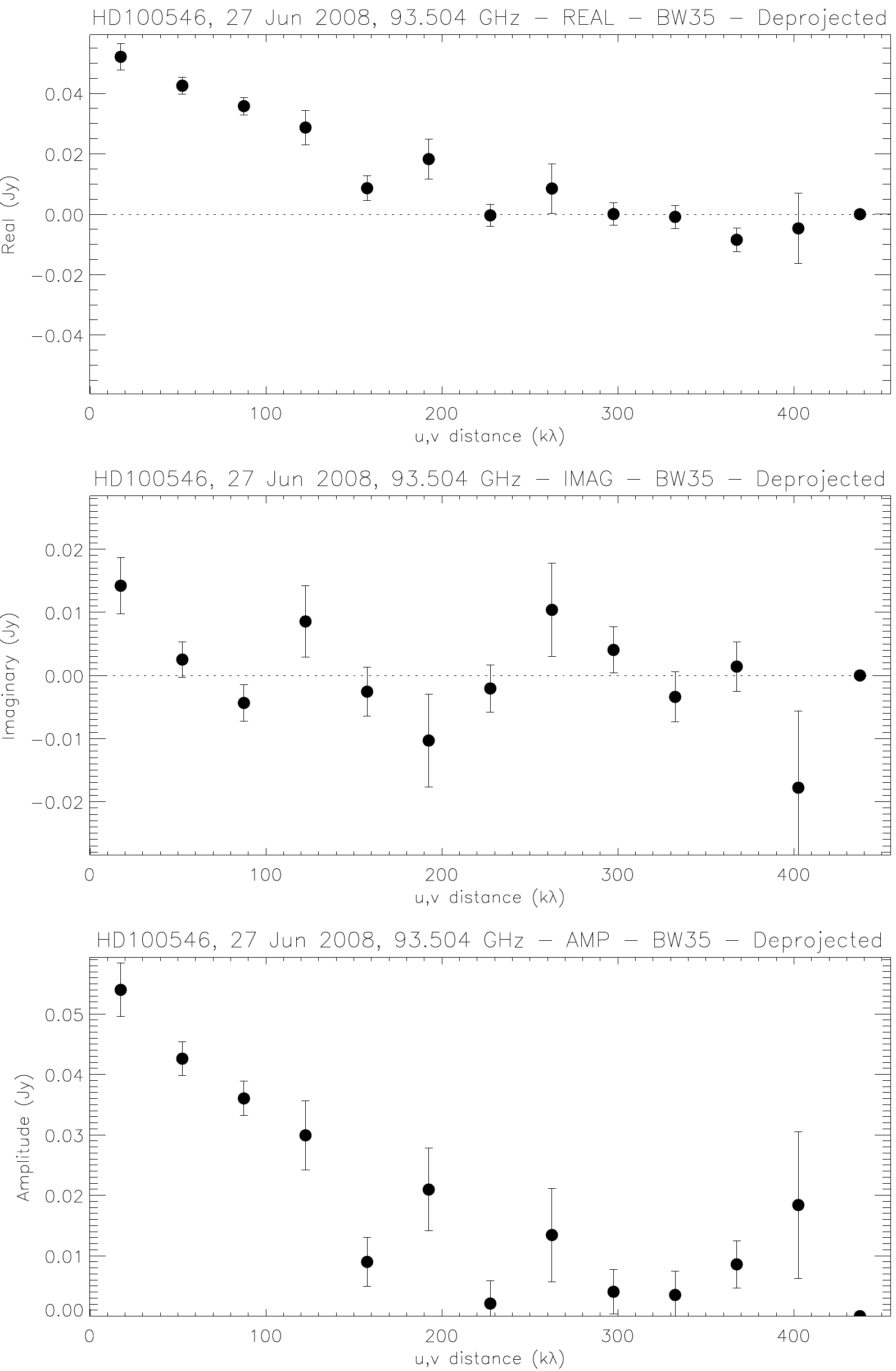}
\caption{Real, imaginary and total visibility amplitudes versus de-projected $u-v$ distance computed 
using the 3~mm data set of 27 June 2008 and combining both frequency sidebands. A bin width of 
35~k$\lambda$ has been used.}
\label{FIG-3mmJun08-Vis}
\end{figure}

The image at 3~mm in Figure~\ref{FIG-3716mm-normcal}-b shows a double peak, a morphology very similar to 
that seen at 7~mm in Figure~\ref{FIG-7mm-normcal}-a and indicative of a gap in the inner disk of HD100546. 
Figure~\ref{FIG-3mmJun08-Vis} shows the deprojected real and imaginary components -- and the resulting
amplitude -- of the visibilities of the combined 93.504 and 95.552~GHz sidebands. As in the 7~mm case seen
in Figure~\ref{FIG-AllCABB-Vis} the real component passes through zero at around 300~k$\lambda$, a null
which confirms the gap seen in the reconstructed image. For the deprojection we have used our own directly 
derived values for the major axis position angle of 140$^{\circ}$ and the disk inclination from face-on 
of 40$^{\circ}$. 

\begin{figure}
\includegraphics[scale=0.385]{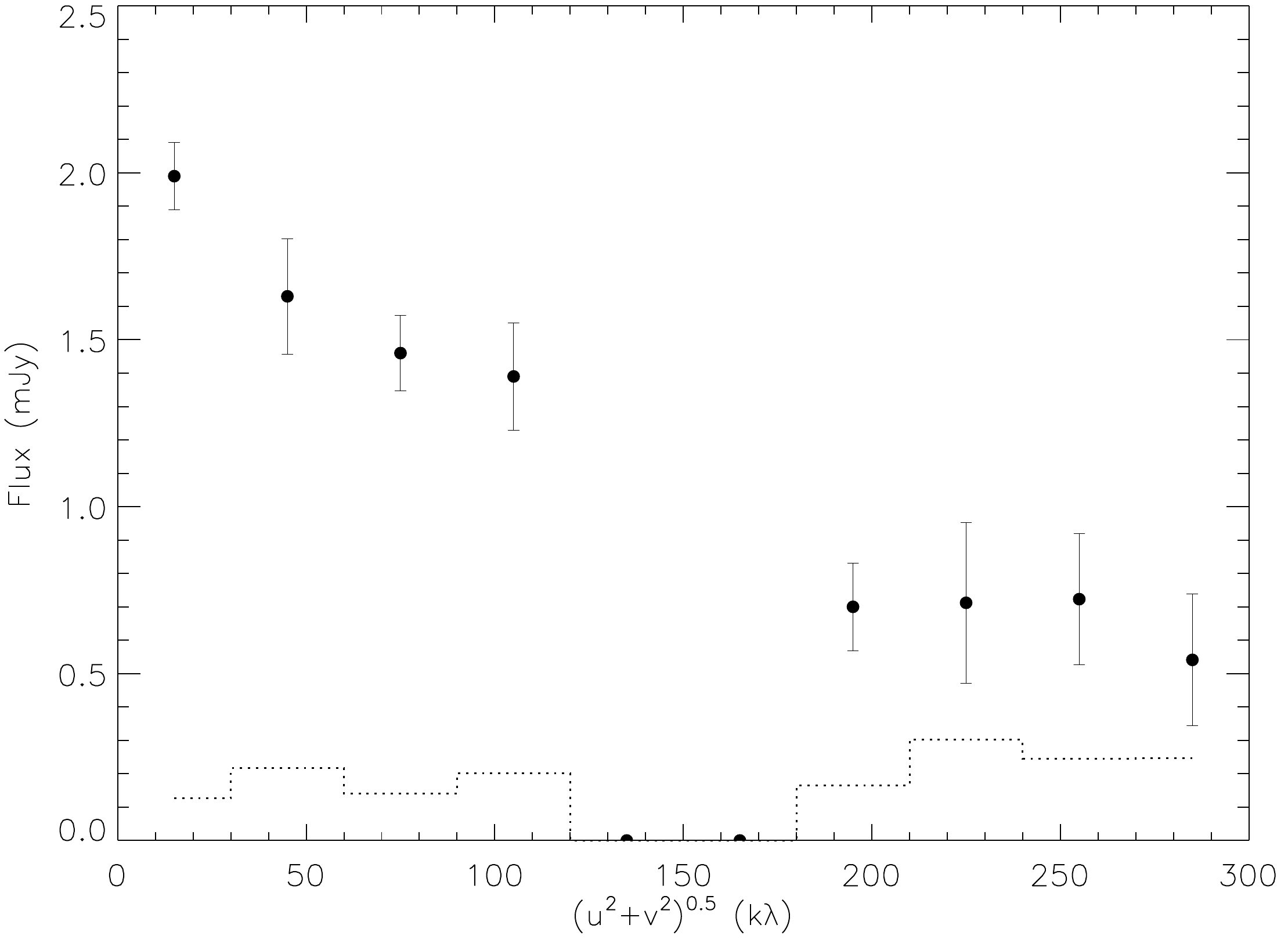}
\caption{Amplitude versus $u-v$ distance of HD100546 at 19.52~GHz from observations taken on 1 May 2007 
in the 1.5C configuration. The dotted line shows the expected amplitude in the case of zero signal. The 
bin size is 30~k$\lambda$.}
\label{HD100-16mm-uvamp}
\end{figure}

Figure~\ref{HD100-16mm-uvamp} is a plot of the 19.52~GHz binned visibility amplitudes as a function 
of baseline distance in k$\lambda$ (not deprojected). The decreasing amplitude with $u-v$ distance is 
a signature that the emission is resolved. An elliptical Gaussian fit in the $u-v$ plane results in 
approximately equal major and minor axes, within a narrow range of FWHM of $\sim$ 0.5\arcsec--0.6\arcsec, 
with no preferred position angle. In other words, the source structure is essentially circular. Although 
the baseline distance extends to about 300~k$\lambda$, where a null appears in similar plots for 3~mm and 
7~mm data, there is instead a suggestion that at 16~mm the amplitudes flatten at a constant level of 
around 0.5~mJy at $\geq$ 200~k$\lambda$. This may indicate an unresolved component to the emission, which 
could feasibly be identified with a free-free emission component discussed in Appendix C.   

\section[]{Molecular gas}

Before an analysis can be made of the dynamics of large-scale molecular gas in the HD100546 disk the systemic 
velocity of HD100546 itself must be known. From their own plus previously published data of narrow photospheric 
absorption lines, Carmona et al. (2011) inferred a radial velocity for HD100546 of 16$\pm$2~km~s$^{-1}$, 
with ranges going down to 14$\pm$2~km~s$^{-1}$ and up to 17$\pm$5~km~s$^{-1}$ (Donati et al. 1997). Using 
their preferred value, the (kinematic) LSR velocity is thus 5.8$\pm$2.0~km~s$^{-1}$. Otherwise, the LSR 
velocity has been reported to be 5.6~km~s$^{-1}$ by Fedele et al. (2013), who assumed the line centre of 
the CO J=10-9 and J=16-15 double-peaked profiles, detected with Herschel, to be the system velocity. Walsh 
et al. (2014) infer the systemic LSR velocity to be 5.7~km~s$^{-1}$ from ALMA CO J=3-2 observations. We 
use the Fedele et al. (2013) value as their data also shows a small distinct peak in both CO lines at 
5.6~km~s$^{-1}$, which presumably represents hot gas close to the star and moving across the line-of-sight. 

\subsection[]{HCO$^+$ J=1$\rightarrow$0}

\begin{figure}
\includegraphics[scale=0.41]{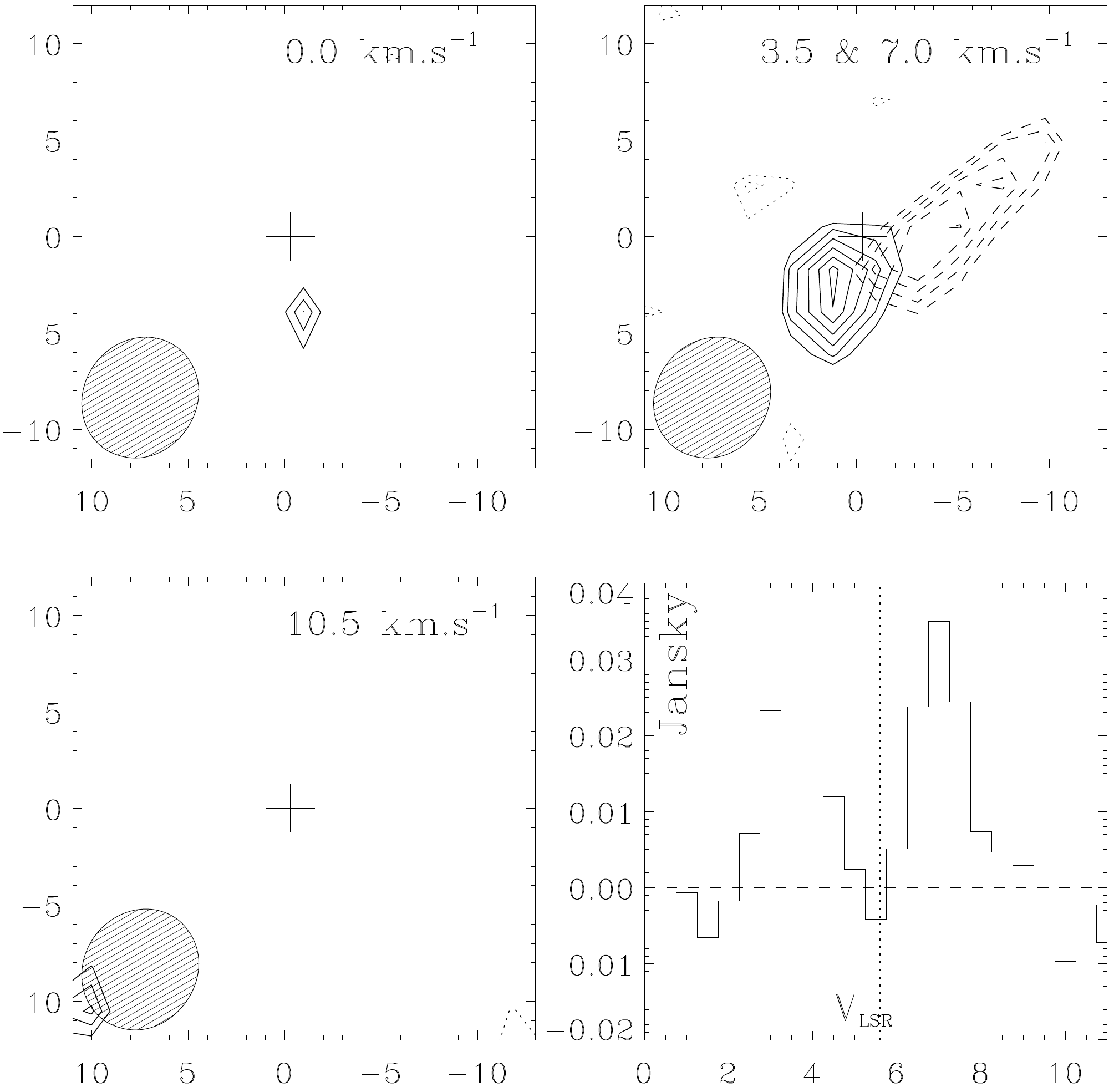}
\vspace{20pt}
\caption{Top row and bottom left: HCO$^+$ J=1 $\rightarrow$ 0 line images of HD100546 taken on 3 October 2007 
in the H75C configuration. Offsets are in arcseconds from the J2000 position, with north up and east to the 
left. The cross marks the expected position of HD100546 after 7.75 years of proper motion. Each image is centred 
at the velocity shown in the top right hand corner, in bins 3.5~km~s$^{-1}$ wide and in steps of 3.5~km~s$^{-1}$. 
The bin size was chosen to take in the entire width of each velocity component of the HCO$^+$ emission line 
seen in the bottom right panel. In the top right panel the 3.5~km~s$^{-1}$ and 7.0~km~s$^{-1}$ components 
are shown in solid and dashed contours respectively. Their positional offset is well aligned with the known 
disk major axis. Contour levels are in terms of the RMS of 7 mJy/beam, with levels at 2.0, 2.25, 2.5, 2.75, 
3.0, 3.25, 3.5$\sigma$. The synthesised beam is $6.5\arcsec\times5.8\arcsec$ at $-35.7^{\circ}$. The 
spectrum of the HCO$^+$ J=1-0 89.2~GHz rotational transition in the bottom right was constructed using 
0.5~km~s$^{-1}$ bins and then Hanning smoothed. The vertical dotted line indicates the systemic velocity
of 5.6~km~s$^{-1}$.}
\label{FIG-hcospec}
\end{figure}

The HCO$^+$ J=1-0 rotational transition at 89.2~GHz was detected with ATCA, shown in Figure~\ref{FIG-hcospec}. 
The detection was made on 3 October 2007 in the compact H75C configuration, which maximised sensitivity 
but minimised spatial resolution. The weakness of the line emission, and the lower sensitivity (e.g. 
higher T$_{\rm sys}$, less antennas) on all other dates when HCO$^+$ was attempted, explains why it was not 
previously detected (Table~\ref{TAB:hd100-obs3mm}). 

Each velocity component of the double-peaked line profile -- at $\sim$ 3.5 and 7.0~km~s$^{-1}$ -- is 
centred near the position of HD100546, but there is a positional offset of the centroids along an axis 
aligned approximately with the known major axis of the disk. When integrated across its entire velocity 
extent the elonagation of the HCO$^+$ emission region is also consistent with this axis, similar to the 
CO~J=3-2 ALMA images in Walsh et al. (2014) and Pineda et al. (2014). Even the 7.0~km~s$^{-1}$ component 
itself appears elongated along this axis, but our confidence in its reality is lower given the contour
levels only extend between 2.0 and 3.5$\sigma$.  

The 3.5~km~s$^{-1}$ component coincides closely with the systemic (LSR) velocity of the dark cloud DC~296.2-7.9 
of 3.2~km~s$^{-1}$, found by us at the position of HD100546 in CO~J=1-0 using the Mopra 22~m telescope 
(bottom panel of Figure~\ref{FIG-NANTEN2}). A slightly different velocity of 3.6~km~s$^{-1}$, also in CO~1-0 
with Mopra, was found by Otrupcek, Hartley \& Wang (2000) at a position about 30\arcmin north. But it is 
highly unlikely this cloud has been detected as the blue-shifted component in our ATCA data, rather than the 
HD100546 disk, as its extended emission would almost surely be resolved out by the interferometer. 

With a systemic velocity of HD100546 of 5.6~km~s$^{-1}$ the 3.5 and 7.0~km~s$^{-1}$ components are blue- 
and red-shifted respectively, and their spatial separation is such that they peak to the south-east and 
north-west. In other words, the south-east and north-west sides of the disk respectively approach and 
recede from Earth. This agrees with the sense of rotation seen in the [O~I] 6300\AA ~line by 
Acke \& van den Ancker (2006), as well as with the ALMA integrated CO~J=3-2 images in Walsh et al. (2014) 
and Pineda et al. (2014).

However, the HCO$^+$ velocity structure is not symmetric about the systemic velocity, with the blue-shifted SE 
side moving 0.7~km~s$^{-1}$ faster than the red-shifted NW side. An assumption of Keplerian rotation allows 
an estimate of the radius from which the emission originates. Using $R=G~M\times(sin~i/v)^2$ with a stellar 
mass of 2.4~M$_{\odot}$ and a disk inclination angle of 45$^{\circ}$, the radius is $\sim$ 540~AU for the 
red side (using a $v$=1.4~km~s$^{-1}$) and $\sim$ 240~AU (with $v$=2.1~km~s$^{-1}$) for the blue side. 
Interestingly, the red-shifted 7~km~s$^{-1}$ integrated emission is also suggestive of a more extended NW side 
of the disk, although we again refer to the relatively low S/N. 

Alternatively, this apparent velocity difference may be accounted for in the context of the warped disk model 
of Quillen (2006), and used by Panic et al. (2010), as the required parameters do overlap with the model. For 
instance, an inclination of 30$^{\circ}$ for the NW red-shifted side instead of 45$^{\circ}$ -- within the 
6$^{\circ}$-15$^{\circ}$ range of the Quillen model with the highly warped region at radii $\geq$ 200~AU -- brings 
the two radii into agreement. A warp of about 10$^{\circ}$ and at $\sim$ 200~AU was inferred by Pineda 
et al. (2014) to explain the non-keplerian features of their ALMA CO~J=3-2 data. 

Wilner et al. (2003) notes that the critical density for collisional excitation of the J=1 rotational 
level is $\sim 6 \times 10^4$ cm$^{-3}$. Thus, assuming collisional excitation is the relevant
mechanism populating the HCO$^+$ rotational levels then it is likely that high gas densities exist at 
radii out to several hundred AU. However, no information is available on whether the J=1 level is in 
thermal equilibrium or is sub-thermally populated, so the actual density cannot be estimated with any 
degree of precision. Even so, judging from the spatial separation of the velocity components and the 
overall image morphology the molecular emission is marginally resolved, and certainly extends to much 
larger radii than the 50--100~AU region from which emission arises from mm--cm sized dust grains. Such 
a size differential between the gas and dust disks is becoming an increasingly more prevalent 
observational phenomenon, e.g. Hughes et al. (2008) and Birnstiel \& Andrews (2014). 

\subsection[]{CO J=4$\rightarrow$3 and J=7$\rightarrow$6}

Following the ATCA detection follow-up data was obtained with the NANTEN2 4~m single-dish sub-millimetre 
telescope, at 4865~m on Pampa la Bola in the Chilean Atacama desert. Spectra were obtained of the CO~J=4-3 
and J=7-6 rotational transitions at respectively 461.040768~GHz and 806.651801~GHz, plus the 
[C~I] $^{3}$P$_{2}$ -– $^{3}$P$_{1}$ 809.341970~GHz transition, on 6-8, 12 and 17-20 December 2007. As 
well as the pointing centred on HD100546 a position offset 38\arcsec east was also observed. Half power 
beam widths and main beam efficiencies were 38\arcsec and 0.50 at 460~GHz, and 26.5\arcsec and 0.45 at 
810~GHz. A dual-channel 460/810~GHz receiver was used, with backends consisting of two acusto optical 
spectrometers with a bandwidth of 1~GHz and a channel resolution of 0.37~km~s$^{-1}$ at 460~GHz and 
0.21~km~s$^{-1}$ at 806~GHz. System temperatures averaged around 670--755~K and 1615--1715~K~at 461~GHz 
and 807~GHz respectively for the on-source observations and about 100~K higher for the offset position. 

\begin{figure}
\includegraphics[scale=0.70]{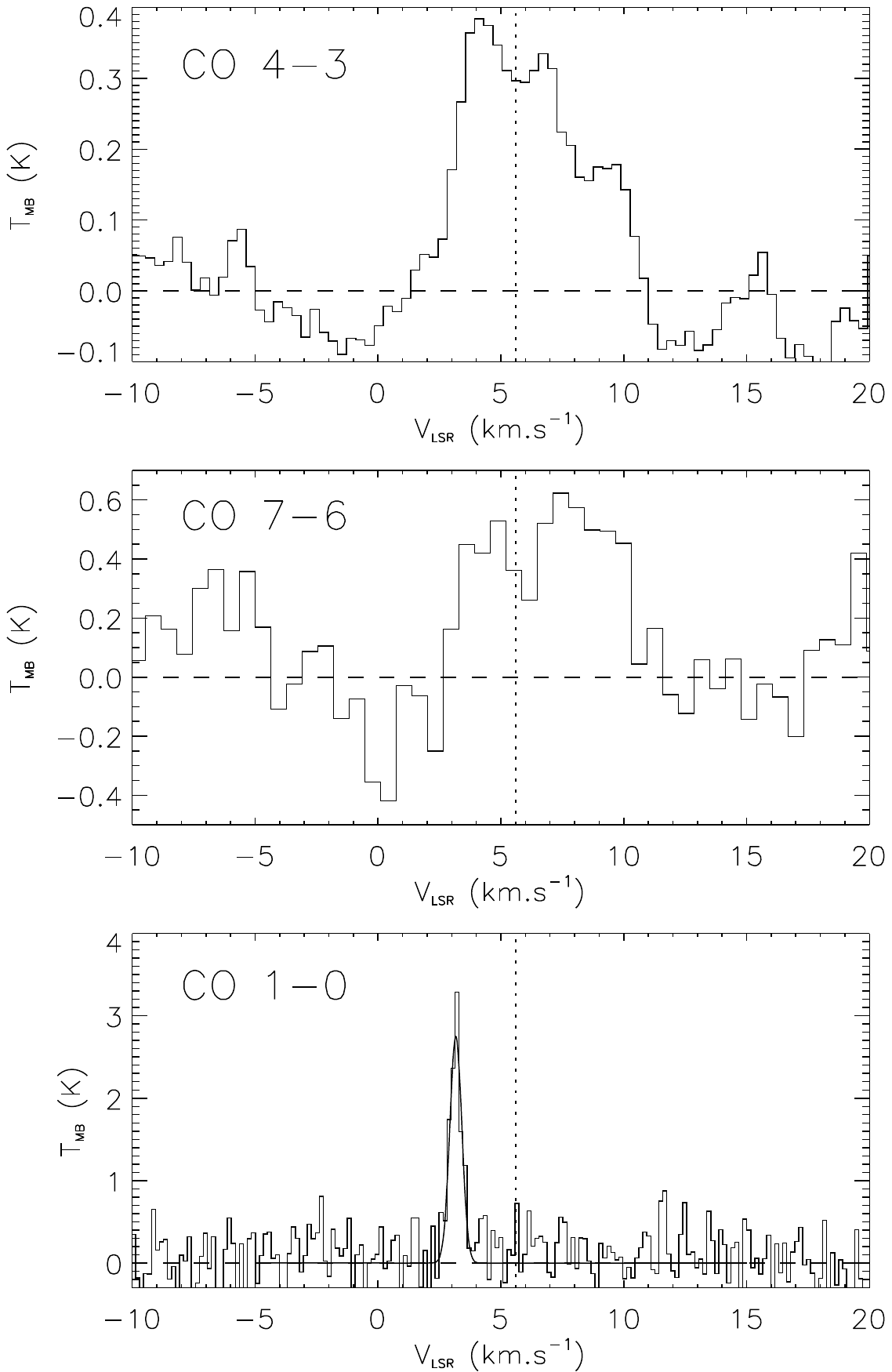}
\caption{NANTEN2 spectra of the CO~J=4-3 (top) and J=7-6 (middle) transitions, and Mopra spectrum of CO~J=1-0 
(bottom), toward HD100546. The CO~7-6 spectrum has been binned across 3 velocity channels. The vertical dotted 
line indicates the systemic velocity of 5.6~km~s$^{-1}$. The Mopra spectrum was obtained on 29 and 30 June 2001
using an SIS receiver and with a beam size of 33\arcsec. See Ladd et al. (2005) for further details on the Mopra
system at that time.}
\label{FIG-NANTEN2}
\end{figure}

Displayed in the top and middle panels of Figure~\ref{FIG-NANTEN2}, the CO~4-3 and 7-6 showed definite 
detections, double-peaked with CO~4-3 velocity components at $\sim$ 4~km~s$^{-1}$ and slightly less than 
7~km~s$^{-1}$. From a single Gaussian fit to each CO line the integrated flux, line centre, FWHM and peak 
height are 2.0$\pm$0.3~K~km~s$^{-1}$, 5.7$\pm$0.2~km~s$^{-1}$, 5.3$\pm$0.4~km~s$^{-1}$ and 0.36$\pm$0.02~K 
for CO~4-3 and 3.2$\pm$0.9~K~km~s$^{-1}$, 7.2$\pm$0.4~km~s$^{-1}$, 5.0$\pm$0.8~km~s$^{-1}$ and 0.60$\pm$0.06~K 
for CO~7-6. To our knowledge this is the first detection of CO~4-3 toward HD100546, and improves the detection 
of CO~7-6 in Panic et al. (2010). The [C~I] line was not detected, as also found by Panic et al. (2010). 

Despite being a single dish of relatively small aperture, the NANTEN2 detections certainly arise from the 
HD100546 disk. Emission from such relatively warm gas, originating from high J levels and with a FWZI of 
$\sim$ 10~km~s$^{-1}$, is highly unlikely to originate from the dark cloud DC~296.2-7.9, which has a FWHM 
and FWZI of only $\sim$ 0.6~km~s$^{-1}$ and 1.7~km~s$^{-1}$ respectively in our Mopra CO J=1$\rightarrow$0 
spectrum (bottom panel of Figure~\ref{FIG-NANTEN2}). Further, the NANTEN2 offset position spectrum showed 
no emission, despite achieving an rms noise level of $\sim$ 0.08~K at the CO~4-3 frequency, good enough for 
a 2-3$\sigma$ detection of a line half as bright as that seen in the on-source spectrum.

The $\sim$ 4 and $\leq$ 7~km~s$^{-1}$ velocity components of our CO~4-3 are in extremely good agreement with 
those found for CO~3-2 by Panic \& Hogerheijde (2009). The S/N of the 7-6 spectrum is good enough to resolve 
the double-peak line profile but insufficient to reliably define the peak positions, except to say they are 
also consistent with the various CO rotational spectra in Panic et al. (2010) and Fedele et al. (2013). There 
is however a distinct $\sim$ 0.5~km~s$^{-1}$ difference in the velocity of the blue wing peak between the 
NANTEN2 CO~4-3 (and thus also APEX CO~3-2) and the ATCA HCO$^+$ 1-0 spectra, whereby the peak emission moves 
toward lower velocities for the single-dish (lower resolution) data. Another difference between the single-dish 
CO and interferometer HCO$^+$ is that the CO has more emission at velocities 'filling in' the region between 
the red and blue wing peaks. 

The CO~4-3 line profile is asymmetric, with the blue wing brighter than the red wing, also in very good agreement
with the CO~3-2 and 6-5 spectra presented in Panic \& Hogerheijde (2009) and Panic et al. (2010). The latter 
interpreted the profile as indicative of a temperature asymmetry between the two sides of the disk, perhaps 
produced by a slight but significant warp between outer and inner regions of the disk. The warp shadows one side 
of the disk, thus depressing the local temperature. They do not explicitly state which side of disk is colder 
or warmer than the other, but their Figure 3 strongly suggests it is the SE side which is warmer. This makes 
sense as it is the blue wing which is enhanced and the SE side is indeed the one approaching the observer. 
Pineda et al. (2014) claim that their ALMA integrated CO~3-2 image and position-velocity structure support 
the existence of a warp, with a transition of around 10$^{\circ}$ in the inner and outer disk inclinations 
occurring at a radius of $\sim$ 200~AU. 

But the Panic et al. temperature asymmetry hypothesis has a problem with the CO~7-6 data presented here.
Panic et al. alluded to such a difficulty on the basis of their own 7-6 spectrum, but had insufficient faith 
in their S/N to pursue it further. Our NANTEN2 detection of the CO~7-6 line confirms its somewhat anomalous 
nature compared to lower J transitions, specifically that the brightness asymmetry between the red and blue 
wings is either non-existent or perhaps even reversed. Certainly the Panic et al. best fit model could not 
fit the CO~7-6 line profile we find. The observed CO~10-9 and 16-15 spectra in Fedele et al. (2013) are also 
essentially symmetric, and were adequately fit with a flat disk model. Also, the predicted profiles in the 
extensive model grid of Bruderer et al. (2012) are symmetric, and thus do not require a warp and its 
associated temperature asymmetry. 

Another interesting facet of the NANTEN2 data is that in both the 4-3 and 7-6 CO lines there is more 
emission at higher red-shifted velocities than in the APEX 3-2 and 6-5 lines in Panic et al. (2010) or 
the HIFI 10-9 and 16-15 lines of Fedele et al. (2013). Although of relatively low S/N compared to the 
APEX and HIFI spectra, the fact that a feature appears at precisely the same velocity of 9-10~km~s$^{-1}$ 
in each NANTEN2 spectrum gives cause to consider that it may be real. If so, then its absence in the other 
spectra, taken at different times, would imply variability in the CO pure rotational line profiles. This 
may not be such a stretch of the imagination as it could otherwise be, given the aforementioned established 
differences in the level of asymmetry between the red and blue sides of line centre, and that the CO~1-0 
ro-vibrational lines also display profile variability (Brittain et al. 2013).

\section[]{Emission origin at centimetre wavelengths}

\subsection[]{Outflow: 3.5 + 6.2 cm and infrared emission}

From the fluxes in Table~\ref{TAB:hd100-obs36cm} the spectral index between 3.5~cm and 6.2~cm is 
1.1$\pm$0.2 on 9 October 2006 and 0.4$\pm$0.3 on 11 July 2007, or 0.8$\pm$0.2 from the mean fluxes on 
the two dates. This is suspiciously close to the predicted value of 0.6 for an ionized spherical wind
or outflow, e.g. Panagia \& Felli (1975), and thus suggestive of such an origin for the HD100546 radio 
emission. If so, this would be consistent with what is known about radio emission of Herbig stars (at 
least for the relatively few for which relevant data exist, e.g. Skinner, Brown \& Stewart 1993). But 
can other data be brought into play to test this conclusion?

Grady et al. (2005) found the Ly$\alpha$ line in HD100546 to be Type III P-Cygni shaped with absorption
to $-350$ km~s$^{-1}$, clear evidence of a high velocity wind. Several hydrogen recombination lines have 
been observed in the near- to mid-IR spectrum of HD100546, including in the ISO-SWS data and other 
ground-based observations by Habart et al. (2006) and Geers et al. (2007). From this and the radio fluxes 
reported here there is clearly ionized gas associated with HD100546, and thus it is possible that some 
of the 16~mm and even 7~mm emission arises from this gas. In order to arrive at the true dust emission 
spectrum this contribution needs to be quantified and the millimetre fluxes subsequently corrected.

\begin{figure}
\includegraphics[scale=0.32]{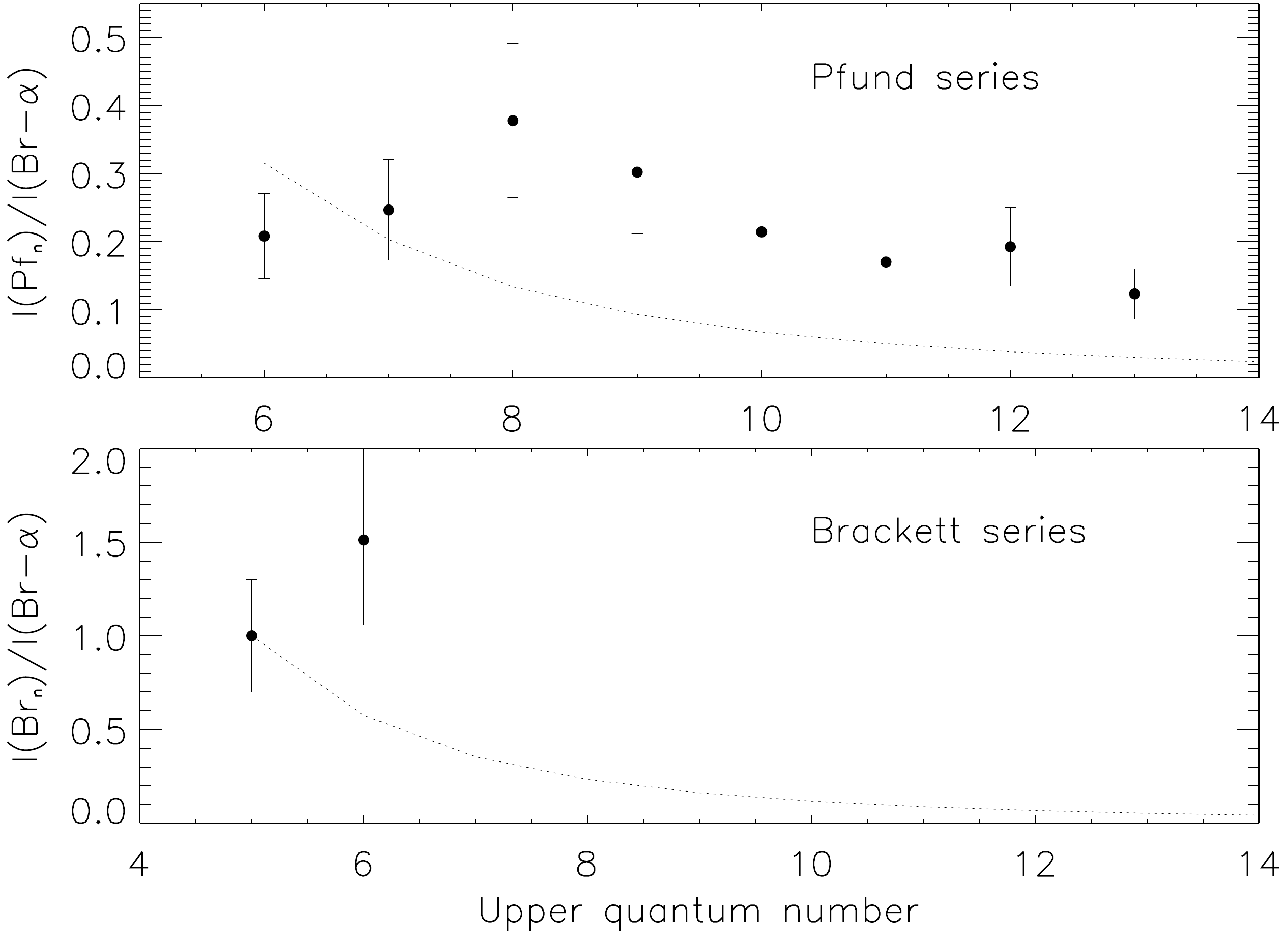}
\vspace{20pt}
\caption{Top: hydrogen recombination lines fluxes as a function of the upper level quantum number. 
Bottom: normalised to the Br-$\alpha$ (n=5--4) transition.}
\label{FIG-hrecomb}
\end{figure}

Despite the aforementioned published spectra of HD100546 containing IR hydrogen recombination lines, 
the integrated fluxes of these lines have not previously been reported. Yet these contain potentially 
important information on the physics at play in the region where accretion and outflow occur. For 
instance, as shown by Nisini et al. (1995) such data can be used as a diagnostic of the mass loss rate.
Therefore, the SWS Highly Processed Data Product of 
Frieswijk et al. (2007)\footnote{ida.esac.esa.int:8080/hpdp/technical\_reports/technote52.pdf} on the 
ISO archive has been used to extract these line fluxes, presented in tabular and graphical formats in 
Table~\ref{TAB-hlines} and Figure~\ref{FIG-hrecomb} respectively.

Figure~\ref{FIG-hrecomb} shows the recombination line fluxes in the Pfund (n$_{\rm lower}$=5) and 
Brackett (n$_{\rm lower}$=4) series, normalised to the Br$\alpha$ line flux. This mode of presentation 
is consistent with similar plots in van den Ancker et al. (2000) and Benedettini et al. (1998, 2001). 
We have applied no extinction correction to the fluxes since the published figures for HD100546 
are all such that A$_{\rm V}$ $\ll$ 1. The dashed line in Figure~\ref{FIG-hrecomb} shows the Case B 
recombination theory expectation from Hummer \& Storey (1987) for an electron temperature and density 
of 10$^{4}$~K and 10$^{4}$~cm$^{-3}$ respectively. 

The most obvious feature in Figure~\ref{FIG-hrecomb} is that the line fluxes (and ratios) do not follow 
the expectation of Case B recombination, suggesting they do not arise in a 'classical' HII region. A 
similar result was also found for several other Herbig AeBe stars by van den Ancker et al. (2000) and 
Benedettini et al. (1998, 2001). Note for instance that the ratios Br$\alpha$/Br$\beta$, Pf$\alpha$/Pf$\beta$ 
and Pf$\alpha$/Pf$\gamma$ are all less than 1. Any amount of extinction would only exacerbate the effect 
(assuming all lines suffer the same amount of extinction). Thus it is almost certain that the recombination 
lines, especially the lower lying $\alpha$, $\beta$ and $\gamma$ transitions, are optically thick. This 
most likely occurs in a density bounded moving flow, be it an accretion inflow or a mass outflow. 

The models of Randich et al. (1991) consider such a scenario, approximating it as a slab of gas with 
a differential expansion velocity between the two sides of 300~km~s$^{-1}$. It is only within their 
considered parameter space of hydrogen gas densities (neutrals plus protons) of 10$^{11}$ and 
10$^{12}$~cm$^{-3}$, and temperatures of 1.0 and 1.25$\times10^{4}$~K, that the Br$\alpha$/Br$\beta$, 
Pf$\alpha$/Pf$\beta$ and Pf$\alpha$/Pf$\gamma$ ratios become less than 1. Such a high density
region must surely be very compact and close to the star, consistent with the H recombination lines 
observed by Habart et al. (2006) and Geers et al. (2007) being spatially unresolved (compared to a 
standard star point spread function), with a FWHM of $\leq$ 10--20~AU.

\begin{table}
\caption{Hydrogen recombination line data for HD100546 from ISO-SWS}
\begin{tabular}{lll}
\hline
   Line   & Wavelength &    Flux         \\
          & ($\mu$m)   &  (W/cm$^2$)     \\ \hline
 6--4 Br$\beta$   & 2.6259 &  5.20e-19  \\
 13--5            & 2.6751 &  4.25e-20  \\
 12--5            & 2.7583 &  6.63e-20  \\
 11--5            & 2.8730 &  5.86e-20  \\
 10--5            & 3.0392 &  7.38e-20  \\
 9--5             & 3.2970 &  1.04e-19  \\
 8--5 Pf$\gamma$  & 3.7406 &  1.30e-19  \\
 5--4 Br$\alpha$  & 4.0523 &  3.44e-19  \\
 7--5 Pf$\beta$   & 4.6538 &  8.49e-20  \\
 6--5 Pf$\alpha$  & 7.4599 &  7.17e-20  \\
\hline
\end{tabular}
\label{TAB-hlines}
\end{table}

The fact that the Br$\alpha$ line is optically thick means that a simple formula can be used to predict 
the radio flux and the mass outflow rate. By combining equations 4 and 15 in Simon et al. (1983) to 
eliminate the mass flow rate, the expression for the 6~cm radio flux becomes 
S$_{\nu}$(mJy)=$1.37\times10^{14}v^{-1/3}F_{L}(Br\alpha)$. The parameters are the terminal wind velocity 
$v$ in units of 100~km~s$^{-1}$, the Br-$\alpha$ line flux in W/m$^2$ and the electron temperature, 
assumed to be 10$^{4}$~K. Using the ISO-SWS observed flux in Table~\ref{TAB-hlines} and a terminal 
velocity of 350~km~s$^{-1}$, found by Grady et al. (2005) in the Ly$\alpha$ line, this gives a predicted 
6~cm flux of 0.32~mJy. Alternatively, a rough estimate of the Br-$\alpha$ line flux in the spectrum of 
Geers et al. (2007) -- assuming it has a FWHM equal to the spectral resolution at 4.05~$\mu$m (where 
$\lambda / \Delta \lambda = 600$) -- is 5.5$\times$10$^{-19}$ W/cm$^{2}$, which predicts a 6~cm flux of 
0.51~mJy. These are in very good agreement with the observed flux shown in Table~\ref{TAB:hd100-obs36cm}. 

Given this good agreement, a rough estimate can be made of the mass outflow rate. Using either the 
Br$\alpha$ line flux or the 6.2~cm radio flux, from equations 15 and 4 in Simon et al. (1983) respectively, 
the inferred mass loss rate from HD100546 is around a few $\times$10$^{-8}$ M$_{\odot}$~yr$^{-1}$.

Based on these results a strong case can be made that the centimetre emission of HD100546 is indeed due 
to a wind with a relatively high mass loss rate. A disk wind has been found by several authors, e.g. 
Grady et al. (2005) in the Si~III and N~I lines with HST STIS. But the magnitude of the outflow rate 
is striking, as a deep spectro-imaging search of HD100546 made by Grady et al. (2005) resulted in no 
indication of a bipolar jet in the Ly$\alpha$ line, unlike HD100546's (putatively) younger cousins 
HD163296 and HD104237 (Devine et al. 2000; Grady et al. 2004).

\subsubsection[]{Accretion rate}

Outflow and accretion are intimately connected processes in young stars, the former essentially being a 
by-product of the latter. Calvet (2004) reports a correlation for T Tauri and FU Orionis stars whereby 
the outflow (mass loss) rate is about a tenth of the accretion (mass inflow) rate. Assuming this also 
holds for Herbig stars then the implied accretion rate for HD100546 is 
a few $\times$10$^{-7}$ M$_{\odot}$~yr$^{-1}$.

Simultaneous accretion and outflow signatures toward HD100546 have been found by many authors, including 
Viera et al. (1999), Deleuil et al. (2004), Grady et al. (2005) and Guimaraes et al. (2006). Though not
all authors estimated an accretion rate $\dot{M}_{acc}$, there have been several values reported. On the
low side, Grady et al. (2005) infer a rate of at most a few $\times$ 10$^{-9}$~M$_{\odot}$~yr$^{-1}$ from 
HST STIS data. However, they did not actually derive their estimate, and it was instead made principally 
from a comparison of its ultraviolet spectrum to that of other Herbig Ae/Be stars with `known' accretion 
rates. 

Much higher accretion rates have been reported before and after the Grady et al. estimate. Talavera 
et al. (1994) derive a rate of $5\times10^{-7}$~M$_{\odot}$~yr$^{-1}$ from the Lyman-$\alpha$ profile 
measured with IUE. Also using IUE data, Blondel \& Djie (2006) used the balance between viscous energy 
produced in the disk and a boundary layer, and the energy radiated by the boundary layer, to estimate 
an accretion rate of $3.44\times10^{-7}$~M$_{\odot}$~yr$^{-1}$. This was for a non-rotating case. In 
the rotating case they instead derived $1.61\times10^{-6}$~M$_{\odot}$~yr$^{-1}$. This is probably too 
high since they used a value of $v~sin~i$ of 250~km~s$^{-1}$, a factor of 3 or 4 larger than values 
reported by other authors. Using equation 14 of Blondel \& Djie (2006) and a $v~sin~i$ of 
65~km~s$^{-1}$ (Guimaraes et al. 2006; Donati et al. 1997) gives a correction to the non-rotating case 
of a factor of $\sim$ 1.5, so resulting in an accretion rate of $\sim$ $5\times10^{-7}$~M$_{\odot}$~yr$^{-1}$, 
consistent with the Talavera et al. (1994) estimate. More recently, an accretion rate of 
log~M$_{\rm acc}$=$-7.23\pm0.13$, or $\sim (6\pm2)\times10^{-8}$~M$_{\odot}$/yr, was determined by 
Pogodin et al. (2012). 

These high values of the mass accretion rate for HD100546 are much more consistent with the value we 
report based on the wind mass loss rate. That they are derived from phenomena widely separated in the 
electromagnetic spectrum, from the ultraviolet through to infrared and radio, and diagnostic of inflow 
and outflow, lends support to their reliability. Being two orders of magnitude higher than the figure 
typically assumed in the literature it also has important implications for the understanding of HD100546 
and its disk. Some of these implications are discussed in the main body of this paper. 

\subsubsection[]{Other possible radio emission mechanisms}

Although there is a strong case for wind emission to be responsible for the radio fluxes of HD100546,
there are other possible mechanisms which should be tested. Skinner, Brown \& Stewart (1993) consider 
a number of such mechanisms in their study of the radio emission from a large sample of Herbig Ae/Be 
stars. These include thermal emission from i) classical HII region, ii) dust, iii) accretion, iv) shocks 
and v) stellar winds. In Table~\ref{TAB:radiomech} we look at what the respective contributions from 
these mechanisms might be. Formulae can be found in the above-mentioned reference, and we refer to that 
work for a discussion of the validity of applying the relevant theory of accretion and shock induced 
radio emission to Herbig stars. In the shock and wind cases, we use a terminal outflow velocity of 
350~km~s$^{-1}$, as found by Grady et al. (2005) in the Ly$\alpha$ line. For the accretion and outflow 
rates we use the several values found in the literature as well as that reported here.  

\begin{table*}
\caption{Possible contributions to the 7~mm, 16~mm, 3.5~cm and 6.2~cm emission of HD100546}
\begin{tabular}{lllllll}
\hline
  Mechanism  & Expected $\alpha$ & \multicolumn{4}{|c|}{Predicted fluxes (mJy)} & Notes \\
             &                & 7~mm & 16~mm  & 3.5~cm & 6.2~cm &       \\
             &                & 44~GHz & 19~GHz & 8.64~GHz & 4.8~GHz &       \\ \hline
 Optically thick HII & 2.0    & 16.9 & 3.14   & 0.65   & 0.20   & Anchored at observed 3.5 cm flux  \\
                     & 2.0    & 31.1 & 5.80   & 1.20   & 0.37   & Anchored at observed 6.2 cm flux  \\  \hline
 Optically thin HII  & $-0.1$ & 684  & 744    & 805    & 854    & n$^2$/V=100 pc$^3$~cm$^{-6}$  \\   \hline
 Accretion           & $-0.1$ & 7.98 & 8.68   & 9.39   & 9.96   & $\dot{M}_{acc}=3\times10^{-9}$ M$_{\odot}$~yr$^{-1}$ \\
                     & $-0.1$ & 160  & 174    & 188    & 199    & $\dot{M}_{acc}=6\times10^{-8}$ M$_{\odot}$~yr$^{-1}$ \\
                     & $-0.1$ & 665  & 723    & 782    & 830    & $\dot{M}_{acc}=2.5\times10^{-7}$ M$_{\odot}$~yr$^{-1}$ \\
                     & $-0.1$ & 1330 & 1446   & 1565   & 1660   & $\dot{M}_{acc}=5\times10^{-7}$ M$_{\odot}$~yr$^{-1}$ \\ \hline
 Shock               & $-0.1$ & 0.021 & 0.023 & 0.025  & 0.026  & $\dot{M}_{wind}=3\times10^{-10}$ M$_{\odot}$~yr$^{-1}$ \\
                     & $-0.1$ & 0.424 & 0.462 & 0.500  & 0.530  & $\dot{M}_{wind}=6\times10^{-9}$ M$_{\odot}$~yr$^{-1}$ \\
                     & $-0.1$ & 1.77  & 1.92  & 2.08   & 2.21   & $\dot{M}_{wind}=2.5\times10^{-8}$ M$_{\odot}$~yr$^{-1}$ \\
                     & $-0.1$ & 3.54  & 3.85  & 4.16   & 4.41   & $\dot{M}_{wind}=5\times10^{-8}$ M$_{\odot}$~yr$^{-1}$ \\ \hline
 Wind                & 0.6  & 0.0044 & 0.0027 & 0.0017 & 0.0012 & $\dot{M}_{wind}=3\times10^{-10}$ M$_{\odot}$~yr$^{-1}$ \\
                     & 0.6  & 0.241  & 0.146  & 0.091  & 0.064  & $\dot{M}_{wind}=6\times10^{-9}$ M$_{\odot}$~yr$^{-1}$ \\
       & 0.6  & {\bf 1.62} & {\bf 0.98} & {\bf 0.61} & {\bf 0.43} & $\dot{M}_{wind}=2.5\times10^{-8}$ M$_{\odot}$~yr$^{-1}$ \\
                     & 0.6  & 4.1    & 2.5    & 1.5    & 1.1    & $\dot{M}_{wind}=5\times10^{-8}$ M$_{\odot}$~yr$^{-1}$ \\ \hline
 Dust          & 2.45 & {\bf 7.79}  & {\bf 1.00}   & {\bf 0.144}   & {\bf 0.034}   & Anchored at 45~mJy at 90~GHz (3~mm) \\
\hline
 Observed fluxes     &      & 8.5$\pm$0.5  & 1.70$\pm$0.10 & 0.65$\pm$0.12  & 0.37$\pm$0.05   &  \\
\hline
\end{tabular}
 \begin{list}{}{}
  \item The calculated $\alpha$ values from the observed fluxes are 2.5$\pm$0.1, 1.9$\pm$0.1, 1.2$\pm$0.1 and 
  0.9$\pm$0.2 between 0.3--0.7~cm, 0.7--1.6~cm, 1.6--3.5~cm and 3.5--6.2~cm respectively, indicating a flattening 
  of the spectrum toward longer wavelengths and thus changing contributions between dust and free-free emission. 
 \end{list}
\label{TAB:radiomech}
\end{table*}
 
At all wavelengths we can rule out emission from both an optically thick or thin `classical' density 
bounded HII region based on several considerations. These include the spectral index constraint, and 
hence the disparity between predicted and observed fluxes, especially at 3.5 and 6.2~cm. 
Similarly, we can also rule out accretion and/or shock emission given the spectral index constraint. 
And even the lowest possible accretion rate produces fluxes far in excess of those observed, whilst 
the lowest and highest outflow rates produce fluxes respectively far below and far above the detected 
values. Clearly the best candidate for the radio emission from HD100546 is free-free emission from an 
ionised wind, with an outflow rate of a few $\times$ 10$^{-8}$ M$_{\odot}$/yr. This is shown in bold in 
Table~\ref{TAB:radiomech}.

\subsection{The SED -- combined dust and free-free emission}

Free-free emission from the wind will certainly dominate the 3.5 and 6.2~cm HD100546 fluxes, but will also 
contribute to the 7 and 16~mm fluxes. On the other hand, thermal dust emission will dominate the 3 to 16~mm
flux, but will also contribute to the 3.5 and 6.2~cm fluxes (also bolded in Table~\ref{TAB:radiomech}). 
Somewhere between the mm and cm regions the dominant emission mechanism switches between dust and free-free 
components, and it is of interest to determine the relative contributions of each process at each frequency. 

Some guide can be gained from Table~\ref{TAB:radiomech}, but a better estimate can be made by fitting
the entire mm--cm SED with two components, fixed with slopes of 2.45$\pm$0.05 for the dust and 0.60$\pm$0.05
for the free-free. An 'anchor' of dust emission of 45.0$\pm$2.5~mJy is used at 90~GHz, whilst a free-free 
'anchor' of 0.325$\pm$0.025 mJy is set at 4.8~GHz, corresponding to a mass loss rate of a 
few $\times$ 10$^{-8}$ M$_{\odot}$/yr. The parameters are varied within their chosen ranges with the 
constraint that the addition of the two emission components remains within the errors bars of the observed 
fluxes. 

A sample result is shown in Figure~\ref{FIG-SEDcompo}, whilst the ranges of the dust (gas) fractional 
contributions at 7~mm, 16~mm, 3.5~cm and 6.2~cm respectively are 0.86$\pm$0.04 (0.14$\pm$0.04), 0.58$\pm$0.07
(0.42$\pm$0.07), 0.25$\pm$0.05 (0.75$\pm$0.05) and 0.10$\pm$0.02 (0.90$\pm$0.02). The gas fractional 
contributions can then be subtracted from the total fluxes in order to arrive at a purely dust emission signal, 
used in the SED model in Figure~\ref{HD100-Alessio}. 

The approximately 40\% contribution of free-free emission at 16~mm may be reflected in the possible unresolved
source component of the binned visibility amplitudes presented in Figure~\ref{HD100-16mm-uvamp}. In that figure
the $\sim$ 0.5~mJy 'plateau' at $u-v$ distances $\geq$ 200~k$\lambda$ is just under a third of the total measured 
16~mm flux of 1.7$\pm$0.1~mJy.


\section[]{Centimetre temporal monitoring}

Monitoring of the centimetre-wave emission from young stars can be another way of establishing the emission
mechanism(s) at particular frequencies. Lommen et al. (2009) and Ubach et al. (2012) found variability from
1.6 through to 6~cm of several T Tauri stars in the Chamaeleon and Lupus molecular clouds, but so far little
has been done on Herbig stars. 

\subsection[]{16~mm}

\begin{figure}
\includegraphics[scale=0.385]{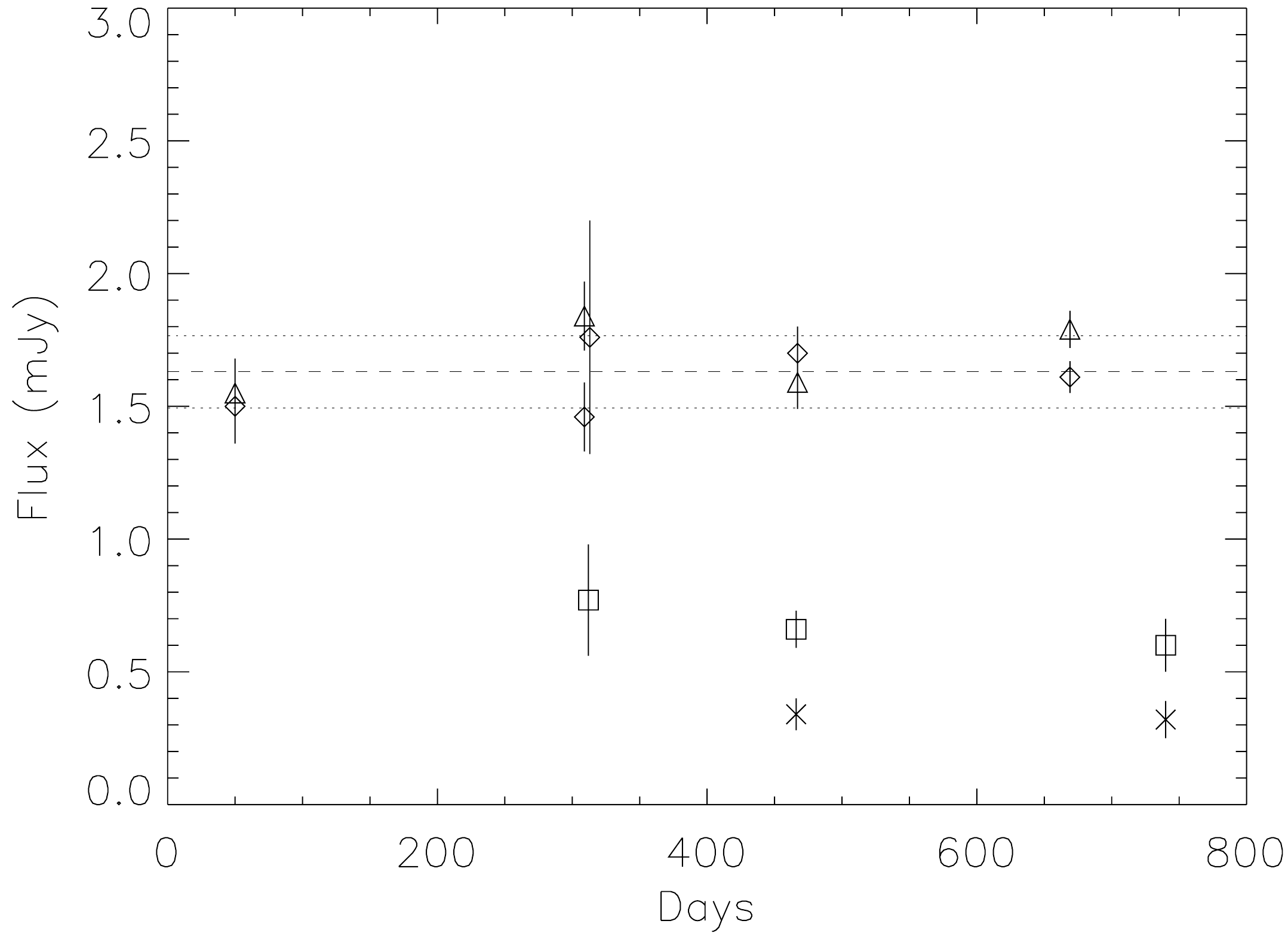}
\caption{Flux versus time plot for HD100546 at 16~mm (open diamonds and triangles), 3.5~cm (open 
squares) and 6.2~cm (asterisks). At 16 mm the first sideband at 18.496~GHz is represented by the 
diamonds and the second sideband at 18.624~GHz (on 3 occasions) and 19.520~GHz (one occasion) by 
the triangles. The dashed and dotted lines represent the mean and standard deviation of all the 
measurements, namely 1.63$\pm$0.14 (see text for further details).}
\label{FIG-temporal}
\end{figure}

Figure~\ref{FIG-temporal} shows the 16~mm emission of HD100546 as a function of time over an interval of a 
few years. The four best measurements -- in terms of $u-v$ plane coverage and sensitivity -- are on 20 
August 2005, 6 May and 10 October 2006, and 1 May 2007. Clearly the emission is stable over timescales 
of months and years. The mean flux and standard deviation from these four measurements for the two sidebands 
is 1.63$\pm$0.14~mJy. 

The 16~mm emission is also stable over days, as demonstrated by the snap-shot observations on 
8, 9 and 10 May 2006. The integration time on each of these days was only $\sim$ 20 minutes. Only 
3 antennas were available on 8 and 9 May, and 4 antennas on 10 May, within the hybrid configuration. 
The mean flux and standard deviation over the 3 days, using both sidebands, is 1.76$\pm$0.44~mJy. 
Despite the precision of the resultant fluxes being relatively poor compared to the other dates, 
there is no statistical difference from day to day, nor between these fluxes and the mean quoted 
above of 1.63$\pm$0.14~mJy. 

As a further check for short-term variability, i.e. hours and minutes, the 20 Aug 2005, 6 May and 
10 October 2006 data sets were broken up into discrete subsets, each with similar integration times, 
down to even single $\sim$ 10~min scans. No statistically significant variability was detected. 
Hence the HD100546 16~mm flux is invariant over timescales of minutes, hours, days, months and 
years. This establishes that the only possible emission processes at 16~mm are thermal dust and/or 
free-free, the respective contributions of which were discusssed in the previous section.

\subsection{3.5 and 6.2~cm}

Also apparent in Figure~\ref{FIG-temporal} is that the flux in the true radio regime -- 3.5 and 
6.2~cm -- is stable over time scales of months and years. Certainly the fluxes measured on 9 May 2006, 
9 October 2006 and 11 July 2007 are consistent with each other. In the following we refer to these 
as the 'stable' or 'quiescent' fluxes, which were concluded in the previous section to be dominated
by free-free emission from an outflowing wind.

\begin{figure*}
\includegraphics[scale=0.80]{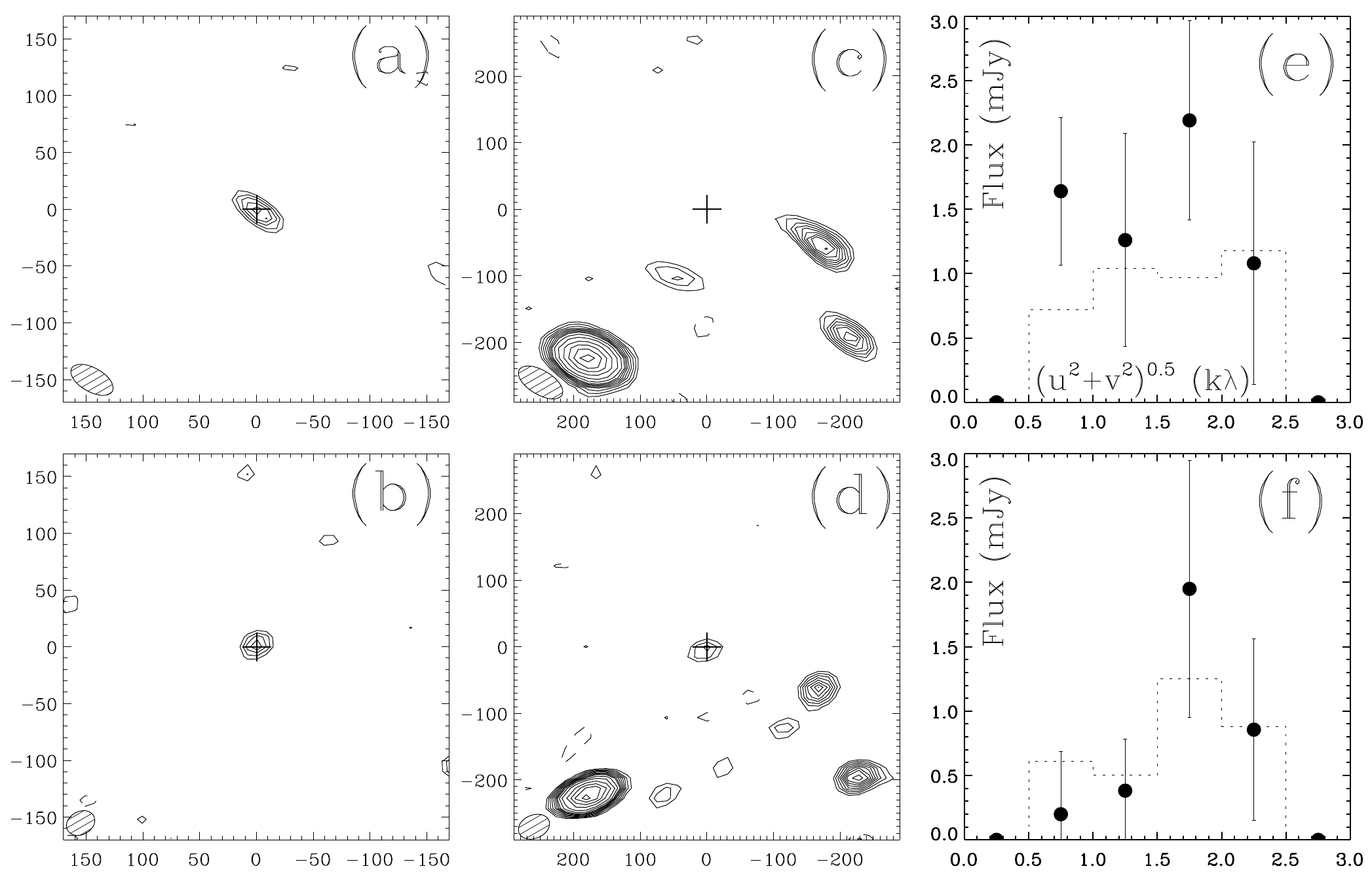}
\caption{Data demonstrating the radio variability of HD100546. Offsets in the images are from
the J2000 co-ordinates, and the cross marks the expected position after 6.77 years of proper
motion.
a) First half of the total 8.64~GHz data set, naturally weighted in the H214 configuration (without CA06) 
on 9 October 2006, with a synthesised beam of 41.3\arcsec$\times$21.0\arcsec at PA=60.5$^{\circ}$. Contours 
are in terms of the RMS of 0.0791~mJy/beam, with levels at 3, 4, 5, 6, 7$\sigma$. 
b) Second half of the total 8.64~GHz data set, naturally weighted in H214 (without CA06) on 9 October 2006, 
with a synthesised beam of 26.3\arcsec$\times$20.3\arcsec at PA=-60.7$^{\circ}$. Contours are in terms of the 
RMS of 0.06798~mJy/beam, with levels at 3, 4, 5, 6, 7$\sigma$. 
c) First half of the total 4.8~GHz data set, naturally weighted in H214 (without CA06) on 9 October 2006, 
with a synthesised beam of 74.8\arcsec$\times$36.4\arcsec at PA=$59.1^{\circ}$. Contours are in terms of the 
RMS of 0.0700~mJy/beam, with levels at 3, 4, 5, 6, 7, 8, 9, 10, 12, 15, 20, 25, 30, 35$\sigma$. 
d) Second half of the total 4.8~GHz data set, naturally weighted in H214 (without CA06) on 9 October 2006, 
with a synthesised beam of 48.8\arcsec$\times$34.9\arcsec at PA=$-67.2^{\circ}$. Contours are in terms of the 
RMS of 0.0637~mJy/beam, with levels at 3, 4, 5, 6, 7, 8, 9, 10, 12, 15, 20, 25, 30, 35$\sigma$.
e) Flux versus $u-v$ distance for first data set (75~mins) at 8.64~GHz on 31 August 2008 in H75. 
f) Flux versus $u-v$ distance for second data set (105~mins) at 8.64~GHz on 31 August 2008 in H75. 
Dashed lines in (e) and (f) represent the expected amplitude in the case of zero signal. Bin size is 
0.5~k$\lambda$. Clearly there is significant signal in (e) but little or no signal in (f), despite the 
latter's seemingly lower noise level.}
\label{HD100-radiovary}
\end{figure*}

However, hidden within this long-term invariance there is a suggestion that the HD100546 radio emission
changes on much shorter timescales, perhaps hourly or even shorter. This was first discovered during the 
measurements on 31 August 2008 (Table~\ref{TAB:hd100-obs36cm}), which indicate that the 3.5~cm flux had 
increased by a factor of $\geq$ 2 over its quiescent value, and then decreased again. See also
Figure~\ref{HD100-radiovary}-e,f.

There were some instrumental and weather-related (i.e. rain) issues on this day, which necessitated 
the three separate observations. But these seemingly had little, if any, adverse impact on the 
quality of the data. The phase behaviour was good even across the $\sim$ 4.5~km path to antenna 6. 
Also, the flux of the gain calibrator B1057-797 is constant for the three data sets, and the flux 
calibrator B0823-500 was detected at its expected brightness. Unfortunately, the sensitivity 
was inadequate to detect HD100546 at 4.8~GHz to assess potential variability at this frequency. But 
the radio jet galaxy (Figure~\ref{FIG-3716mm-normcal}-f) in the field was at a constant flux, again 
suggesting that the overall data quality is good.

To further investigate the possibility of short term radio variability the 9 May and 9 October 2006 
and 11 July 2007 data sets were separated into discrete subsets, as was done at 16~mm. On these dates 
the observations were conducted in overall better conditions. No statistically significant variation 
over $\geq$ 30~min intervals at 8.64~GHz was found on 9 May and 9 October 2006, whilst at 4.8~GHz the 
sensitivity on 9 May was inadequate to detect HD100546. 

However, variations at 4.8~GHz on as short a timescale as 40~mins could be discerned on 9 October, with 
the flux increasing by a factor of up to $\sim$ 3 above the quiescent value of 0.3--0.4~mJy (and as stated 
above with little or no concomitant change in the 8.64~GHz flux). See Figure~\ref{HD100-radiovary}-a,b,c,d.
There is also a suggestion that the 4.8~GHz flux can decrease from its quiescent value. For instance, on 
9 October the fluxes measured (in the $u-v$ plane) for the two halves were $0.17\pm0.08$ and $0.57\pm0.07$~mJy 
at 4.8~GHz (shown in Figure~\ref{HD100-radiovary}-c,d respectively), whilst at 8.64~GHz they were 
$0.64\pm0.09$ and $0.63\pm0.09$~mJy (Figure~\ref{HD100-radiovary}-a,b). Self-evidently the 'apparent' or 
'instantaneous' spectral index also changes, being 2.3$\pm$0.5 and 0.2$\pm$0.2 respectively. Similarly, 
on 11 July 2007 consecutive thirds of the data had 4.8~GHz fluxes of $0.34\pm0.07$, $0.06\pm0.08$ and 
$0.39\pm0.08$~mJy, but were stable at $0.42\pm0.07$, $0.45\pm0.08$ and $0.47\pm0.09$~mJy at 8.64~GHz, 
with 'apparent' spectral indices of $0.4\pm0.3$, $3.4\pm1.3$ and $0.3\pm0.3$ respectively.

The compact nature of the IR hydrogen recombination line emission suggests that any radio emission -- 
presuming a cause and effect connection -- would similarly be spatially unresolved. However, if there 
was extended structure associated with the radio emission of HD100546 then we might expect the apparent 
flux to vary during a 12~hour synthesis track as the beam size and shape (and thus spatial resolution) 
changes. The wind from HD100546 may well be extended, as Grady et al. (2005) found the (bipolar) Si~III 
disk wind to be extended over at least 100~AU, or 1 arcsecond at HD100546's distance. But this could 
only feasibly be relevant in the case of the 11 July 2007 data taken in a 6~km E--W configuration, 
although even then the extended structure would probably have to be larger than a few arcseconds in 
size. For the different 6.2~cm fluxes on 9 October 2006, or the different 3.5~cm fluxes on 31 August 
2008, taken in compact hybrid configurations, the extended emission would need to be larger than about 
40 and 70 arcseconds respectively. This is highly unlikely, and indeed we see no convincing evidence for
source extension in the amplitudes versus $u-v$ distance for the 6~km configuration data on 11 July 2007.

Numerous examples of variability have been seen for HD100546, occurring over different timescales of 
minutes, hours, days, months and years. Examples include accreting and outflowing gas profiles (Pogodin 
1995; Grady et al. 1997; Vieira et al. 1999; Deleuil et al. 2004; Giumaraes et al. 2006)), optical 
polarization (Yudin \& Evans 1998; Clarke et al. 1999), and Algol-like $uvby$ photometric minima 
coincident with a H$\alpha$ line profile variation (Vieira et al. 1999). 

\subsubsection[]{Variability mechanisms}

The aforementioned variability phenomena in other regions of the electromagnetic spectrum of HD100546 
could feasibly give rise to radio fluctuations. However, rigorously pursuing the origin for the variable 
radio emission is beyond the scope of this paper, and may not (yet) be entirely warranted by the paucity 
of the data. Even so, that the 4.80--8.64~GHz spectral index exhibits excursions both above and below the 
value of $\sim$ 0.6 for the quiescent fluxes suggests that either the dominant physical process for the radio 
emission changes or the opacity between the radio emitting region and Earth changes, or perhaps even both.
But whatever mechanism is behind the observed radio fluctuations of HD100546 it must also account for 
the instances of a decline in the 6.2~cm emission whilst the 3.5~cm flux remains constant or changes only 
very little (i.e. within the senstivity limit of our data). Presumably it is this characteristic which 
will be most discriminatory of the relevant process and/or the physical conditions within the region 
causing the changes. 
 
By analogy with T Tauri stars (TTS) and other Herbig stars, the review article of Gudel (2002) and the 
Herbig star radio survey of Skinner, Brown \& Stewart (1993) provide some guidance to what might be 
occurring in HD100546, and we briefly mention some possible processes. 

Radio flares and longer term variations have been observed for several TTS, identified with non-thermal
gyrosynchrotron emission. Magnetospheric disk accretion in TTS is a source of x-ray emission and has 
also been identified with radio variability. HD100546 is an x-ray source, but not especially bright in 
comparison to other Herbig stars, and in any case the origin of x-ray emission in these stars is still 
controversial (e.g. Hamidouche, Wang \& Looney 2008; Stelzer et al. 2006; Skinner et al. 2004). Potentially 
counting against a gyrosynchrotron-like process for HD100546 is that a magnetic field has not been detected, 
despite several attempts. A possible detection of a weak field of around 90$\pm$30~G was reported by Hubrig 
et al. (2009), but was not confirmed in previous and subsequent studies by Donati et al. (1997), Wade et al. 
(2007), Hubrig et al. (2013) and Bagnulo et al. (2012). The latter authors found an upper limit of $\le$ 
50~G, much lower than the $\geq$ several hundred Gauss fields of TTS.

If it is not obvious that a magnetic process is behind the HD100546 radio variability then other 
possibilities are accretion- or shock-related emission. In the former case, enhanced accretion could 
be a result of discrete events such as infalling comet-like bodies, as has already been postulated for 
HD100546 (e.g. Grady et al. 1997). Such enhanced accretion would also presumably elevate the outflow 
rate, which may either impact surrounding material or develop an instability and thus produce a shock. 
In both cases the spectral index is expected to be flat, i.e. $-0.1$, which could be consistent with 
some of the observed radio variations, or at least with the fact that the changes are relatively small 
at both radio frequencies. But the magnitude of the changes expected for accretion seem much larger 
than those observed, since even for a modest increase in the stable accretion rate the expected 
emission rises proportionally. A proportional increase of emission with outflow rate is also the case 
for a shock process, but at least in this case the magnitude of shock-induced radio emission is in 
better agreement with the observed radio fluxes. 

Perhaps instead it is as simple as a scenario where the mechanism responsible for the quiescent radio
fluxes of HD100546, namely free-free wind emission, is also the one behind the variations. In this case
a relatively small increase in the outflow rate has a proportionally larger effect on the radio emission 
(which scales to the 4/3 power of the outflow rate). Potentially supportive of this idea is that the IR 
hydrogen recombination line fluxes -- also diagnostic of the wind -- are possibly different between the 
ISO measurements reported here and those of Geers et al. (2007). As calculated above, the difference -- 
indicated by the Br$\alpha$ line -- is within the range observed in the 6.2~cm radio emission. 

\section[]{References for stellar and disk parameters in Table~3}

References are given here for the stellar and disk properties of each transition disk system included in 
Table~\ref{TAB:ExoBD-Limits}. \\

\noindent
Stellar properties (distance, age, mass and accretion rate) are taken from: \\
Costigan et al. (2014) for RY~Tau, UX~Tau~A, BP~Tau \\
Garufi et al. (2014) for FT~Tau \\
Manara et al. (2014) for LkH$\alpha$~330, DM~Tau, LkCa~15, GM~Aur, RX~J1615, SR~21, WSB~60, DoAr~44 \\
Mendigutia et al. (2014) for HD142527 \\
Ingleby et al. (2013) for BP~Tau, DM~Tau, GM~Aur, LkCa~15 \\
Salyk et al. (2013) for CI~Tau \\
Donehew \& Brittain (2011) for CQ~Tau, MWC~758 \\
Mendigutia et al. (2011) for CQ~Tau, RY~Tau \\
Pogodin et al. (2012) for HD100546, HD135344B \\
Espaillat et al. (2010) for LkCa~15, DoAr~44, UX~Tau~A \\ 
Isella et al. (2009) for DM~Tau, LkCa~15, RY~Tau, GM~Aur, SR~24~S \\
Najita et al. (2007) for UX~Tau~A \\
Blondel \& Tjin~A~Djie (2006) for HD169142, MWC~758, CQ~Tau\\ 
Garcia Lopez et al. (2006) for HD169142, HD135344B, HD142527, CQ~Tau \\
Natta et al. (2006) for SR~24S, SR~21, WSB~60\\
Calvet et al. (2005) for DM~Tau, GM~Aur \\
Calvet et al. (2004) for RY~Tau \\
White \& Ghez (2001) for GM~Aur, DS~Tau, LkCa~15, BP~Tau, CI~Tau, DM~Tau \\
Gullbring et al. (1998) for BP~Tau, DS~Tau, GM~Aur \\
Hartmann et al. (1998) for BP~Tau, DS~Tau, GM~Aur, CI~Tau, DM~Tau, LkCa~15 \\ 
Valenti et al. (1993) for BP~Tau, CI~Tau, DM~Tau, DS~Tau, GM~Aur \\

\noindent
Disk properties (visibility nulls, radius in large dust, small dust and gas tracers), plus some stellar properties, 
are taken from: \\
Canovas et al. (2015), Tsukagoshi et al. (2014), Romero et al. (2012) for Sz~91 \\
Follette et al. (2015), Bruderer et al. (2014), van der Marel et al. (2013), Brown et al. (2012), 
Geers et al. (2007) for Oph~IRS48 \\
Huelamo et al. (2015), Olofsson et al. (2011) for T~Cha \\
Perez et al. (2015), Avenhaus et al. (2014b), Cassasus et al. (2013), Fukugawa et al. (2013), Cassasus et al. (2012), 
Verhoeff et al. (2011), Ohashi et al. (2008), Cassasus \& Wright et al. (in preparation, for first visibility null 
at 34 GHz) for HD142527 \\
van der Marel et al. (2015) for SR~21, HD135344B, LkCa~15, RX J1615-3255, SR~24~S, J1604-2130 \\
Isella et al. (2014), Thalmann et al. (2014), Andrews et al. (2011b), Pietu et al. (2007, 2006) for LkCa~15 \\
Osorio et al. (2014), Quanz et al. (2013b), Honda et al. (2012) for HD169142 \\
Perez et al. (2014) for HD135344B, SR~21 \\
Pietu et al. (2014) for DS~Tau \\
Zhang et al. (2014), Mathews et al. (2012), Mayama et al. (2012) for J160421.7-213028 \\
Follette et al. (2013), Pontoppidan et al. (2008), Eisner et al. (2009) for SR~21 \\
Garufi et al. (2013), Muto et al. (2012), Lyo et al. (2011), Pontoppidan et al. (2008) for HD135344B \\
Isella et al. (2013), Pontoppidan et al. (2011), Brown et al. (2008) for LkH$\alpha$~330 \\
Maaskant et al. (2013) for HD169142, HD135344B, Oph~IRS48 \\
Takami et al. (2013), Isella, Carpenter \& Sargent (2010a), Akeson et al. (2005) for RY~Tau \\
Trotta et al. (2013), Banzatti et al. (2011) for CQ~Tau \\
Cieza et al. (2012), Orellana et al. (2012) for J1633.9-2422 \\
Dong et al. (2012), Hashimoto et al. (2013, 2012) for PDS~70 \\
France et al. (2012) for BP~Tau, DM~Tau, GM~Aur, HD135344B, LkCa~15, UX~Tau~A \\
Tanii et al. (2012) for UX~Tau~A \\
Andrews et al. (2011a) for MWC~758, HD135344B, LkH$\alpha$~330, SR~21, UX Tau~A, SR~24~S, DoAr~44, LkCa~15, 
RX~J1615-3255, GM~Aur, DM~Tau, WSB~60 \\
Espaillat et al. (2011) for DM~Tau, GM~Aur, LkCa~15, RY~Tau, UX~Tau~A \\
Gr\"afe et al. (2011) for DM~Tau, GM~Aur \\
Guilloteau et al. (2011) for BP~Tau, CI~Tau, CQ~Tau, DM~Tau, FT~Tau, GM~Aur, LkCa~15, MWC~758 \\
Andrews et al. (2010) for SR~24~S \\
Espaillat et al. (2010) for LkCa~15, UX~Tau~A, DoAr~44, GM~Aur, DM~Tau \\
Isella et al. (2010b), Isella et al. (2008) for MWC~758 \\
Merin et al. (2010) for RX~J1615.3-3255 \\
Andrews et al. (2009) for SR~21, WSB~60, DoAr~44 \\
Brown et al. (2009) for LkH$\alpha$~330, SR~21~N, HD135344B \\
Hughes et al. (2009), Dutrey et al. (2008) for GM~Aur \\
Brown et al. (2007) for LkH$\alpha$~330, SR~21~N, HD135344B, T~Cha \\
Eisner et al. (2004) for MWC~758, CQ Tau \\


\label{lastpage}

\end{document}